\documentclass[preprint2]{aastex6}

\newcommand{\kms}{km~s$^{-1}$}
\usepackage{captcont}
\usepackage{amsmath}
\shortauthors{Long et al.}
\usepackage{CJKutf8}

\usepackage{verbatim}

\begin{document}
\begin{CJK*}{UTF8}{gbsn}

\title{An ALMA Survey of CO isotopologue emission from Protoplanetary Disks in Chamaeleon I}
\author{Feng Long(龙凤)\altaffilmark{1,2}, Gregory J. Herczeg(沈雷歌)\altaffilmark{1}, Ilaria Pascucci\altaffilmark{3,9}, Emily Drabek-Maunder\altaffilmark{4}, Subhanjoy Mohanty\altaffilmark{4}, Leonardo Testi\altaffilmark{5}, Daniel Apai\altaffilmark{3,6,9}, Nathan Hendler\altaffilmark{3}, Thomas Henning\altaffilmark{7}, Carlo F. Manara\altaffilmark{8}, Gijs D. Mulders\altaffilmark{3,9}}

\altaffiltext{1}{Kavli Institute for Astronomy and Astrophysics, Peking University, Yi He Yuan Lu 5, Haidian Qu, 100871 Beijing, China}
\altaffiltext{2}{Department of Astronomy, School of Physics, Peking University, 100871 Beijing, China}
\altaffiltext{3}{Lunar and Planetary Laboratory, The University of Arizona, Tucson, AZ 85721, USA}
\altaffiltext{4}{Imperial College London, London SW7 2AZ, UK}
\altaffiltext{5}{ESO/European Southern Observatory, Garching bei M{\"u}nchen, Germany}
\altaffiltext{6}{Steward Observatory, The University of Arizona, Tucson, AZ 85721, USA}
\altaffiltext{7}{Max Planck Institute for Astronomy, Heidelberg, Germany}
\altaffiltext{8}{Scientific Support Office, Directorate of Science, European Space Research and Technology Centre (ESA/ESTEC), Noordwijk, The Netherlands}
\altaffiltext{9}{Earths in Other Solar Systems Team, NASA Nexus for Exoplanet System Science}

\begin{abstract}
The mass of a protoplanetary disk limits the formation and future growth of any planet. Masses of protoplanetary disks are usually calculated from measurements of the dust continuum emission by assuming an interstellar gas-to-dust ratio.  To investigate the utility of CO as an alternate probe of disk mass, we use ALMA to survey $^{13}$CO and C$^{18}$O J = $3-2$ line emission from a sample of 93 protoplanetary disks around stars and brown dwarfs with masses from 0.03 -- 2 M$_{\odot}$ in the nearby Chamaeleon I star-forming region.  We detect $^{13}$CO emission from 17 sources and C$^{18}$O from only one source.  Gas masses for disks are then estimated by comparing the CO line luminosities to results from published disk models that include CO freeze-out and isotope-selective photodissociation. Under the assumption of a typical ISM CO-to-H$_2$ ratios of $10^{-4}$, the resulting gas masses are implausibly low, with an average gas mass of $\sim$ 0.05 M$_{Jup}$ as inferred from the average flux of stacked $^{13}$CO lines. The low gas masses and gas-to-dust ratios for Cha I disks are both consistent with similar results from disks in the Lupus star-forming region.  The faint CO line emission may instead be explained if disks have much higher gas masses, but freeze-out of CO or complex C-bearing molecules is underestimated in disk models. The conversion of CO flux to CO gas mass also suffers from uncertainties in disk structures, which could affect gas temperatures.  CO emission lines will only be a good tracer of the disk mass when models for C and CO depletion are confirmed to be accurate.
\end{abstract}

\section{Introduction} \label{sec:intro}
The formation and migration of planets within protoplanetary disks likely contribute to the rich diversity of exoplanet architectures (see review by \citealt{WinnFabrycky2015}).  The initial disk mass, the evolution of the gas-to-dust ratio, the evolution of gas surface density distribution with time and the gas dispersal timescale all influence how many and what types of planets can form in a stellar system \citep[e.g.,][]{pollack1996,boss1997,Mordasini2012,alexander2014}. Statistical studies reveal possible correlations between stellar mass and planet occurrence rates, including a higher frequency of small planets in close orbits around M-dwarfs than around FGK stars \citep{howard2012,Mulders2015} and a higher occurrence rate of giant planets around solar-type stars than around M-dwarfs \citep{Johnson2010}, although this correlation may not apply to hot Jupiters \citep{obermeier2016}. While these relationships likely originate in correlations between stellar mass and disk properties, linking disk properties to the outcome of planet formation requires an unbiased census of initial disk masses and the evolution of disk mass and structure with time.

Recent sub-mm surveys of disks demonstrate that the dust mass in disks is strongly correlated with stellar mass in individual regions with ages between 1--10 Myr \citep{andrews2013,mohanty2013,ansdell2016,ansdell2017,barenfeld2016,pascucci2016}. The correlation of dust mass with stellar mass is steeper in the 5--10 Myr Upper Sco association than in the younger regions, which when combined with a much lower fraction of stars with disks in Upper Sco \citep{LuhmanMamajek2012} suggests that dust in disks evolve quickly and with a dependence on the mass of the central star.  However, the estimates of dust mass rely on the dust opacity, which depends on the dust grain size distribution and composition \citep[e.g.,][]{Beckwith2000}, and on the assumed dust temperature \citep[e.g.,][]{andrews2013,pascucci2016,vanderPlas2016}.  In addition, the total disk mass is usually converted from dust mass using the gas-to-dust ratio of $\sim$~100 for the interstellar medium \citep{Bohlin1978}. During disk evolution, gas and dust may decouple from each other, as seen in differences in their spatial distributions in some disks \citep[e.g.,][]{isella2007,Andrews2012,vandermarel2013,perez2015}.

Independent gas mass measurements are needed as observational evidence to better understand gas and dust evolution.  Because H$_2$ does not emit at cold temperatures, CO isotopologues are widely used to probe the gas content of disks.  CO chemistry has been studied extensively and the pure rotational CO lines are readily detectable at millimeter wavelengths \citep{henning2013}. However, as with dust, the conversion of a CO line luminosity to a total gas mass is plagued by uncertainties, primarily in the gas-phase CO-to-H$_2$ abundance ratio and in isotopologue ratios.  Even for C$^{18}$O, line opacities can be high is some disks \citep[e.g.][]{yu2017}.

CO gas depletes through freeze-out at the disk mid-plane \citep[e.g.,][]{Dutrey1997,vanZadlehoff2001}, reducing the CO-to-H$_2$ abundance ratio. CO gas is also photodissociated in the warm surface layer of the disk \citep[e.g.][]{aikawa2002,kamp04,gorti2011,walsh12}, in which the less abundant  C$^{18}$O and $^{13}$CO are selectively dissociated by ultraviolet radiation, while $^{12}$CO may effectively self-shield. This isotope-selective photodissociation therefore modifies the abundance ratios of the CO isotopologues and the conversion from rare isotopologues to the CO gas mass \citep{vanDishoeck1988,lyons05,smith2009}.

When CO freeze-out and isotope-selective photodissociation are implemented into chemical models of disks, the conversion between line luminosities and disk mass change by as much as two orders of magnitude \citep[e.g.][]{woitke2009,miotello2014}. \citet{miotello2016} (hereafter, MvD16) provides conversion factors from CO isotopologue line luminosity to disk mass, calculated from thermo-chemical models of static disks that include freeze-out and isotope-selective photodissociation, as implemented from the DALI code \citep{Bruderer2013,miotello2014}. In an earlier, simplified approach, \citet{WilliamsBest2014} (hereafter, WB14) developed parametric models of disk properties, constraining the CO freeze-out at T $<$ 20 K and estimating isotope-selective photodissociation by a reduction of C$^{18}$O abundance. The gas mass is then derived by comparing the observed CO isotopologue line luminosity with their simulated line luminosity.

A possible missing ingredient in these static disk models is carbon depletion into complex molecules, which may then freeze out.  Carbon depletion, in excess of that estimated from freeze-out or photosdissociation, is inferred from the comparison of HD and CO gas mass \citep{bergin2013,Favre2013,schwarz16,McClure2016} and from observations of neutral C and CO emission \citep{kama2016}.  To account for these discrepancies, elemental carbon may be sequestered from CO into more complex carbon chains or CO$_2$, which is then locked up in ice grains \citep{Favre2013,du2015,kama2016b}. Similar depletion processes also apply to oxygen volatiles, as suggested from water observations \citep[e.g.][]{du2017}. The depletion of CO and other volatiles may be accelerated by vertical mixing, because the turbulence in the midplane is much lower than at the disk surface \citep{krijt2016,xu2016}.

In this work, we use ALMA to survey $^{13}$CO and C$^{18}$O line emission from disks in the young ($\sim$ 2 Myr, \citealt{Luhman2004}), nearby ($188\pm12$ pc, see Appendix \ref{gaia})\footnote{The companion paper \citet{pascucci2016} used a pre-Gaia distance of 160 pc.  In this paper, all luminosities and disk masses are re-calculated using this updated distance.  The stellar masses adopted from \citet{pascucci2016} are not recalculated in this paper, since most of our sample are low mass stars on the vertical Hayashi tracks.} Chamaeleon I star-forming region, hereafter Cha I.  Our results and interpretations are broadly consistent with the analysis of CO emission from disks in the Lupus star-forming region by \citet{ansdell2016} and recent interpretation of those data by \citet{miotello2016b}. We first describe our ALMA observations and data reduction in Section \ref{sec:observe}. The methods developed for measuring the line fluxes and upper limits are presented in Section \ref{sec:method}. The CO gas properties and disk mass inferred from the weak CO emission and disk models are described in Section \ref{sec:detection}--\ref{sec:mass}. We compare our results with other star-forming regions and discuss disk evolution and its implications for planet formation theories in Section \ref{sec:lupus} and \ref{sec:implication}.  We summarize our findings in Section \ref{sec:conclusion}.

\section{ALMA observations} \label{sec:observe}
We use ALMA Band 7 to survey the Class II protoplanetary disks in Cha I in the Cycle 2 program 2013.1.00437 (PI: I. Pascucci). The sample was split into shallow observations of 54 stars with spectral type (SpTy) equal or earlier than M3, hereafter referred to as the Hot sample, and deeper observations of 39 stars with later SpTy, referred to as the Cool sample. The sample, observation calibrator setups, and results of the dust continuum are described in detail by \citet{pascucci2016}. One object, 2MASS J11183572-7935548, is likely a member of $\epsilon$ Cha \citep{Luhman2008,Murphy2013,Lopez2013} and has been excluded from the global analyses of disks in this work.

In this paper, we focus on observations from one baseband, which was split into segments centered at 330.6 GHz to target the $^{13}$CO J = 3-2 and at 329.3 GHz to target the C$^{18}$O J = 3-2. The correlator was configured to cover both lines, each with a bandwidth of 117.2 MHz and a channel separation of 0.122 MHz (0.11 km s$^{-1}$). The three other spectral windows were configured for continuum observations. For each target, two sets of observations were executed for a total integration time of 24 s per source for the Hot sample and 120 s per source for the Cool sample. The observations were performed in good weather conditions with a precipitable water vapor (PWV) of $\sim$ 0.6--0.9 mm.

The ALMA data are calibrated using the \textit{Common Astronomy Software Application} (CASA) package and following the data reduction scripts provided by NRAO, including flux, phase, bandpass, and gain calibrations. The absolute flux scale has a systematic uncertainty of 10$\%$. The two executions of each object are then concatenated using the \textit{concat} task after flux calibration. Continuum emission is subtracted in the uv-plane from the spectral windows containing CO isotopologue lines using \textit{uvcontsub} task. The spectral line data cubes are then created with the \textit{clean} algorithm using the continuum-subtracted visibilities.  Considering the weakness of line emission and the low detection rate in our sample, natural weighting is chosen in \textit{clean} to favor shorter baselines and hence achieve a higher signal-to-noise ratio (S/N).

In general, longer baseline visibilities trace smaller scale structures and shorter baseline visibilities trace extended emission from larger scale.  Since our primary objective is to measure the total flux from the disk, all baselines are included in \textit{clean}.  This choice also provides consistency with the dust analysis of the same sample by \citet{pascucci2016}.
Excluding the visibilities with baselines $<$ 40 k$\lambda$ would avoid large-scale ripples from any extended emission but will also lower the calculated line fluxes in most cases.  In alternate reductions that exclude baselines shorter than $< 40$ k$\lambda$, the flux differences are within uncertainty in most cases, although the two sources 2MASS J11075792-7738449 and 2MASS J11095340-7634255 would be much fainter.  The $^{13}$CO detection rates would also not be affected.

The resulting synthesized beam for the CO data cube is 0.$\arcsec$7$\times$0.$\arcsec$5  for both samples, the same as the continuum data. The CO transitions are imaged at a spectral resolution of 0.25 km s$^{-1}$, and reach the 1$\sigma$ noise rms of 30 (43) mJy beam$^{-1}$ and 107 (126) mJy beam$^{-1}$ for $^{13}$CO (C$^{18}$O) in the Cool and Hot samples, respectively. The slightly higher noise in C$^{18}$O may be caused by placing the C$^{18}$O line closer to the band edge than $^{13}$CO, resulting in a lower instrument response.

\section{Measuring CO Line Fluxes and Upper Limits } \label{sec:method}
Only a few strong $^{13}$CO detections are easily identified from line profiles. Therefore, we develop a method to uniformly identify weak detections and measure line fluxes, uncertainties, and upper limits from the continuum-subtracted cleaned images. We first stack the images and spectra from the full sample to identify the emission region and velocity range used for uniform flux measurements for each source. The flux uncertainties are calculated by using the full spectral images outside the CO emission range, which yields lower uncertainties and better consistency within the Hot and Cool samples than if only the CO spectral channels would have been used. For sources with CO detections, the extraction regions are adjusted to calculate final fluxes that include the full spatial extent of the emission.  The following subsections describe each step in this method.

\begin{figure*}[t]
    \includegraphics[width=0.95\textwidth]{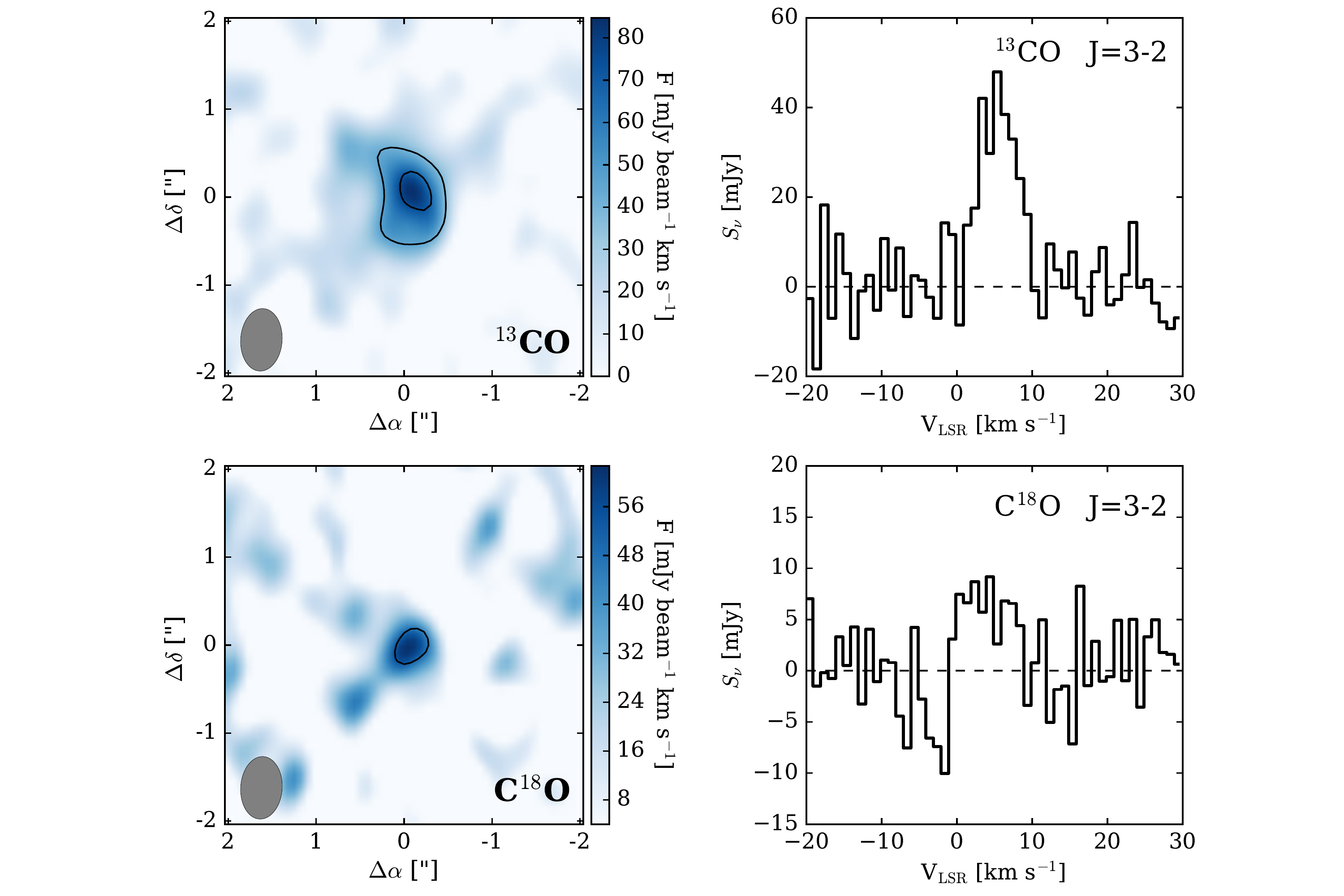}
    \caption{Left: The stacked image of $^{13}$CO and C$^{18}$O from all 93 sources, weighted by rms and corrected for offsets, with 3$\sigma$ and 5$\sigma$ contours in solid black line. The typical beam size is shown in the left corner. Right: The stacked spectrum for $^{13}$CO and C$^{18}$O from all 93 sources, binned to 1 km s$^{-1}$. \label{fig:stacking}}
\end{figure*}

\subsection{Defining the Extraction Region} \label{sec:extraction region}
We obtain initial CO images by integrating over the velocity range of 0 -- 10 km s$^{-1}$ over which emission is detected (see Figure \ref{fig:stacking}, right panels). The images are shifted to the centroid position of the continuum emission by applying the phase-center offsets from \citet{pascucci2016}, weighted by the noise (not noise squared, so that each observation contributes equally to the final stacked image), and then stacked into a single image.  For sources that were not detected in continuum emission, a phase-center (position) offset of $-0\farcs3$ in right ascension and $0\farcs0$ in declination were adopted from the median phase-center offset of continuum detections from \citet{pascucci2016}.

The stacked images for $^{13}$CO and C$^{18}$O are shown in the left panels of Figure \ref{fig:stacking}, with the peak S/N per beam of 5.8 and 3.5, respectively. Based on the 3$\sigma$ contours, we adopt extraction radii of $\sim$ 0.$\arcsec$6 for $^{13}$CO and 0.$\arcsec$3 for C$^{18}$O, both centered at the continuum position. Stacked images from the continuum-detected sources yield similar spatial extent but higher peak S/N.  Although $^{13}$CO and C$^{18}$O should be emitted from the same disk emission area, the marginal detection in C$^{18}$O leads us to choose a smaller extraction radii (with a total aperture size comparable to one beam size) to optimize the detection rate of weak signals.

The spectra are then extracted from these circular apertures centered at the measured or expected continuum source position from \citet{pascucci2016}.  Because only a few sources show a clear spectral profile, we also stack all spectra, weighted by rms and binned by four channels to a spectral resolution of 1 km s$^{-1}$. The upper right panel of Figure \ref{fig:stacking} shows that the stacked  $^{13}$CO emission covers from 0 -- 10 km s$^{-1}$, which is adopted as the extraction velocity range for flux calculations. The centroid velocity of $^{13}$CO emission of 5.4 km s$^{-1}$ is consistent with the radial velocity measurements of an average LRSK value of $\sim$ 5 km s$^{-1}$ in Cha~I sources \citep[e.g.][]{vankempen2009,Nguyen2012}. Since the stacked C$^{18}$O line lacks sufficient S/N to measure the spectral profile, we assume that C$^{18}$O and $^{13}$CO have the same emission velocity.

\begin{figure*}[t]
 \centering
  \includegraphics[width=0.95\textwidth]{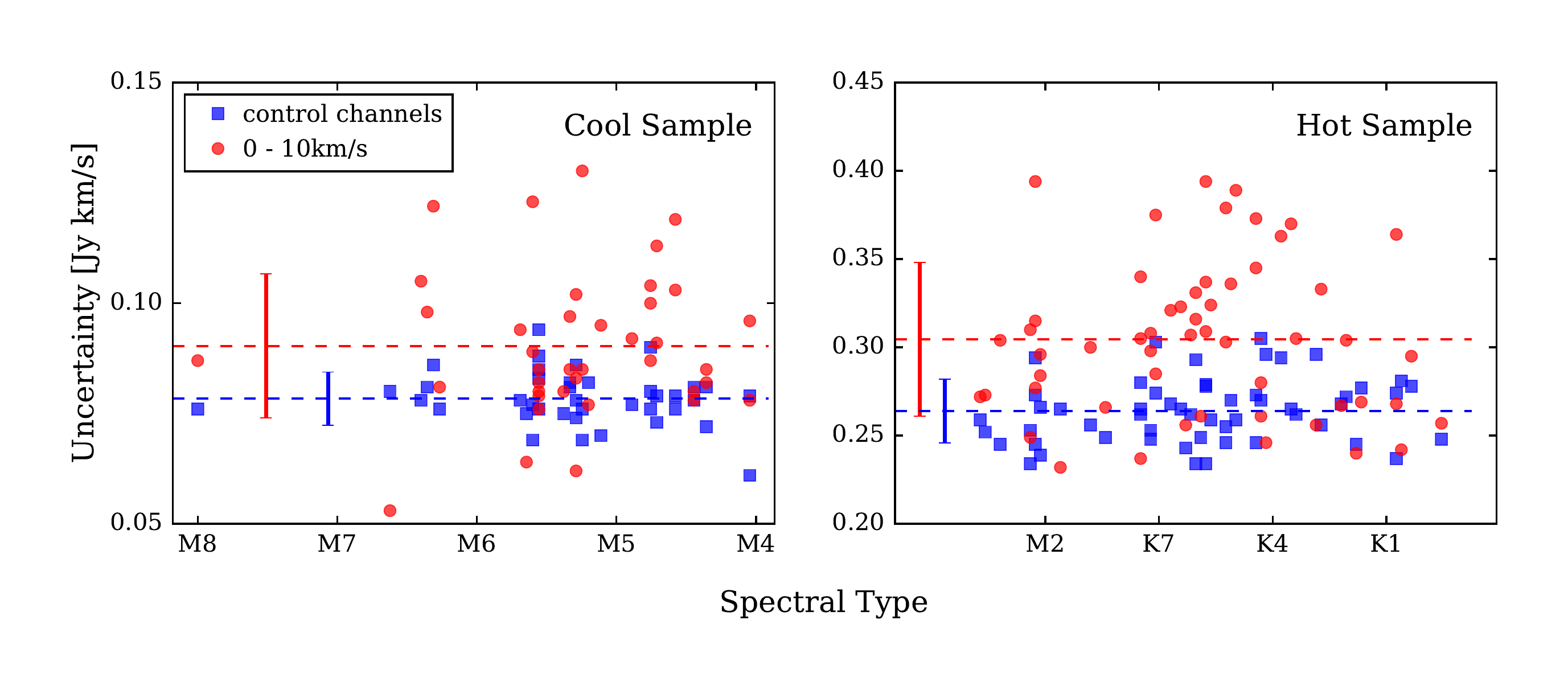}
  \caption{The comparison of uncertainties from 0 -- 10 km s$^{-1}$ and mean value of six control channels for Cool (Left) and Hot (Right) sample. The mean (dashed line) and 1$\sigma$ standard deviation (solid vertical line) are marked for 0 -- 10 km s$^{-1}$ range in red and control channels in blue. \label{fig:uncertainty} }
\end{figure*}

\begin{deluxetable}{rccccc}
\tablecaption{Uncertainty comparison\label{tab:uncertainty}}
\tablehead{
\colhead{Samples} &
\multicolumn{2}{c}{Mean} &
\colhead{} &
\multicolumn{2}{c}{Standard deviation} \\
\cline{2-3}
\cline{5-6}
\colhead{} & \colhead{Hot} & \colhead{Cool} & \colhead{} & \colhead{Hot} & \colhead{Cool} }
\startdata
0 -- 10 km s$^{-1}$   & 0.304 & 0.090 & & 0.043 & 0.016 \\
Control channels & 0.264 & 0.078 & & 0.018 & 0.006 \\
\enddata
\tablecomments{The mean value and standard deviation of uncertainties in the Hot and Cool sample, in unit of Jy km s$^{-1}$ }
\end{deluxetable}

\subsection{Measuring the uncertainties in fluxes}

The spatial and spectral extraction regions defined in \S \ref{sec:extraction region} will be applied in \S \ref{sec:significance} to measure fluxes and upper limits in the $^{13}$CO line in the 0--10 \kms\ spectral range and $0\farcs6$ radii apertures ($0\farcs3$ for C$^{18}$O). In other studies, the uncertainty is usually measured from the noise level in the same velocity channels as the flux measurements.  In this section, we demonstrate that this approach yields large inconsistencies across the sample.  We instead develop a method to use the full spectral range to assess uncertainties that are uniform within the hot and the cool sample.

Uncertainties in fluxes are measured by calculating the standard deviation of fluxes in randomly distributed apertures with the same extraction area and 10 \kms\ velocity widths.  The velocity ranges used for uncertainty calculations are spread across the full spectral band in six independent ranges\footnote{The six velocity ranges are -45 $\sim$ -35, -30 $\sim$ -20, -15 $\sim$ -5, 15 $\sim$ 25, 25 $\sim$ 35, 35 $\sim$ 45 km s$^{-1}$ }, hereafter called control channels, and from 0 -- 10 \kms, called the emission channels. In each 10 \kms\ velocity range from the control channels, fluxes are extracted from twenty circular apertures, each placed at random central location across the image.  In the 0 -- 10 \kms\ emission channels, the aperture centers are always located at a distance of two radii from the source center.

The mean value and standard deviation of uncertainties from each sample are shown in Figure \ref{fig:uncertainty} and Table \ref{tab:uncertainty}.  When only the 0 -- 10 km s$^{-1}$ range is used, the standard deviation of the uncertainties is 2.5 times higher than if the control channels are used. The large scatter of uncertainties in the 0 -- 10 km s$^{-1}$ channels would render any uniform approach invalid, either by missing detections due to an anomalously high flux uncertainty or by yielding false detections when the uncertainty is underestimated.  The uncertainty difference may be the result of the \textit{clean} process during which the emission structure can introduce additional pattern noise spreading in these channels, although in a few cases the noise is lower in the emission channels. Therefore, we adopt the mean value from the six control channels as the flux uncertainty for each source to provide better consistency in the calculated flux uncertainties between objects.

\begin{figure*}[t]
    \includegraphics[width=0.51\linewidth]{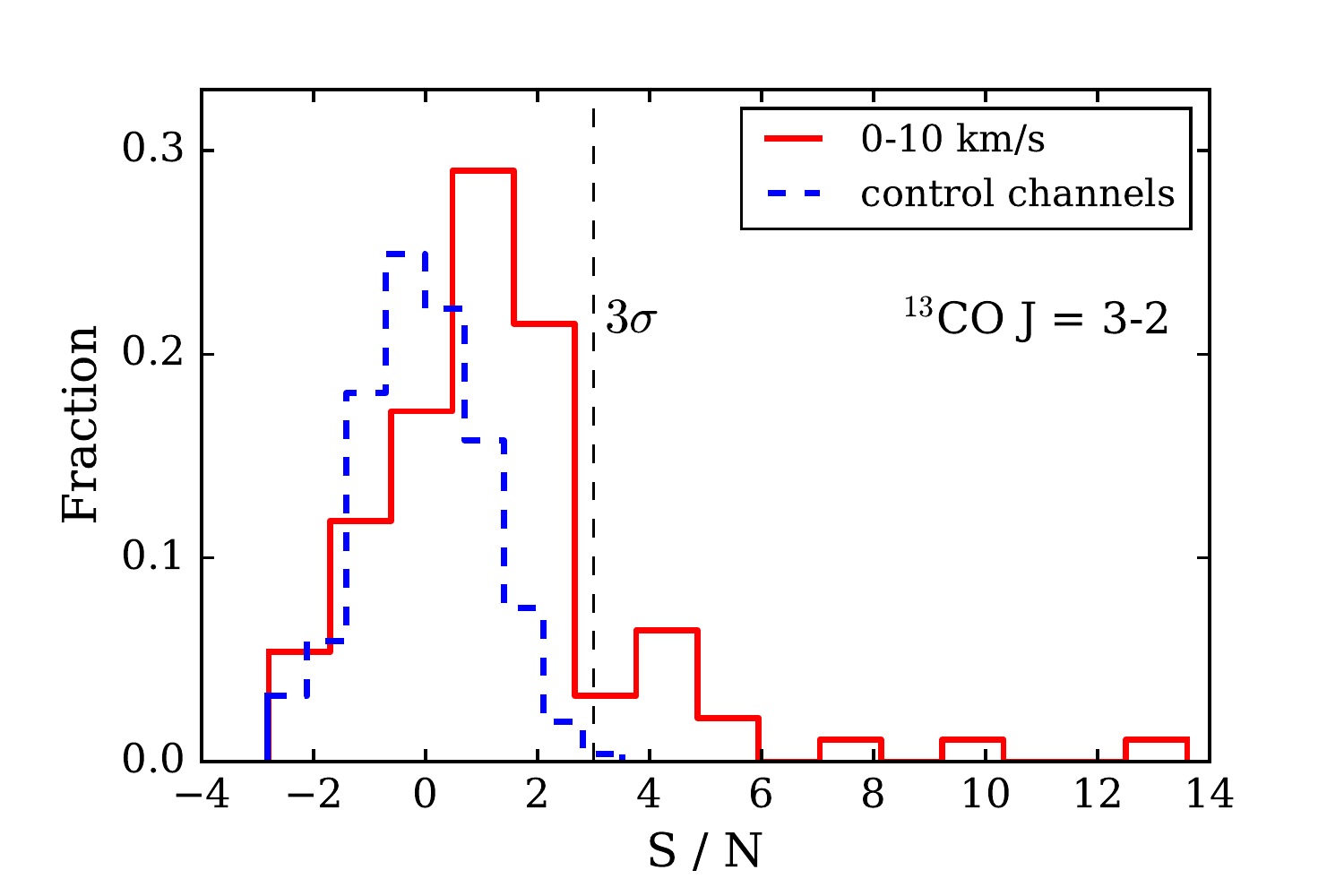}
    \includegraphics[width=0.51\linewidth]{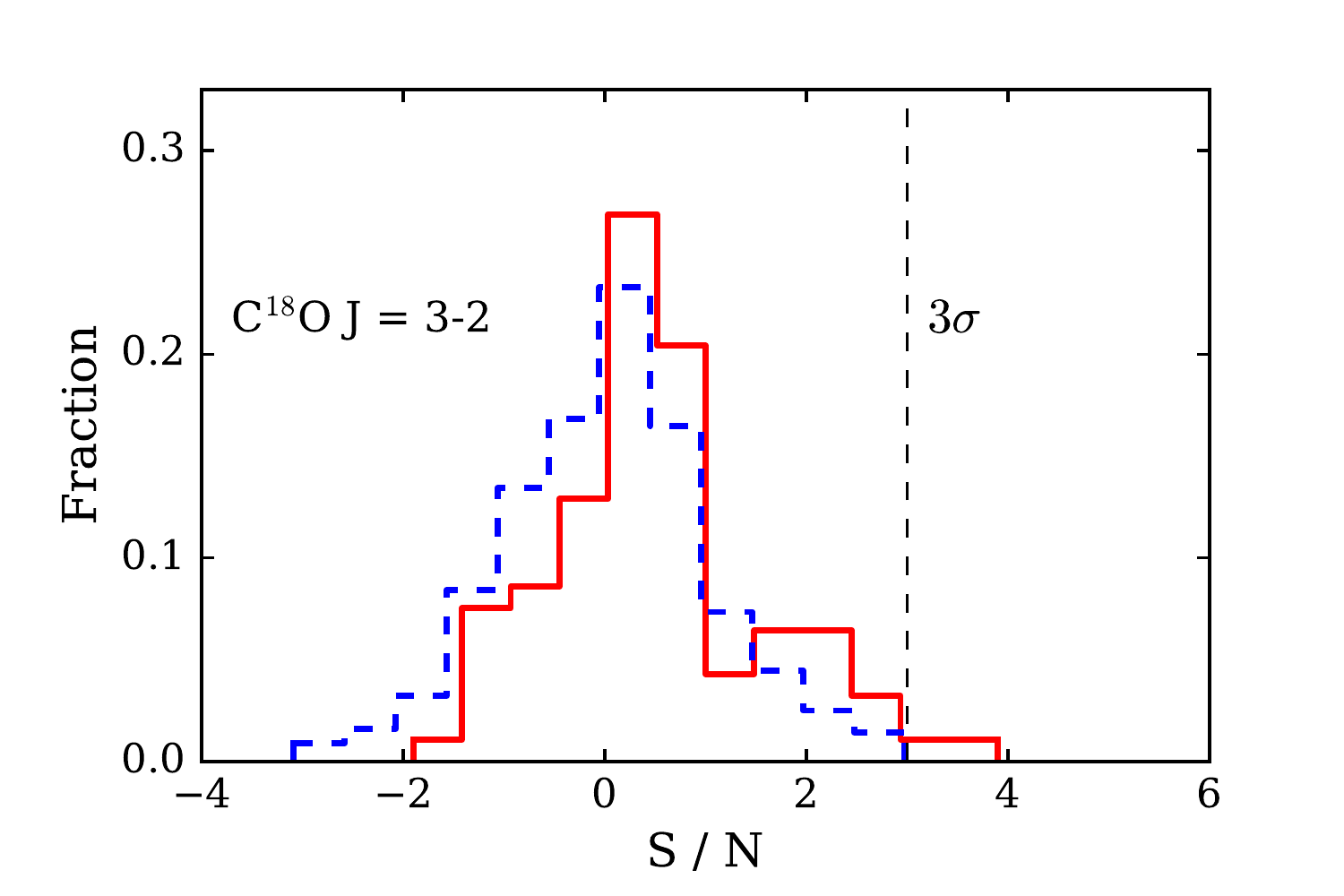}
  \caption{The normalized histogram of ratios of flux to uncertainty (S/N) for $^{13}$CO (Left) and C$^{18}$O (Right). Dashed blue line is for the control channels and solid red line for 0 - 10 km s$^{-1}$ channels. The vertical dash lines indicate the 3$\sigma$ significance in $^{13}$CO and in C$^{18}$O. \label{fig:3sigma} }
\end{figure*}

\subsection{Quantifying the Significance of Detections} \label{sec:significance}
To simulate a false detection probability and to determine the confidence level for detections, we compare the flux measurements centered on the CO lines with background fluxes from the six control channels for each of the 93 targets. The background fluxes are calculated within the measured (or expected) continuum center in each 10 km s$^{-1}$ velocity range from the control channels with a $0\farcs6$ radius aperture in $^{13}$CO data cube ($0\farcs3$ for C$^{18}$O), resulting in a total sample of 558 fluxes. The uncertainty for each target is adopted as discussed above.  The normalized histograms of flux-to-uncertainty ratios (S/N) in 0 -- 10 km s$^{-1}$ and in the control channels are shown in Figure \ref{fig:3sigma}. The 558 data points from the control channels in $^{13}$CO and in C$^{18}$O are distributed in a Gaussian-like shape. Only one of 558 $^{13}$CO and two of 558 C$^{18}$O data points have S/N larger than 3, consistent with the expectation for the frequency of a 3$\sigma$ deviation from the mean in a Gaussian distribution. Therefore, we define $^{13}$CO and C$^{18}$O detections as any source exceeding 3$\sigma$ significance from the 0 -- 10 km s$^{-1}$ emission channels, respectively.  With this cutoff, detections have a confidence level of $\sim 99.8$\%.

Fluxes are then measured in the $^{13}$CO line using a $0\farcs6$ radius aperture ($0\farcs3$ radius aperture for C$^{18}$O), and with uncertainties applied from the analysis of the control channels. The distribution of S/N in  $^{13}$CO line includes a long tail exceeding 3$\sigma$. In the distribution of C$^{18}$O fluxes, only one 4$\sigma$ outlier stands out as a significant detection. Furthermore, while the distributions in the control channels are centered at zero significance, the peak positions of the distributions in 0 -- 10 km s$^{-1}$ range fall at 1.5$\sigma$ and 0.5$\sigma$ in $^{13}$CO and C$^{18}$O, respectively. Many sources likely have $^{13}$CO fluxes that are just below our detection limits and would have been detected with 2--4 times better sensitivity.

\begin{figure*}[t]
    \includegraphics[width=0.95\textwidth]{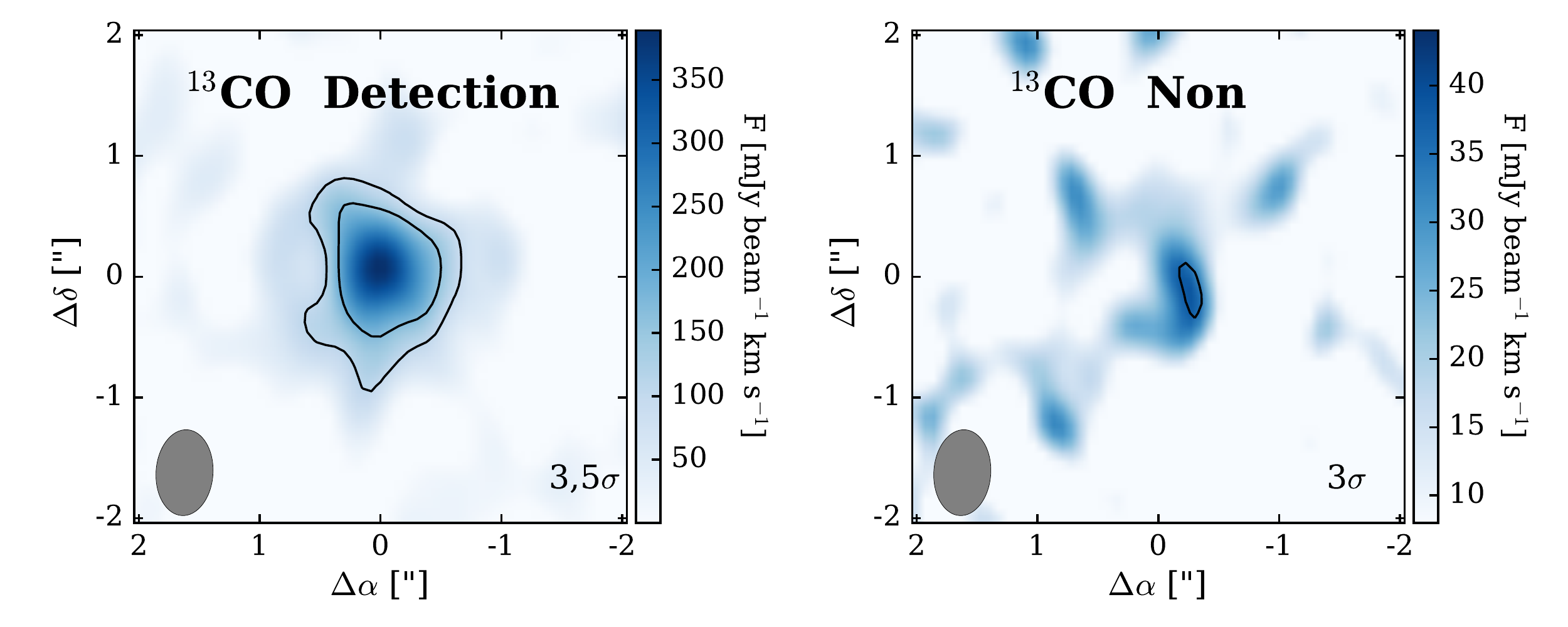}
    \caption{The stacked $^{13}$CO image for $^{13}$CO detections (Left) and non-detections (Right), with 3$\sigma$ and 5$\sigma$ contours in solid black line. The typical beam size is shown in the left corner. \label{fig:stacking_13co}}
\end{figure*}

\subsection{CO Line Fluxes and Detections} \label{sec:flux and detections}
Following the methods described in \S \ref{sec:extraction region}--\ref{sec:significance}, fluxes and uncertainties are measured consistently across our sample over a uniform spatial area and velocity range.  Table \ref{tab:co_flux_all} lists the measured fluxes and uncertainties for both $^{13}$CO and C$^{18}$O emission from each source, along with the flux-to-uncertainty ratios for $^{13}$CO when a smaller ($0\farcs3$) extraction radius is used to avoid missing weak emission or small disks.  In the stacked $^{13}$CO image, the flux would be a factor of 2.5 lower if the $0\farcs3$ extraction radius were used instead of the $0\farcs6$ radius.
Because the uncertainty increases with larger aperture size, in several cases the flux is significant when extracted from the smaller aperture and below significance when extracted from the larger aperture.  The opposite may also occur when the emission is spread over large areas.  Therefore, we classify sources as significant $^{13}$CO detections if the flux exceeds 3$\sigma$ significance when extracted from both apertures.  Sources are considered tentative $^{13}$CO detections if the flux exceeds $\sim$ 3$\sigma$ only in one aperture or is nearly 3$\sigma$ in both apertures.  2MASS J11100010-7634578 has $^{13}$CO fluxes that are less than 3$\sigma$ in both apertures for our uniform CO flux measurement but is considered a significant detection because the emission is spatially extended beyond the $0\farcs6$ radius aperture.  In addition, some sources with weak emission and a narrow line width may be missed in our analysis when 0 -- 10 km s$^{-1}$ velocity range is uniformly used, since the uncertainty would increase when more channels are included.

For the sources classified as detections, the method adopted above to measure fluxes uniformly across the sample needs to be adjusted to account for different spatial sizes and line profiles of the CO emission. To determine the size of the $^{13}$CO emission region, we generate intensity maps (zero-moment maps) by integrating over the 0 -- 10 km s$^{-1}$ velocity range.  We then measure the fluxes by increasing the circular aperture size in $0\farcs15$ radius increments on the intensity map until the flux flattens. Since the increasing flux in a larger aperture could either be attributed to real disk signal or noise fluctuations, the real disk size is determined after inspecting the intensity map. For each individual detection, the resulting source size is used as the extraction aperture for flux measurements and is shown as dashed circle on intensity map in Appendix \ref{fig-zoo}.  In the extracted spectra, most $^{13}$CO detections display weak emission across the 0 -- 10 km s$^{-1}$ velocity range as shown in Appendix \ref{fig-zoo}.  Two exceptions, 2MASS J11004022-7619280 and 2MASS J11075792-7738449, exhibit narrow line emission with FWHM $\sim$ 1.5 km s$^{-1}$. For each detection, the final flux and uncertainty measurements are extracted from these updated spatial and spectral region of the $^{13}$CO emission, as listed in Table \ref{tab:flux_mass}.

For the $^{13}$CO non-detections, which are below 3$\sigma$ significance in both apertures, the 3$\sigma$ contour size in the stacked image is smaller than the beam size.  It is therefore reasonable to choose the $0\farcs3$ radius aperture (comparable to one beam size) for non-detections to lower the $^{13}$CO upper limits. However, such a small aperture may miss emission from large disks with low surface density. While we adopt upper limits for $^{13}$CO non-detections from $0\farcs6$ radius aperture results, we also discuss the lower upper limits obtained from smaller extraction regions when results depend on fluxes.

We also excluded the shortest baselines and repeated this full analysis.  Most fluxes and uncertainties are similar, with two exceptions: the extended nebulosity near 2MASS J11075792-7738449 (see Section \ref{sec:J11075792}) and the large disk around 2MASS J11095340-7634255.

\section{Analysis of CO Gas Detections }\label{sec:detection}
In this section, we provide an overview of the general properties of the $^{13}$CO and C$^{18}$O emission, including fluxes and spatial distributions.  We then investigate possible correlations with dust mass in the disk, stellar mass and accretion rate.

\subsection{The general properties of CO detections}
Out of 93 disks in our sample, 12 are classified as significant detections and 5 as tentative detections in $^{13}$CO J = 3--2.
The source 2MASS J11075792-7738449 shows extended $^{13}$CO emission and extremely high $^{13}$CO flux, which is likely associated with the reflection nebulosity and is excluded from the full sample analysis (see also Appendix \ref{sec:J11075792}). Only the strongest continuum source in our sample, 2MASS J11100010-7634578, is detected in both $^{13}$CO and C$^{18}$O.  Another source, 2MASS J11065939-7530559, is tentatively detected in C$^{18}$O at 3$\sigma$ but is rejected due to the absence of $^{13}$CO emission. All of the significant and tentative $^{13}$CO detections are also detected in continuum emission at 887$\mu$m \citep{pascucci2016}.

For the only detection in C$^{18}$O, 2MASS J11100010-7634578, the flux ratio of $^{13}$CO to C$^{18}$O line emission is $\sim 1-2$, much lower than expected from their isotopologue ratio of $\sim 8$ \citep{wilson1994}. This indicates that $^{13}$CO is likely optically thick (see discussion in Appendix \ref{sec:J11100010}). No other source is detected in C$^{18}$O, which implies that the typical $^{13}$CO emission is usually not as optically thick as suggested by this one source.  The disk around 2MASS J11095340-7634255, which is bright in $^{13}$CO, has a C$^{18}$O flux upper limit that is 5--7 times fainter, consistent with the $^{13}$CO being optically thin.

For a more universal assessment, we stack the $^{13}$CO and C$^{18}$O spectral images for all $^{13}$CO detections (excluding 2MASS J11100010-7634578), applying the same approach as described in \S \ref{sec:extraction region} and weighting by noise rather than noise squared.  The mean signals are 574$\pm$31 mJy km s$^{-1}$ in $^{13}$CO and 68$\pm$16 mJy km s$^{-1}$ in C$^{18}$O, which is consistent with most $^{13}$CO emission being optically thin.
The stacked image of 75 sources that individually lack $^{13}$CO emission yields a noise-weighted average line flux of 70$\pm$15 mJy km s$^{-1}$ in $^{13}$CO and 27$\pm$8 mJy km s$^{-1}$ in C$^{18}$O. The images of $^{13}$CO non-detections, stacked separately in the Hot and Cool samples, yield fluxes of 59$\pm$15 mJy km s$^{-1}$ and 31$\pm$13 mJy km s$^{-1}$, respectively. The stacking results confirm the analysis in \S \ref{sec:significance} that some weak detections may be just beyond our detection limits. The stacked $^{13}$CO images for $^{13}$CO detections and non-detections are shown in Figure \ref{fig:stacking_13co}.

\begin{figure*}[t]
    \includegraphics[width=0.5\linewidth]{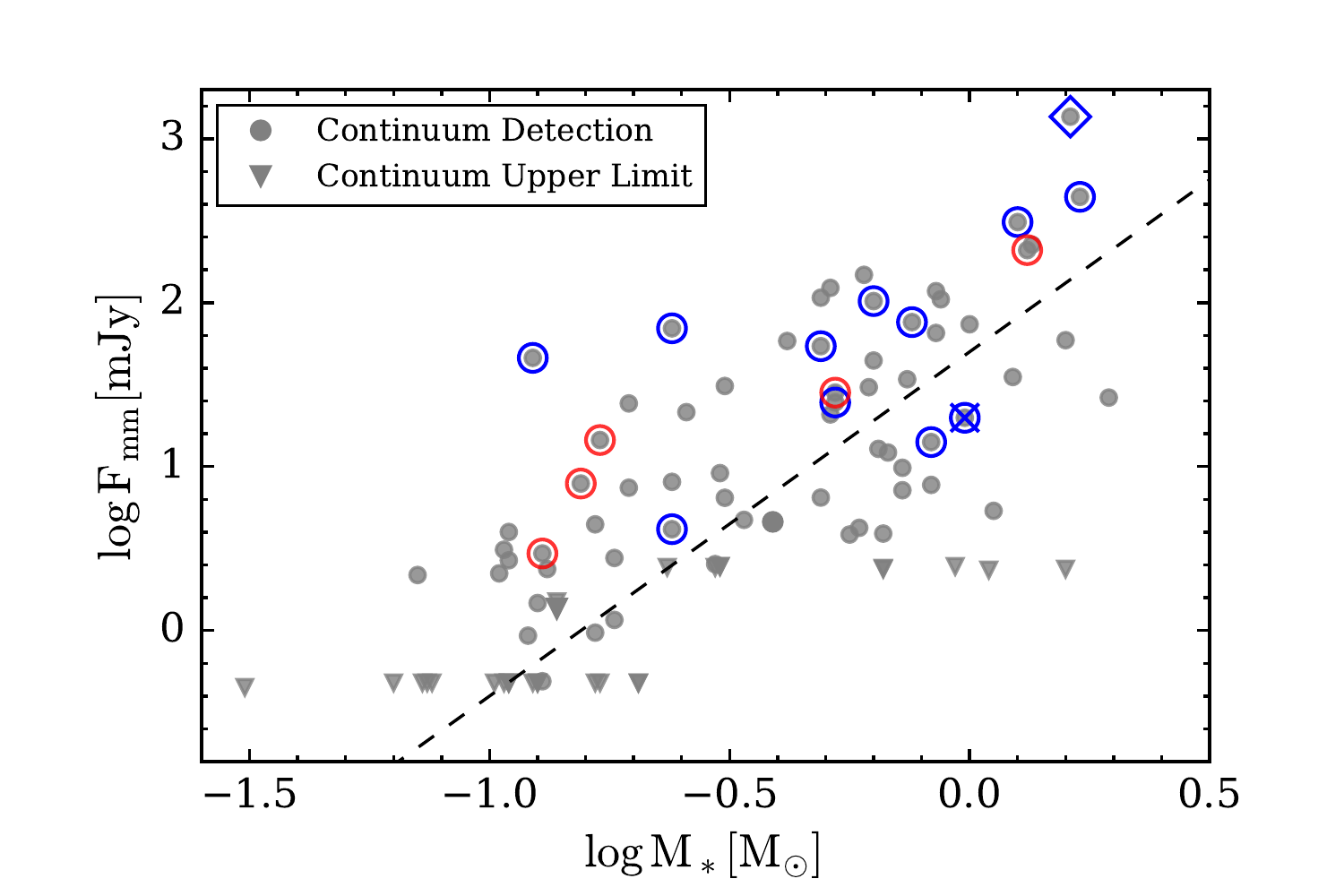}
    \includegraphics[width=0.5\linewidth]{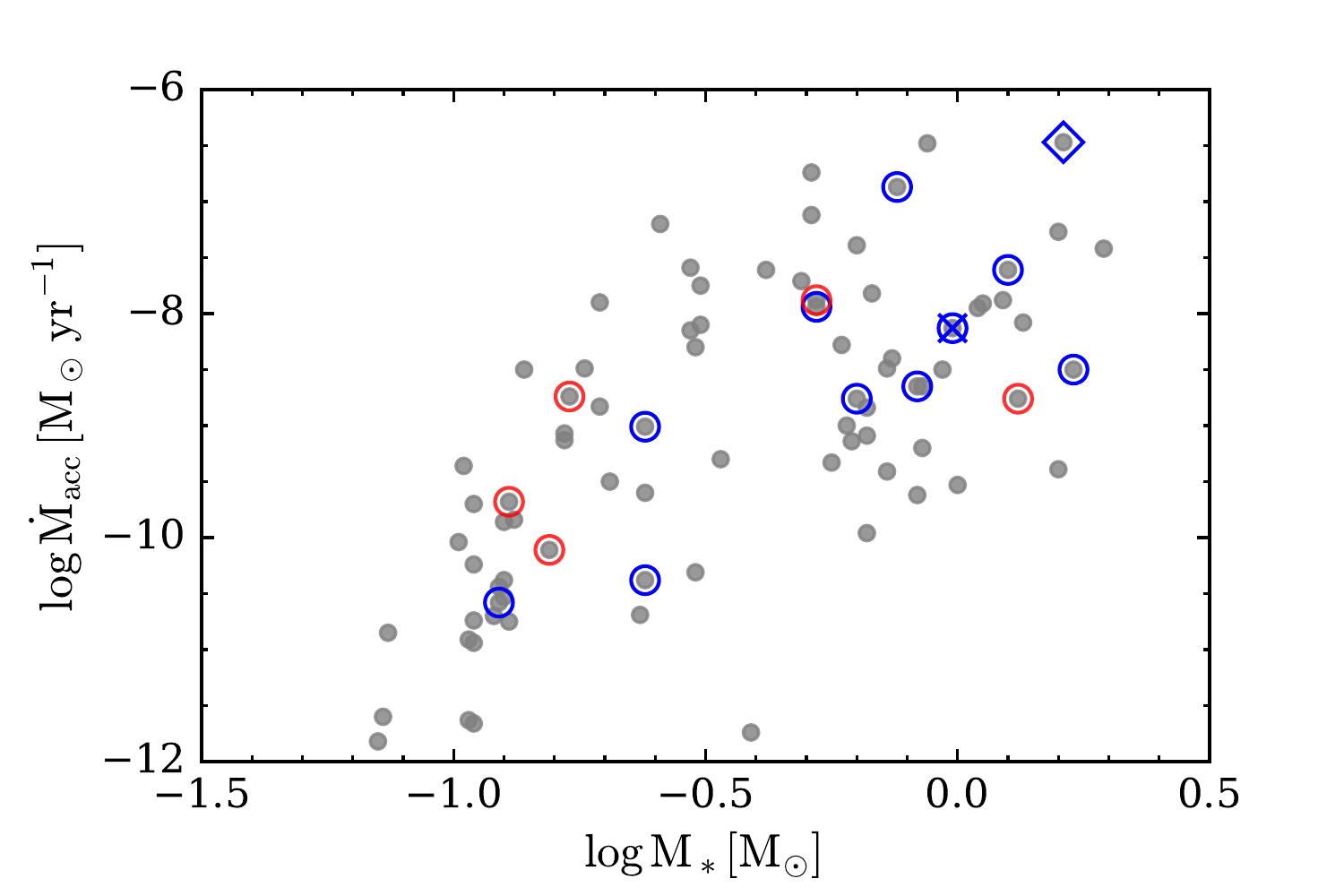}
  \caption{Left: 887$\mu$m continuum flux densities ($F_{\rm mm}$) vs. stellar mass ($M_*$), data and best fit relation plotted as dashed line [log(F$_{\rm mm}$/mJy)=1.9($\pm$0.2)$\times$log(M$_*$/M$_\odot$)+1.6($\pm$0.1)] are adopted from \citet{pascucci2016}, also see Table \ref{tab:co_flux_all}. Gray dots are continuum detections, while downward triangles stand for upper limits for non-detections. Significant (12) and tentative (5) $^{13}$CO detections are marked with blue and red, respectively. The strongest continuum source, 2MASS J11100010-7634578, detected in both $^{13}$CO and C$^{18}$O, is noted with diamond. 2MASS J11075792-7738449, with off-source $^{13}$CO emission, is marked with an `X' . Right: Accretion rate for 82 sources \citep{manara2016a,manara17}
 versus stellar mass \citep{pascucci2016} \label{fig:Fmm_Ms}}
\end{figure*}

\subsection{Correlations of star and disk properties with CO line fluxes} \label{sec:co flux relation}
The gas mass should depend on the initial mass of the disk, which is expected to vary with stellar mass, and on the mass loss rate from the disk due to viscous accretion onto the star and to mass loss in photoevaporative and MHD winds \citep[e.g.][]{Hartmann1998,alexander2014,armitage15}.  In the viscous accretion disk, the viscous accretion rate should scale with the gas mass.  As a consequence, if the $^{13}$CO emission is an accurate diagnostic of the gas mass, then we should expect the $^{13}$CO flux\footnote{The fluxes may be multiplied by 4$\pi$d$^2$ to convert into luminosity.} to correlate with the dust mass (if dust traces disk mass), the mass of the central star and the accretion rate onto the star.
In this subsection, we investigate whether any of these parameters are correlated with the $^{13}$CO line fluxes.

In this analysis, continuum flux densities and stellar masses\footnote{The stellar masses were calculated assuming a distance of 160 pc, but are similar to the masses that would be calculated for the updated 188 pc distance.} of our sample are adopted from \citet{pascucci2016}.  The stellar masses were calculated based on evolutionary tracks from non-magnetic models of \citet{feiden16} and \citet{baraffe15}. For multiple star systems, the adopted stellar mass is the mass of the primary, following \citet{pascucci2016}. Accretion rates are obtained from \citet{manara2016a,manara17} and are adjusted for updated luminosities and stellar radii increased for the adopted distance and for the masses used by \citet{pascucci2016}.

Figure \ref{fig:Fmm_Ms} (left) shows the dust emission versus stellar mass, highlighting the $^{13}$CO and C$^{18}$O detections. Overall, sources with CO emission are located over a wide parameter space in both stellar mass and continuum flux. Within any single stellar mass bin, CO detections are more likely from sources with strong continuum emission. The strongest continuum emitters in both the Hot and Cool samples are also detected in $^{13}$CO. When accounting for both significant and tentative detections, the deeper Cool sample has a similar $^{13}$CO detection rate to the Hot sample. However, if we consider only the significant detections, nearly 75$\%$ come from the Hot sample, which again suggests that higher-mass stars tend to have brighter CO emission.

\begin{figure*}[t]
    \includegraphics[width=0.5\linewidth]{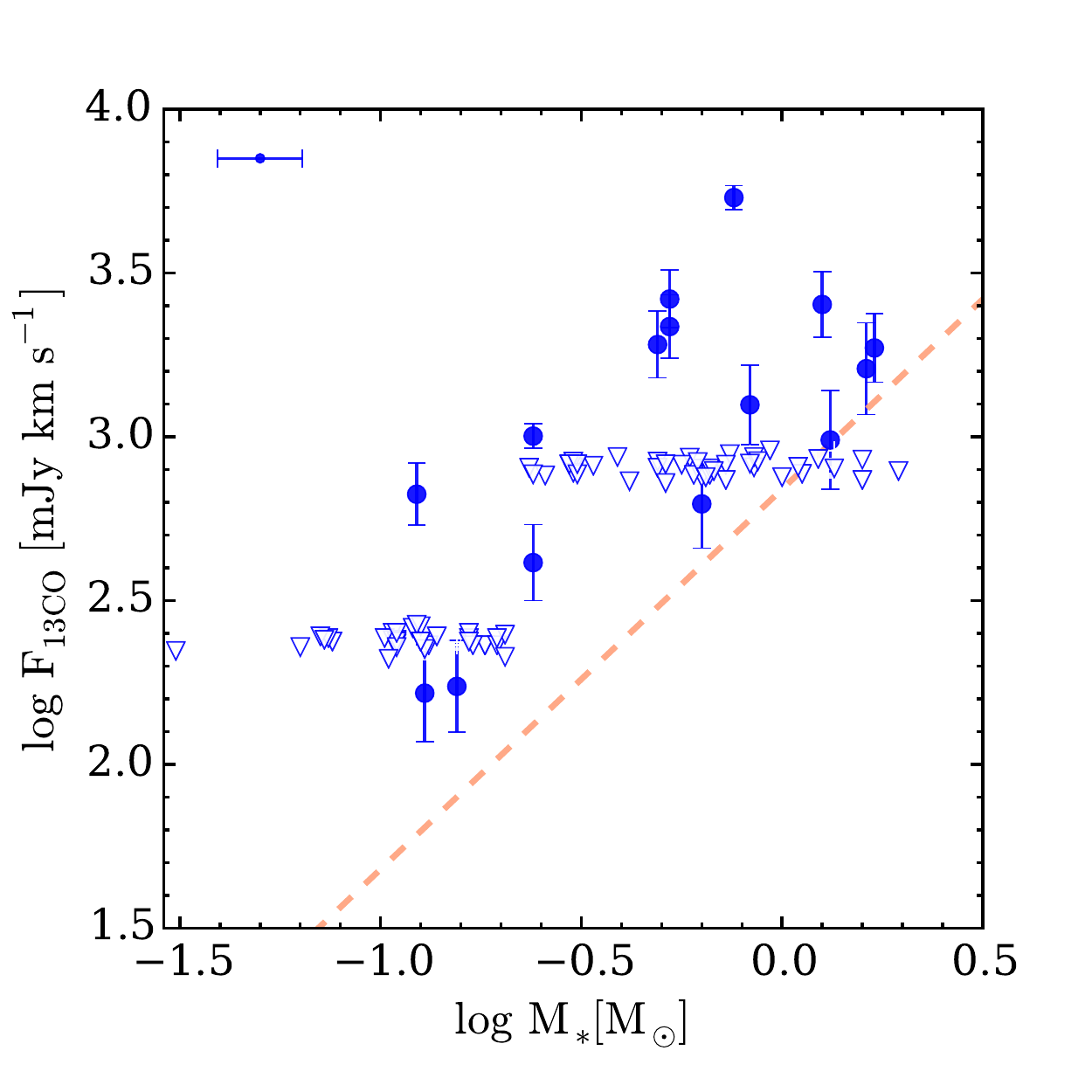}
    \includegraphics[width=0.5\linewidth]{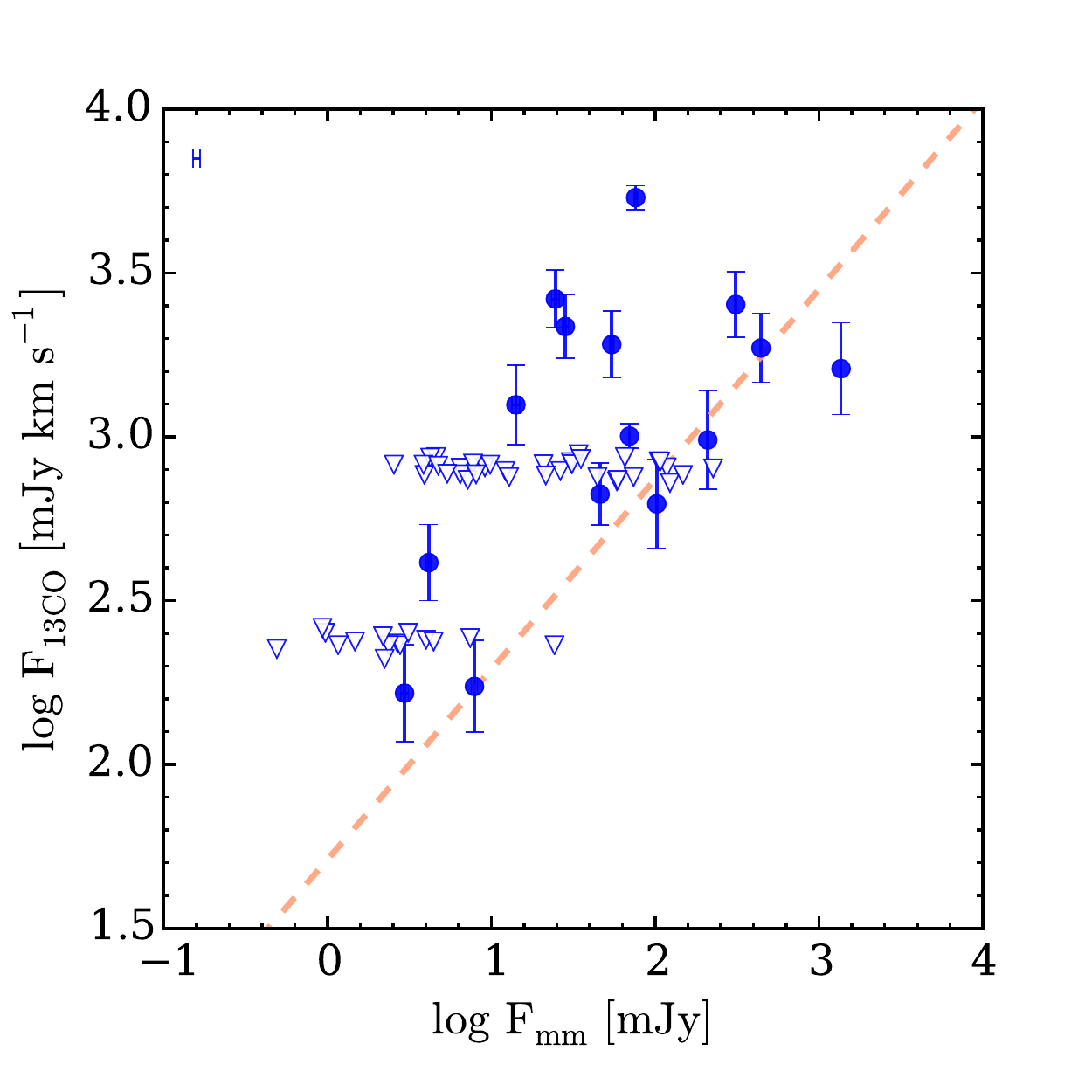}
  \caption{Left: the $^{13}$CO flux as a function of stellar mass; Right: the $^{13}$CO flux as a function of 887$\mu$m continuum flux density for continuum detections. The $^{13}$CO detections are shown in solid dots, while the $3\sigma$ upper limits are shown as triangles for $^{13}$CO non-detections. The best fit results from Bayesian analysis (dashed lines) are calculated using the measured fluxes and uncertainties for all points, including the non-detections (shown as $3\sigma$ upper limits only for visualization). The best fits therefore appear lower than the measured data points on the plot and are consistent with results of stacked emission from the non-detections. The median error bar in log($M_{*}$) and log($F_{mm}$) is shown in the top left corner in each panel. \label{fig:Fco_vs} }
\end{figure*}

Figure \ref{fig:Fco_vs} demonstrates the correlation of $^{13}$CO flux with respect to stellar mass. The best-fit linear relationship between the log of the stellar mass and the log of $^{13}$CO flux is $\log(F_{13CO}/{\rm mJy})=1.16 (\pm0.13)\times \log(M_*/M_\odot)+2.84(\pm 0.18)$,
calculated using a Bayesian analysis, as follows.  To use the measured line fluxes and uncertainties in the full sample,  we adopt the model as $F_{13CO}=\beta\cdot M_* ^\alpha$ with $\alpha$ and $\beta$ sampled uniformly in prior range.  The $\chi^2$ likelihood function is then used to assess the fit quality, accounting for uncertainties in both $F_{13CO}$ and $M_*$ and for the intrinsic scatter in the relationship. The posterior parameter space is sampled in a Markov Chain Monte Carlo (MCMC) algorithm with \textit{emcee} \citep{mcmc2013}. We refer to corner maps in Appendix \ref{sec:corner_plot} for the posterior analysis results. The linear relation is then recovered by converting the best-fit model to log space.
For the non-detections, we use the extracted fluxes and uncertainties instead of 3 times the uncertainty upper limits \citep[also see discussion in ][]{mohanty2013}.

In an alternative approach, the fit is constrained by $3\sigma$ upper limits rather than the measured data points and uncertainties. In this case, the Bayesian linear regression method \textit{Linmix} \footnote{from https://github.com/jmeyers314/linmix} \citep{Kelly2007} yields a best fit with a slope of $1.14\pm0.27$ and a lower intercept of $2.50\pm0.20$. However, because the stacking demonstrates that the upper limits include some signal, we prefer using the full information.  The slope is consistent and thus robust in both methods, but the intercept differs due to different treatment in upper limits and likelihood function.

In these fits\footnote{ In these fitting methods, the best-fit results and uncertainties are likely dependent on the assumption of Gaussian scatter and prior distribution in Bayesian approach, especially when high fraction of data are upper limits. We therefore adopt a third fitting approach -- the \textit{cenken} routine in R NADA package \citep{feigelson2012}, which uses the nonparametric Akritas-Thiel-Sen estimator \citep{akritas1996} and makes no initial assumption about the data distribution, to test the reliability of our results. This linear regression method including the censored data information but leaving out the measurement errors, provides the fitting result consistent with parametric modeling with slope and intercept of 1.07 and 2.58, respectively.},
the power law of 1.16 between stellar mass and $^{13}$CO flux is shallower than the power law of $1.9-2.1$ between stellar mass and continuum flux from \citet{pascucci2016}\footnote{The correlation in stellar mass and continuum flux was calculated using the method of \citet{Kelly2007}. We also fit the stellar mass and continuum flux relationship with model $F_{cont}=\beta\cdot M_* ^\alpha$, which yields a slope of 1.45 and an intercept of 1.87. The fitted correlations in stellar mass and continuum flux from the two methods are reasonably consistent, when only 30$\%$ data are non-detections. As noted in \textit{Linmix}, when there is non-detection, the maximum-likelihood estimate including upper limits may not be valid. When large fraction of data points are censored, the resulted correlation may likely not represent the true relation. }.
The relationship between $^{13}$CO flux and stellar mass may be steeper  if higher sensitivity were achieved, especially for disks around low-mass stars. Since 80$\%$ of our sample is not detected in $^{13}$CO, more sensitive observations, especially for the low-mass end, are needed to better constrain this relation. The slope between CO flux and stellar mass may also be shallower because the $^{13}$CO emission is more likely to be optically thick in more massive disks. The treatment of binarity likely introduces some additional error that is not accounted for in our analysis, since the appropriate mass may be the total system mass for circumbinary disks, or the mass of the secondary if the disk is around the less massive component.

The same Bayesian analysis applied to 887$\mu$m continuum fluxes, adopted from \citet{pascucci2016}, and $^{13}$CO fluxes for sources detected in continuum yields a best-fit result, $\log(F_{\rm 13CO}/{\rm mJy}) = 0.58 (\pm0.08)\times \log(F_{\rm mm}/{\rm mJy})+1.71 (\pm0.16).$

The right panel of Figure~\ref{fig:Fmm_Ms} identifies the CO detections in our sample in a plot of accretion rate versus stellar mass.  The only detection in both $^{13}$CO and C$^{18}$O lines, 2MASS J11100010-7634578, is indeed a strong accretor.  However, in any stellar mass bin, accretion rates of $^{13}$CO detections are within the range of accretion rates of $^{13}$CO non-detections. A Kolmogorov-Smirnov two-sample test\footnote{http://docs.scipy.org/doc/scipy-0.14.0/reference/stats.html} of accretion rates in CO detections and CO non-detections yields a 74\% probability, indicating statistically similar parent distributions.  The same Bayesian analysis applied to stellar accretion rate and CO flux yields a slope of only 0.2. From the full sample analysis, there is at most a weak correlation between the gas emission from the disk and the accretion rate. The lack of a correlation is also consistent with the analysis of gas mass, measured from $^{13}$CO and C$^{18}$O emission, and accretion rate for disks in Lupus \citep{manara2016b}.

\begin{figure}[t]
    \centering
        \includegraphics[scale=0.6]{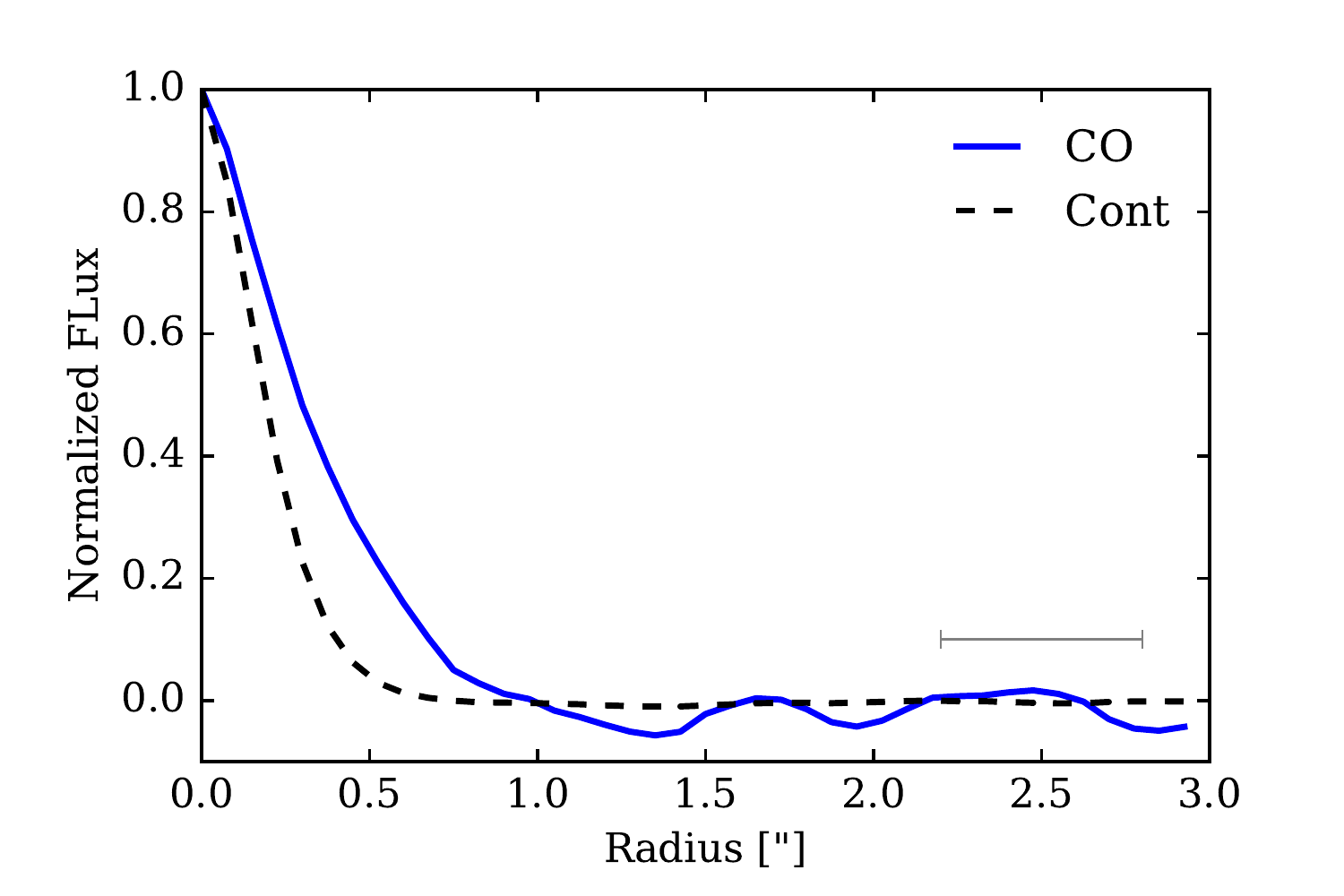}
\caption{Azimuthally averaged $^{13}$CO J=3-2 and continuum emission normalized to the peak. The typical beam size is shown at the right corner. \label{fig:profile_size}}
\end{figure}

\subsection{Morphology of CO emission} \label{sec:morphology}
The 887 $\mu$m continuum images and $^{13}$CO intensity maps for each detection are shown in Appendix \ref{fig-zoo}. The radial profiles in the stacked continuum and $^{13}$CO intensity maps from the full sample, as shown in Figure \ref{fig:profile_size}, indicate that $^{13}$CO emission is slightly more extended than dust emission, with a radial profile FWHM 30$\%$ wider in $^{13}$CO emission. Similarly, many well-known disks, such as HD163296, V4046 Sgr and Elias 2-27, have $^{13}$CO emission that is 2--3 times extended than the sub-mm dust emission \citep[e.g.,][]{isella2007,Rosenfeld2013,Perez2016}. The $^{13}$CO emission for the tentative detections is typically unresolved in our beam with FWHM of $\sim 0\farcs 6$.  The brightest $^{13}$CO disk, 2MASS J11095340-7634255 has emission on $2^{\prime\prime}$ scales, and would be a factor of two fainter if baselines shorter than 40 k$\lambda$ were excluded.

The $^{13}$CO gas emission is spatially resolved in two transition disks, 2MASS J10581677-7717170 and J11173700-7704381 (see also \S \ref{sec:transition disk}). A lower gas surface density inside the dust cavity is often found from high-resolution observations in transition disks \citep{Bruderer2014,zhang2014,perez2015,vandermarel2015,vandermarel2016}. In 2MASS J10581677-7717170, which shows an inner dust cavity from sub-mm imaging \citep{pascucci2016}, the $^{13}$CO emission is stronger in the dust ring than in inner dust cavity, which may be caused either by a gas density drop \citep{vandermarel2015} or insufficient sensitivity. The $^{13}$CO emission is also azimuthally asymmetric, with an emission deficit to the west of the disk and 20--50$\%$ higher flux in the north peak. Even though the dust cavity is not resolved in 2MASS J11173700-7704381, the deficit of $^{13}$CO emission is observed in the inner disk.  Two emission clumps with peak fluxes differ by 30$\%$ are also seen in the NE-SW direction. However, due to the limited sensitivity, gas structures presented here may not fully represent the gas distribution in these systems. Higher resolution and more sensitive observations are therefore needed to recover the gas distribution in detail in these transition disks.

The $^{13}$CO moment 1 map (velocity map) and line spectrum for each detection are also shown in Appendix Figure \ref{fig-zoo}.  Throughout the Cha I sample, two stars, 2MASS J11004022-7619280 and 2MASS J11095340-7634255 show clear double-peaked line profiles in the $^{13}$CO spectrum, with distinguishable red and blue-shifted components in velocity maps that indicate Keplerian rotating disks. The narrow velocity range (3.5 -- 6 km s$^{-1}$) in 2MASS J11004022-7619280 suggests that the disk is viewed nearly face-on. Another source, 2MASS J11100010-7634578, has azimuthally asymmetric $^{13}$CO emission, with possible absorption at one location that may confuse the detection of Keplerian rotation (see Appendix \ref{sec:J11100010}). Other sources probably lack sufficient S/N to display the pattern of Keplerian rotation.

\subsection{CO Emission from Transition Disks} \label{sec:transition disk}
Transition disks have cavities in their dust distributions, as identified from a deficit of mid-IR emission and from high-resolution images of sub-mm continuum emission (see reviews by \citealt{WilliamsCieza2011} and \citealt{Espaillat2014}). Nine disks in our sample were classified as likely transition disks, of which eight candidates were identified indirectly from their mid-IR spectral energy distributions \citep{Kim2009,manoj2011} and one transition disk was identified directly from sub-mm imaging \citep{pascucci2016}. Four of these transition disks, 2MASS J10581677-7717170, 2MASS J11083905-7716042, 2MASS J11173700-7704381 and the $\epsilon$ Cha member 2MASS J11183572-7935548, are detected in $^{13}$CO emission (including two that are spatially resolved, see \S \ref{sec:morphology}). None of the transition disks are detected with C$^{18}$O emission. The fraction of transition disks (or candidates) with detected with $^{13}$CO (4/9) is higher than the fraction of non-transition disks with $^{13}$CO detections (13/84).

In the comparison sample of Lupus (see also \S \ref{sec:lupus}), 12 disks were classified as transition disk candidates from mid-IR spectroscopy, with six confirmed by sub-mm dust imaging \citep{ansdell2016}. Most (11/12) transition disks in Lupus were detected in $^{13}$CO, and six of them were also detected in C$^{18}$O \citep{ansdell2016}. In both the Cha I and the Lupus sample, the $^{13}$CO detection rate is higher in the transition disk sample than in the full sample.

\section{CO Gas Masses of Protoplanetary Disks in Cha I} \label{sec:mass}
In this section, we first describe our methods to convert the $^{13}$CO and C$^{18}$O line luminosities and upper limits into a gas mass for disk in our sample.  We then discuss the range of gas masses and inferred gas-to-dust ratios, as well as the correlation of gas mass with stellar mass. These conversions provided by disk models may severely underestimate the total gas mass in the disk. Large uncertainties in the CO-to-H$_2$ ratio and the gas temperature structure are discussed in detail in Section 6.2.

Much of this analysis uses the stacked fluxes and the best fit line between the stellar mass and $^{13}$CO flux to describe a typical source in the sample.  The correlations and some interpretations may be biased to the few disks that are detected in $^{13}$CO emission.  The use of stacked fluxes should provide a more accurate estimate of the median source, although even these measurements may be skewed by outliers.  A more robust analysis would require a much deeper survey with a high detection rate.

\subsection{Converting line luminosities to gas mass} \label{sec:converting}

The amount of gas in protoplanetary disks provides important constraints on disk evolution and planet formation. Optically thin tracers are needed to probe down to the disk midplane.  If the CO isotopologues are optically thin, as expected for low-mass disks, the uncertainty of converting a detected CO flux into disk mass resides in the [CO]/[H$_2$] ratio and isotopic ratios.  Corrections for the line excitation and optical depth are also uncertain, especially when only individual lines and transitions are considered.

Physical-chemical models are needed to evaluate these processes and to reduce uncertainties, but it is time-consuming to individualize the models to the specific (and often unknown) properties of each source in a survey. To provide a simple conversion factor from the observed line luminosity to gas disk mass, MvD16 developed hundreds of full thermal-chemical models that include isotope-selective photodissociation and CO freeze-out. The models calculate the expected emission from disks with gas masses from 10$^{-5}$ -- $10^{-1}$ $M_\odot$ and a range of radial and vertical structures. In MvD16, stellar luminosities in models are calculated for 1 $L_{\odot}$ for T Tauri stars and 10 $L_{\odot}$ for Herbig Ae stars. Most of the $^{13}$CO detections in our Cha I sample have stellar luminosities between 0.1--1 L$_{\odot}$.  CO line luminosities are fainter by $\sim 25\%$ for disk models with 0.1 $L_{\odot}$ stellar luminosity compared with T Tauri models \citep{miotello2016b}, which would lead to gas masses that are larger, but still within the uncertainties. In this analysis, we adopt the T Tauri models with stellar luminosity of 1 $L_{\odot}$. Other sources of uncertainty are introduced because the accretion rates in the MvD16 models are higher than measured accretion rates in Cha I \citep{manara2016a,manara17}, and because the MvD16 models calculate models on full disks and do not consider the complicated physical structures that are likely present in all disks, including inner cavities.

In the low-mass disk regime from MvD16 disk models, CO line luminosities scale linearly with gas disk masses.  Since most of our CO detections are only detected in $^{13}$CO and have low line luminosities, we fit simple functions of disk mass with line luminosity using the median value in each mass bin from MvD16 model grids (see Appendix \ref{fig:MvD16_fit}). Our fitting functions are slightly different from \citet{miotello2016b} in the choice of transition mass and in the treatment of disk inclinations in model grids.  The gas mass derived using our fitted functions can be converted to the gas masses of \citet{miotello2016b} by dividing our gas mass by factors of 1.6 for disks with an inclination angle of 10$\degr$ and by 1.2 for an inclination angle of 80$\degr$, respectively (see details in Appendix \ref{sec:MvD16_fit}). The measured line luminosities or upper limits are adopted for gas mass estimates in the fitted functions of MvD16 model grids, for the $^{13}$CO detections or non-detections in our Cha I sample, respectively.
For the sources detected only in $^{13}$CO, gas masses are calculated from the fitted functions of $^{13}$CO line luminosities. Upper and lower limits in gas mass are constrained by $^{13}$CO line flux uncertainties and C$^{18}$O upper limits. For the one source detected in both lines, we compare the gas mass and upper/lower mass boundaries from two separate fittings, and adopt the final gas mass with a lower limit from C$^{18}$O and an upper limit from $^{13}$CO.
For the $^{13}$CO non-detections, the upper limits are adopted from the smaller one of the gas mass upper limits, as estimated from $^{13}$CO and C$^{18}$O upper limits of line luminosities.  The gas masses and upper/lower limits for our sample are listed in Table \ref{tab:flux_mass}.

An alternate effort by WB14 used parameterized model grids to estimate gas disk mass. The effects of CO freeze-out and CO photodissociation are explored under different surface density distributions, gas temperature profiles, gas disk masses, and disk geometry. The isotope-selective photodissociation is estimated by reducing the C$^{18}$O abundance by a factor of 3.
Tests of a few well-studied disks (e.g., DM Tau, GG Tau) suggest that this reduced C$^{18}$O abundance yields results that are consistent with results from chemical models \citep{Dutrey1997}. To conduct a direct comparison of the two models, we derive the gas mass from WB14 model grids by adopting the same approach described in detail in \citet{ansdell2016}. We search for model grids of simulated line luminosities within uncertainties of our observed $^{13}$CO and C$^{18}$O line luminosities or constrained by upper limits. Gas masses for $^{13}$CO detections are adopted as the mean value of the log M$_{{\rm gas}}$ from the acceptable models to reduce the effects of extremely large or small grid points. The upper (M$_{{\rm max}}$) and lower (M$_{{\rm min}}$) boundaries are also set by the maximum and minimum M$_{\rm gas}$ in the accepted model grids. For sources without detected $^{13}$CO emission, only upper limits for gas mass are provided. The gas masses and upper limits derived from WB14 are also listed in Table \ref{tab:flux_mass}. The high gas mass upper limits derived from WB14 are from disk models with low atmospheric temperatures and rapidly decreasing temperature profiles (e.g., T$_{atm,1au}$=500 K, q = 0.65).  If these specific model grids are excluded, then the upper limits on gas mass would not be much higher than gas masses inferred from $^{13}$CO detections.

\begin{figure*}[t]
    \includegraphics[width=0.5\linewidth]{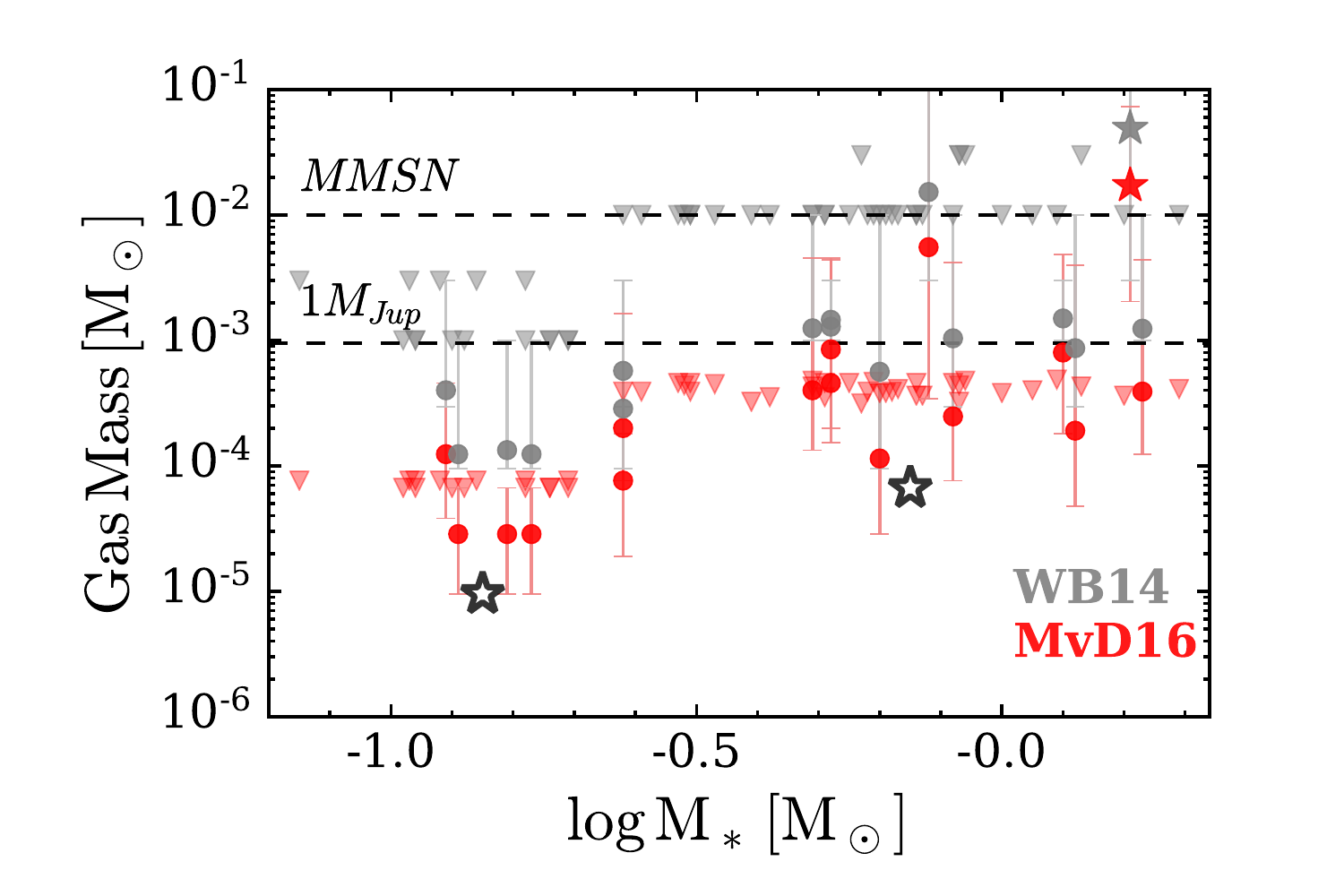}
    \includegraphics[width=0.5\linewidth]{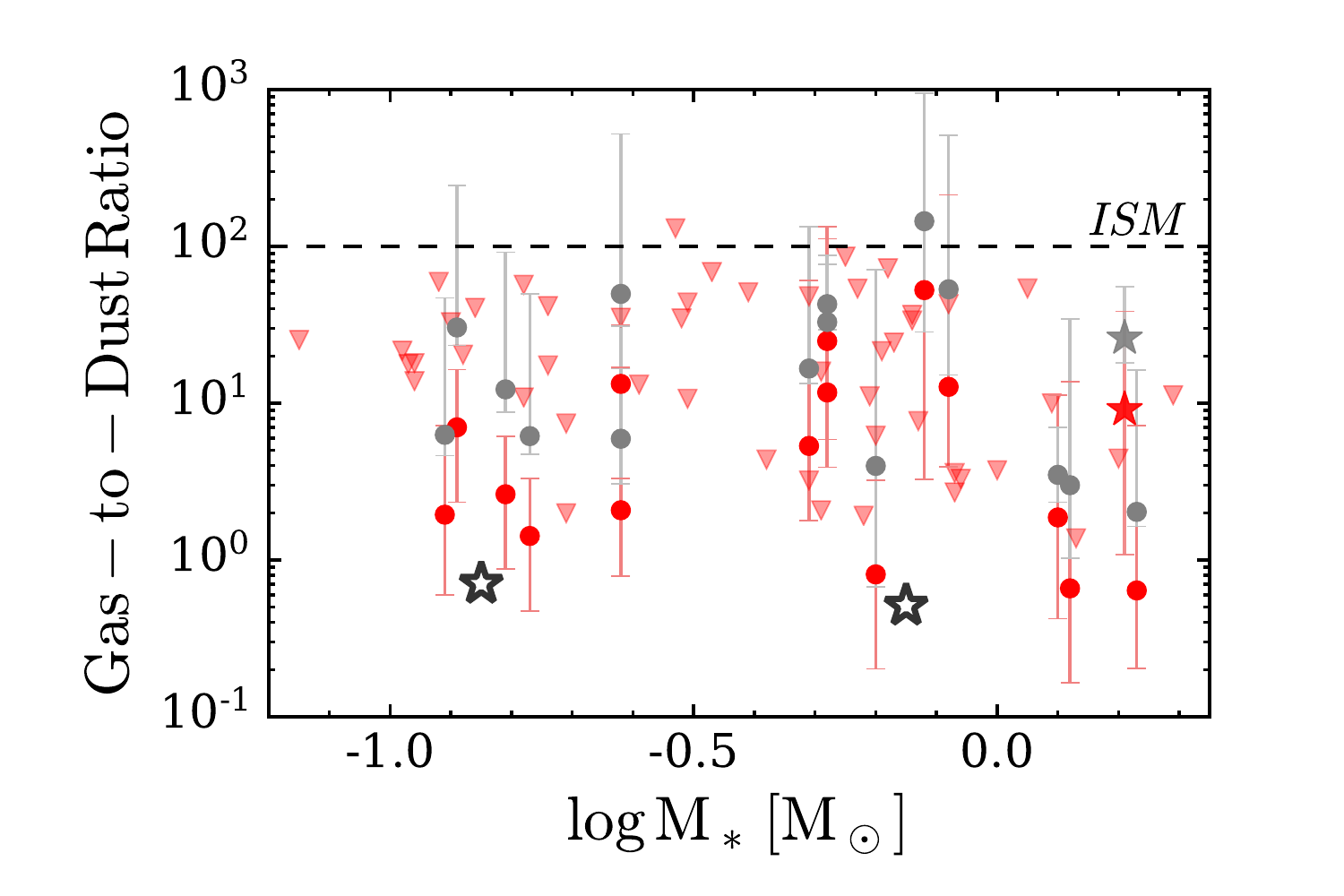}
\caption{Left: Gas masses vs. Stellar mass. Gas masses or upper limits are derived from \citet{WilliamsBest2014} (in gray) and from \citet{miotello2016} (in red) using estimated line fluxes or upper limits for $^{13}$CO detections and non-detections, respectively. The locations for 1 M$_{Jup}$ and MMSN are also marked. Black stars represent the average gas mass in Hot and Cool sample, based on the results from stacking the CO images. Right: Gas-to-dust ratio vs Stellar mass. The ISM value of gas-to-dust ratio is labeled out.  The only detection in both lines, 2MASS J11100010-7634578, is marked with a solid star. \label{fig:Mgas} }
\end{figure*}

\subsection{Gas masses and gas-to-dust ratios}
The derived gas masses and upper limits for continuum detections in our Cha I sample are shown in the left panel of Figure \ref{fig:Mgas}, from both WB14 and MvD16 models.  Gas masses inferred from CO detections are very low, spanning from 0.03--1 M$_{\rm Jup}$ (10-300 M$_{\rm Earth}$). The two exceptions, 2MASS J11095340-7634255 and J11100010-7634578, have gas masses similar to the Minimum Mass Solar Nebula (MMSN, 0.01 M$_{\odot}$), the mass of materials required to form the solar planetary system \citep{kusaka1970,weidenschilling1977}. We also stack the Hot and Cool sample separately in $^{13}$CO and calculate the gas mass in MvD16 fitted functions. The stacked line fluxes correspond to gas masses of $\sim$ 0.07 M$_{\rm Jup}$ (22 M$_{\rm Earth}$) for the Hot sample and $\sim$ 0.01 M$_{\rm Jup}$ (3 M$_{\rm Earth}$) for the Cool sample.

Gas masses from MvD16 models are generally lower than from WB14 model grids, mainly due to a wider gas temperature range adopted in WB14 models (as discussed by MvD16). If additional depletion factors in the C$^{18}$O abundance for lower mass stars are included in WB14 models, as suggested by the models of MvD16, then the parameterized models would yield even higher gas masses.

The inferred gas-to-dust ratios are shown in the right panel of Figure \ref{fig:Mgas}, with the dust mass calculated by \citet{pascucci2016} assuming a fixed dust temperature of 20 K and then scaled to the 188 pc Gaia distance. Since gas mass upper limits derived from WB14 are biased towards one set of disk models (see discussion in \S \ref{sec:converting}), gas-to-dust ratios based on WB14 gas mass are only shown for CO detections. The gas-to-dust ratios are spread over two orders of magnitude for sources detected with CO emission, and most sources have lower ratios than typical ISM value ($\sim$ 100). The median value of gas-to-dust ratio is $\sim 4$ when the gas mass is evaluated from MvD16 and $\sim 15$ when evaluated from WB14. In most CO non-detections, gas-to-dust ratios are also lower than typical ISM value. The stacked gas masses in Hot and Cool samples, when applied with the mean dust mass in each sample, yield an average gas-to-dust ratio of $\sim$ 1, two orders of magnitude lower than the ISM value.

The choice of dust temperature affects the dust mass measurements \citep[see discussion in]{andrews2013,vanderPlas2016,pascucci2016}. If $T_{{\rm dust}}$ scales with $L_*$, we would obtain lower gas-to-dust ratios in very low mass stars than when a constant $T_{{\rm dust}}$ = 20 K is assumed, as cooler temperature leads to higher dust mass. However, $T_{{\rm dust}}$ is independent of $L_*$, if dust disk size scales linearly with stellar mass \citep{hendler2017}. The dependence of disk radii on stellar mass is not yet well quantified in unbiased samples. The low gas masses and low gas-to-dust ratios are consistent with the results from a small sample in Taurus \citep{WilliamsBest2014} and a nearly complete Class II disk sample in Lupus \citep{ansdell2016,miotello2016b}.  We will discuss the implications of the low gas masses and gas-to-dust ratios in \S \ref{sec:implication}.

While the scaling relation between dust disk mass and stellar mass has been measured in multiple star-forming regions \citep{andrews2013,mohanty2013,barenfeld2016,ansdell2016,pascucci2016}, the scaling relation between gas disk mass and stellar mass is less understood and only limited to the Lupus Clouds \citep{miotello2016b}.

Due to the high rate of CO non-detections in our Cha I sample, we establish an empirical correlation between the gas mass and the stellar mass based on the relation of CO line flux with stellar mass and the conversion of CO line flux to gas mass. We adopt the best fit of $^{13}$CO line flux with stellar mass in \S \ref{sec:co flux relation}, and then convert the $^{13}$CO line fluxes to gas masses based on the fitted functions of the median $^{13}$CO line luminosity with gas mass from MvD16 models in \S \ref{sec:converting}. The best-fit gas mass and stellar mass relation is given as log(M$_{\rm gas}$/M$_{\rm Jup}$)= 1.27 ($\pm$0.14)$\times$log(M$_*$/M$_\odot$) -- 0.76 ($\pm$0.11).
This fit only provides an average gas mass in a given stellar mass bin and is highly dependent on the robustness of the relation of $^{13}$CO flux and stellar mass.  The real relation may be steeper, since $^{13}$CO emission is more likely to be optically thick in more massive disks and lines from smaller disks are likely to be fainter. \citet{miotello2016b} found a power law index of 0.63 between disk gas mass and stellar mass in Lupus clouds with 34 detections of $^{13}$CO emission and 10 C$^{18}$O detections, and the rest non-detection upper limits. More CO detections from high sensitivity datasets and more accurate gas mass estimates are needed to evaluate the robustness of the established correlation between gas mass and stellar mass.

\section{Discussion}
Sub-millimeter continuum surveys of protoplanetary disks have all found that stellar masses and dust disk masses are strongly correlated, with large scatter \citep[see review by][]{WilliamsCieza2011}. The disks in the 1--3 Myr-old regions of Taurus, Lupus and Cha I separately lead to indistinguishable relationships between the dust disk mass and stellar mass \citep{andrews2013,ansdell2016,pascucci2016}. In this section, we compare the CO gas properties of Cha I found here to those obtained in the ALMA survey of Lupus disks, which had a similar experimental design that targeted $^{13}$CO and C$^{18}$O emission\footnote{The ALMA continuum survey of the Upper Sco Association \citep{barenfeld2016} observed the $^{12}$CO line, which probes the disk surface area instead of bulk gas mass.}.   We then discuss the implications for the weak CO emission detected from both Cha I and Lupus.

\begin{figure*}[t]
    \includegraphics[width=0.5\linewidth]{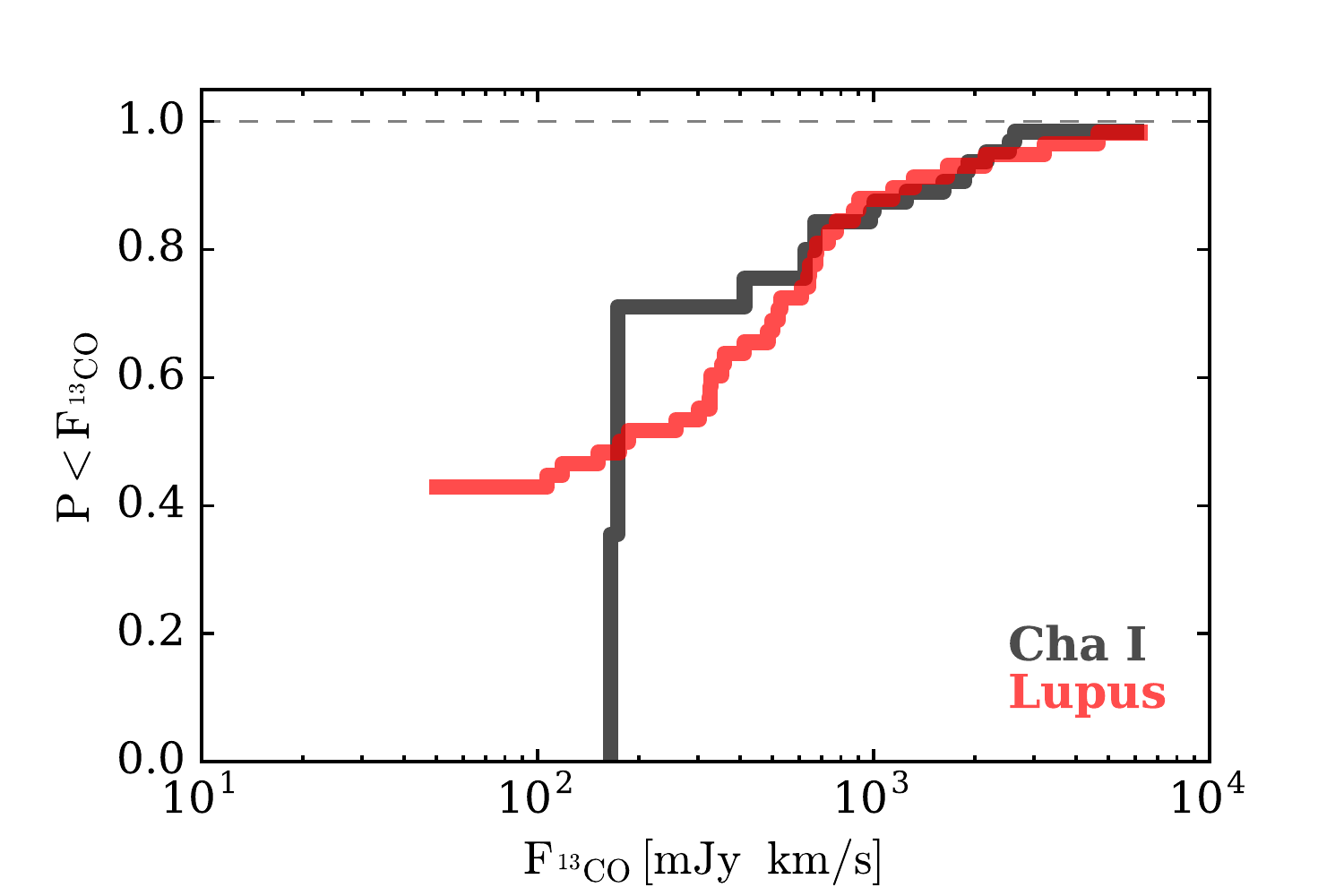}
    \includegraphics[width=0.5\linewidth]{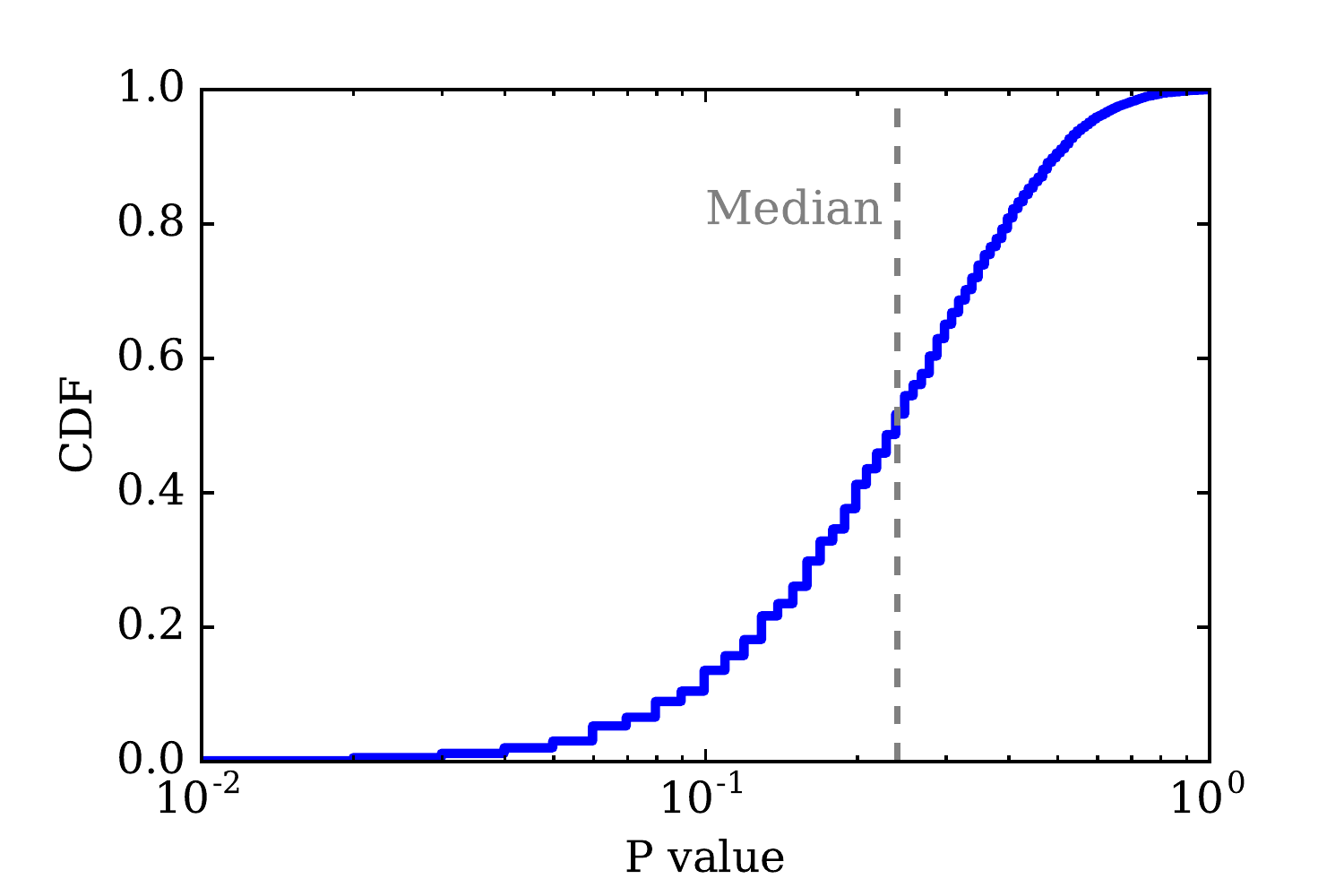}
\caption{Left: The $^{13}$CO flux cumulative distributions for Cha I (black) and Lupus (red) for sources detected with sub-mm continuum. CO fluxes in Lupus sample are scaled to Cha I distance. The distributions are calculated using the Kaplan-Meier estimator to consider upper limits. Right: Comparison of $^{13}$CO flux distribution in Lupus and Cha I. The cumulative distribution is shown for 10$^4$ runs of probability that CO fluxes in two sub-samples with the same stellar mass distribution are drawn from the same parent distribution.   \label{fig:Lupus} }
\end{figure*}

\subsection{Comparison to CO emission from the Lupus Molecular Cloud} \label{sec:lupus}
The ALMA survey of Lupus was implemented with a strategy similar to our Cha I survey.  The Lupus sample was split into M-dwarfs, with a sensitivity comparable to our Cool Sample in Cha I (M3--M8), and AFGK stars, with a sensitivity 2 -- 3 times deeper than the Hot Sample (G2-M3) in Cha I. Twenty sources in Lupus lack optical/near-IR spectral types, are likely highly obscured or very low mass objects, and are excluded from all comparisons presented here. Stellar and CO luminosities of Lupus objects are adjusted to the updated Gaia distance of $157 \pm 7$ pc (see Appendix \ref{gaia}).  The average age of stars with disks in Lupus ($\sim 2$ Myr) is indistinguishable from the average age of those in Cha I.

The sub-millimeter continuum detection rate is higher in Lupus (58/69) than in Cha I (65/92). \citet{pascucci2016} found that a single $M_{dust}$ to $M_{*}$ relation can explain the dust disk masses in both clusters within measurement uncertainties, with a similarity that is improved with the updated distances. The sub-mm dust emission suggests that Lupus and Cha I are in the same stage of disk evolution.  In contrast, the $^{13}$CO detection rate is much lower in Cha I (17$\%$, or 15/91)\footnote{2MASS J11183572-7935548, a likely member of $\epsilon$ Cha, and 2MASS J11075792-7738449, with offset $^{13}$CO emission from extended nebulosity, are excluded.} than in Lupus (48$\%$, or 33/69).  Because most of these detections are near the sensitivity limits, such a large difference may be explained by the differences in depth and sample.  When the Lupus sample is scaled to the Cha I distance and assigned the sensitivity of our survey, the $^{13}$CO detection rate in Lupus would decrease to 24$\%$, similar to our detection rate for Cha I. When separated into stars of spectral type K, M0--M3, and M4--M6, the detection rate in each subsample is only slightly higher in Lupus than in Cha I, which may be explained by the different methodologies.

The $^{13}$CO flux cumulative distributions for Cha I and Lupus, scaled to Cha I distance,  are shown in the left panel of Figure \ref{fig:Lupus}.
The distributions are calculated using the Kaplan-Meier estimator in the R NADA package to include $3\sigma$ upper limits. The mean $^{13}$CO flux is slightly higher in Lupus than in Cha I. The two distributions are generally consistent at the higher flux end.  The large discrepancy in the low flux regime is mainly due to different sensitivities in the two regions. We also perform the two-sample test with the \textit{cendiff} routine including censored data in NADA package, which estimates the probability of two samples sharing the same parent distribution. We first constrain the stellar mass distribution by assigning same number sources in each stellar mass bin in both clusters, following the same approach in \citet{andrews2013} and \citet{ansdell2016}. A two-sample test of $^{13}$CO flux with censored data in the two sub-samples with same stellar mass distribution is then calculated. The cumulative distribution of statistical results (P value) from 10$^4$ iterations is shown in the right panel of Figure \ref{fig:Lupus}, with the median P value of 0.25, indicating that the $^{13}$CO flux distribution in Cha I and Lupus are statistically indistinguishable.

However, the conclusions from this comparison would change if the lower value of upper limits for $^{13}$CO non-detections from the $0\farcs3$ radius aperture were adopted. The $^{13}$CO cumulative distributions for the two samples would be consistent at high fluxes, but the distribution in the Cha I sample would flatten out at low fluxes. The same two-sample test in 10$^4$ iterations results in a median P value of 0.003, which would suggest ruling out the hypothesis that the disk masses from both regions are statistically indistinguishable.

Because of the high fraction of non-detections in both samples, the differences in the beam sizes, sensitivities, and extraction methods prevent any firm conclusion on the similarity of CO gas properties in the disks in Lupus and Cha~I.

\subsection{CO depletion and implications for planet formation} \label{sec:implication}
CO line emission has been previously found to be fainter than expected from the dust emission, assuming the canonical ISM gas-to-dust and [CO]/[H$_2$] ratios \citep[e.g.,][]{Dutrey2003,chapillon2008,rodriguez2015,pericaud17}.  This conclusion, from small and biased samples, was recently strengthened in a large and unbiased sample of disks in the Lupus star-forming region \citep{ansdell2016,miotello2016b}, and is confirmed here for a complete survey of disks from the young (2-3 Myr) Cha~I star forming region. The derived low CO-based gas mass in 2-3 Myr old disks has three possible explanations: (1) the conversion of CO line flux to CO gas mass is incorrect, (2) the total gas mass in these disks are very low, or (3) the C depletion into CO and/or complex ices is underestimated, thereby altering the CO-to-H$_2$ ratio.  We discuss each of these possibilities in the following subsections.

\subsubsection{Conversion of CO line flux to CO gas mass}
The conversion of CO line flux to CO gas mass requires the application of physical-chemical models, along with the canonical assumption that the CO-to-H$_2$ ratio is $\sim 10^{-4}$. Since each individual disk cannot be modeled, we rely upon results from grids of disk models (WB14 and MvD16) that were calculated over ranges of disk and stellar properties.  As discussed in \S \ref{sec:converting}, these grids are not complete.  Relative to our sample, the stellar mass and mass accretion rates in these models are high. Mini-grids in \citet{miotello2016b} suggest that these results are robust to difference in stellar mass (stellar luminosity), though the uncertainties induced by changing the accretion rates are not explored.

The CO gas masses obtained from conversions using the WB14 and MvD16 grids are both very low, despite very different modeling approaches and implementation of isotopic-selective photodissociation.  However, these conversions should be confirmed with further tests using independent disk models and dedicated analyses of disks with well-known structures.  The uncertainties in the disk models include the temperature structure of the gas and the physical structure of the disk.  Model grids cannot be expected to account for the many disk structures that must exist in our sample, and in any case our knowledge of these structures is very limited.  If the gas surface density peaks sharply near the star, most of the gas will be optically thick and will not contribute significantly to the measured fluxes \citep{yu2016,yu2017}. Observations of multiple CO lines at high spatial resolution could help to resolve this potential error.

\subsubsection{Very low total gas masses}
If gas masses are very low, the disk dispersal timescales for these disks would be implausibly short.  Under the assumption that the $^{13}$CO emission is accurately converted into a disk mass, a typical 0.7 $M_\odot$ star in Cha I would have a disk gas mass of $\sim 10^{-4}$ M$_\odot$.  With a typical accretion rate of 2$\times$10$^{-9}$ M$_\odot$ yr$^{-1}$ \citep{manara2016a,manara17}, disks would disperse within $\sim 10^5$ yr, a very short time relative to the current age of $\sim 2$ Myr and the average disk dispersal timescale of 2 -- 5 Myr \citep[e.g.,][]{haisch2001,hernandez2008,fedele2010}. Most gaseous disks would have already disappeared.

However, if the conversions between CO emission and CO gas mass are somehow correct, and the C is not depleted, then gas masses for the 2 Myr old disks in Cha I and in Lupus are very low and insufficient to form a $\sim$ 1 M$_{Jup}$ giant planet. An early start and completion of giant planet formation would be required to explain the abundance of giant planets in the observed planet population \citep{cumming2008,WinnFabrycky2015}.  The low gas-to-dust ratios implied by the low gas masses may facilitate the formation of planetesimals through either gravitational collapse or streaming instability \citep{youdin2002,youdin2005,bai2010}.  The remaining mass in disks would be sufficient to continue forming  super-Earth and Neptune-mass planets, potentially explaining their prevalence around solar-type stars \citep{howard2010,Mayor2011,petigura2013,muldes2015b}.

\subsubsection{CO depletion into ices}
These gas masses are also much lower than those obtained from HD emission \citep{bergin2013,McClure2016}, although the HD line was only detected for a few sources. The CO emission may not be an accurate tracer of the gas mass \citep[see also][]{miotello2016b}.  A low CO abundance, and therefore weak emission, may result from most C turning into more complex carbon chains and freezing out into ices.  Since the gas mass depends inversely on [CO]/[H$_2$], a higher gas mass would be obtained if the [CO]/[H$_2$] abundance ratio were lower.  The few detections of HD emission from disks (with 3 detections from 7 objects in the sample) suggests that the gas mass may be ten times higher than that inferred from modeling the observed $^{13}$CO emission and provides gas-to-dust ratios with median values consistent with 100 \citep{bergin2013,McClure2016}.  In addition, the relationship between disk mass and mass accretion rate in Lupus supports predictions from viscous accretion, if gas-to-dust ratios of $\sim 100$ \citep{manara2016b} -- however, modeling results of dust disk masses and accretion rates in Cha I may be more challenging to explain with simple viscous accretion (Mulders et al., in prep.).

These results suggest a constant ISM gas-to-dust ratio in disks and in turn implies that gas masses from $^{13}$CO are severely underestimated.  On the other hand, direct absorption line measurements indicate a [CO]/[H$_2$] abundance ratio consistent with the ISM in the warm disk surface around RW Aur \citep{france2014}.  The CO depletion may vary both with disk radius and disk height. Active carbon chemistry leads to the formation of carbon chains or CO$_2$ that freeze out subsequently to lock up the carbon and oxygen elements in the solids and thus lower the [CO]/[H$_2$] abundance ratio \citep{Favre2013,du2015,kama2016,kama2016b,yu2016,xu2016}.  This interpretation was independently suggested also as the favored explanation for the low $^{13}$CO fluxes in Lupus in the recent analysis by \citet{miotello2016b}. Additionally, the depletion of oxygen is also suggested from water observations and simulations as freeze-out of volatiles followed by grain growth and settling \citep{krijt2016,du2017}, and has significant effects on hydrocarbon abundances \citep{bergin2016}.

If carbon and/or oxygen depletion is more severe than is currently estimated in the chemical models, a higher gas mass and gas-to-dust ratio would be recovered. The ISM gas-to-dust ratio can be achieved if CO abundance is depleted by a factor of 10--100, in which case gas masses in disks around the higher mass stars would be similar to the MMSN.  In this case, measurements of CO depletion is required to calibrate the CO-based gas mass. A high C abundance in ices of complex molecules should affect the abundances in any planets that form within these disks, similar to the processes suggested for the CO and H$_2$O ice lines by \citet{oberg11}.

\section{Conclusions} \label{sec:conclusion}
In this paper, we present an ALMA survey of $^{13}$CO and C$^{18}$O line emission in a large sample (93, complete down to 0.1 $M_\odot$) of protoplanetary disks in the nearby ($\sim$ 188 pc) and young ($\sim$ 2 Myr) Chamaeleon I star-forming region. We develop a uniform method to identify detections and measure line fluxes, uncertainties and upper limits. This method is optimized for analysis of surveys dominated by weak signals and upper limits. The gas masses of these disks are then estimated using the $^{13}$CO and C$^{18}$O $J=3-2$ lines to understand how rapidly disk evolves.  Our main conclusions are as follows:

\begin{itemize}

\item We detect $^{13}$CO emission from 17 out of 93 sources (15 of 92 sources in Cha I with a disk origin), consisting of 12 significant detections and 5 tentative detections. Only one disk, 2MASS J11100010-7634578, is detected in both $^{13}$CO and C$^{18}$O lines. The sources with $^{13}$CO detections have a wide spread in stellar mass, sub-mm continuum flux and accretion rate.  The detection rates and line fluxes from stacked observations suggest that in most cases the measured $^{13}$CO emission is optically thin on spatial scales of $\sim 0.5-1.0$ arcsec, although the only source with C$^{18}$O detection is optically thick in $^{13}$CO (line flux ratio $\sim$ 1--2).

\item Gas masses are estimated by adopting the parametric models (WB14) and full-chemical models (MvD16). Even though gas masses derived from WB14 are generally higher than from MvD16, the CO isotopologue emission constrain the M$_{\rm gas}$ around 1 M$_{\rm Jup}$ in the CO detections. The average gas masses derived from the stacked $^{13}$CO line fluxes are 0.07 M$_{\rm Jup}$ and 0.01 M$_{\rm Jup}$ for Hot and Cool sample, respectively. If these gas masses are correct, then the derived gas-to-dust ratios would be 1--10, much lower than the standard ISM value of 100. These tiny disk masses inferred from the CO fluxes and chemical model grids imply dispersal timescales of disks through viscous accretion that are implausibly short, as estimated from measured accretion rates in \citet{manara2016a,manara17}.

\item The low gas masses and low gas-to-dust ratios in Cha I, as derived from disk models with CO isotopologue emission, confirm similar results from a disk survey of the 1--3 Myr Lupus star-forming region.  However, whether the disks from Cha I have similar or weaker CO fluxes than the disks in Lupus is uncertain because of observational differences and low detection rates.

\item  The conversions of CO flux to CO gas mass depend on the accuracy of the disk models, which have not been adapted to the full range of stellar luminosities, accretion rates, and disk structures.  The gas masses may be severely underestimated if CO-to-H$_2$ abundance ratio is lower than the ISM value, which may be caused by C and/or O depletion and lock-up, or if CO freeze-out is underestimated.  Additional observations of CO and other carbon-bearing species are needed to gain a comprehensive understanding on CO chemistry and thus the related gas masses.

\end{itemize}

\acknowledgements
We appreciate the thorough and constructive comments from the anonymous referee, which served to improve the rigor and clarity of our results.  We thank Fred Ciesla, Sebatiaan Krijt, Johan Olofsson, Germano Sacco, and Laszlo Szucs for contributions to the proposal that was awarded this data.  F.L. thanks Ran Wang and Chaoli Zhang for help learning CASA and Zhiyuan Ren and Jinghua Yuan for help with python plotting. F.L. and G.J.H also thank Emma Yu and Neal Evans for discussion about C depletion and Uma Gorti for a discussion on gas temperatures. FL and GJH are supported by general grant 11473005 awarded by the National Science Foundation of China.  I.P. acknowledges support from an NSF Astronomy \& Astrophysics Research Grant (ID: 1515392). C.F.M gratefully acknowledges an ESA Research Fellowship.  The results reported herein benefitted from collaborations and/or information exchange within NASA's Nexus for Exoplanet System Science (NExSS) research coordination network sponsored by NASA’s Science Mission Directorate.

This paper makes use of the following ALMA data: ADS/JAO.ALMA\#2013.1.00437.S. ALMA is a partnership of ESO (representing its member states), NSF (USA) and NINS (Japan), together with NRC (Canada) and NSC and ASIAA (Taiwan), in cooperation with the Republic of Chile. The Joint ALMA Observatory is operated by ESO, AUI/NRAO and NAOJ.

\bibliographystyle{aasjournal}
\bibliography{ref}

\begin{thebibliography}{}
\expandafter\ifx\csname natexlab\endcsname\relax\def\natexlab#1{#1}\fi

\bibitem[{{Aikawa} {et~al.}(2002){Aikawa}, {van Zadelhoff}, {van Dishoeck}, \&
  {Herbst}}]{aikawa2002}
{Aikawa}, Y., {van Zadelhoff}, G.~J., {van Dishoeck}, E.~F., \& {Herbst}, E.
  2002, \aap, 386, 622

\bibitem[{{Akritas} \& {Bershady}(1996)}]{akritas1996}
{Akritas}, M.~G., \& {Bershady}, M.~A. 1996, \apj, 470, 706

\bibitem[{{Alexander} {et~al.}(2014){Alexander}, {Pascucci}, {Andrews},
  {Armitage}, \& {Cieza}}]{alexander2014}
{Alexander}, R., {Pascucci}, I., {Andrews}, S., {Armitage}, P., \& {Cieza}, L.
  2014, Protostars and Planets VI, 475

\bibitem[{{Andrews} {et~al.}(2013){Andrews}, {Rosenfeld}, {Kraus}, \&
  {Wilner}}]{andrews2013}
{Andrews}, S.~M., {Rosenfeld}, K.~A., {Kraus}, A.~L., \& {Wilner}, D.~J. 2013,
  \apj, 771, 129

\bibitem[{{Andrews} {et~al.}(2012){Andrews}, {Wilner}, {Hughes}, {Qi},
  {Rosenfeld}, {{\"O}berg}, {Birnstiel}, {Espaillat}, {Cieza}, {Williams},
  {Lin}, \& {Ho}}]{Andrews2012}
{Andrews}, S.~M., {Wilner}, D.~J., {Hughes}, A.~M., {et~al.} 2012, \apj, 744,
  162

\bibitem[{{Ansdell} {et~al.}(2017){Ansdell}, {Williams}, {Manara}, {Miotello},
  {Facchini}, {van der Marel}, {Testi}, \& {van Dishoeck}}]{ansdell2017}
{Ansdell}, M., {Williams}, J.~P., {Manara}, C.~F., {et~al.} 2017, ArXiv
  e-prints, arXiv:1703.08546

\bibitem[{{Ansdell} {et~al.}(2016){Ansdell}, {Williams}, {van der Marel},
  {Carpenter}, {Guidi}, {Hogerheijde}, {Mathews}, {Manara}, {Miotello},
  {Natta}, {Oliveira}, {Tazzari}, {Testi}, {van Dishoeck}, \& {van
  Terwisga}}]{ansdell2016}
{Ansdell}, M., {Williams}, J.~P., {van der Marel}, N., {et~al.} 2016, \apj,
  828, 46

\bibitem[{{Anthonioz} {et~al.}(2015){Anthonioz}, {M{\'e}nard}, {Pinte}, {Le
  Bouquin}, {Benisty}, {Thi}, {Absil}, {Duch{\^e}ne}, {Augereau}, {Berger},
  {Casassus}, {Duvert}, {Lazareff}, {Malbet}, {Millan-Gabet}, {Schreiber},
  {Traub}, \& {Zins}}]{anthonioz2015}
{Anthonioz}, F., {M{\'e}nard}, F., {Pinte}, C., {et~al.} 2015, \aap, 574, A41

\bibitem[{{Armitage}(2015)}]{armitage15}
{Armitage}, P.~J. 2015, ArXiv e-prints, arXiv:1509.06382

\bibitem[{{Bai} \& {Stone}(2010)}]{bai2010}
{Bai}, X.-N., \& {Stone}, J.~M. 2010, \apj, 722, 1437

\bibitem[{{Baraffe} {et~al.}(2015){Baraffe}, {Homeier}, {Allard}, \&
  {Chabrier}}]{baraffe15}
{Baraffe}, I., {Homeier}, D., {Allard}, F., \& {Chabrier}, G. 2015, \aap, 577,
  A42

\bibitem[{{Barenfeld} {et~al.}(2016){Barenfeld}, {Carpenter}, {Ricci}, \&
  {Isella}}]{barenfeld2016}
{Barenfeld}, S.~A., {Carpenter}, J.~M., {Ricci}, L., \& {Isella}, A. 2016,
  \apj, 827, 142

\bibitem[{{Beckwith} {et~al.}(2000){Beckwith}, {Henning}, \&
  {Nakagawa}}]{Beckwith2000}
{Beckwith}, S.~V.~W., {Henning}, T., \& {Nakagawa}, Y. 2000, Protostars and
  Planets IV, 533

\bibitem[{{Bergin} {et~al.}(2016){Bergin}, {Du}, {Cleeves}, {Blake}, {Schwarz},
  {Visser}, \& {Zhang}}]{bergin2016}
{Bergin}, E.~A., {Du}, F., {Cleeves}, L.~I., {et~al.} 2016, \apj, 831, 101

\bibitem[{{Bergin} {et~al.}(2013){Bergin}, {Cleeves}, {Gorti}, {Zhang},
  {Blake}, {Green}, {Andrews}, {Evans}, {Henning}, {{\"O}berg}, {Pontoppidan},
  {Qi}, {Salyk}, \& {van Dishoeck}}]{bergin2013}
{Bergin}, E.~A., {Cleeves}, L.~I., {Gorti}, U., {et~al.} 2013, \nat, 493, 644

\bibitem[{{Bohlin} {et~al.}(1978){Bohlin}, {Savage}, \& {Drake}}]{Bohlin1978}
{Bohlin}, R.~C., {Savage}, B.~D., \& {Drake}, J.~F. 1978, \apj, 224, 132

\bibitem[{{Boss}(1997)}]{boss1997}
{Boss}, A.~P. 1997, Science, 276, 1836

\bibitem[{{Bruderer}(2013)}]{Bruderer2013}
{Bruderer}, S. 2013, \aap, 559, A46

\bibitem[{{Bruderer} {et~al.}(2014){Bruderer}, {van der Marel}, {van Dishoeck},
  \& {van Kempen}}]{Bruderer2014}
{Bruderer}, S., {van der Marel}, N., {van Dishoeck}, E.~F., \& {van Kempen},
  T.~A. 2014, \aap, 562, A26

\bibitem[{{Chapillon} {et~al.}(2008){Chapillon}, {Guilloteau}, {Dutrey}, \&
  {Pi{\'e}tu}}]{chapillon2008}
{Chapillon}, E., {Guilloteau}, S., {Dutrey}, A., \& {Pi{\'e}tu}, V. 2008, \aap,
  488, 565

\bibitem[{{Comer{\'o}n}(2008)}]{comeron2008}
{Comer{\'o}n}, F. 2008, {The Lupus Clouds}, ed. B.~{Reipurth}, 295

\bibitem[{{Cumming} {et~al.}(2008){Cumming}, {Butler}, {Marcy}, {Vogt},
  {Wright}, \& {Fischer}}]{cumming2008}
{Cumming}, A., {Butler}, R.~P., {Marcy}, G.~W., {et~al.} 2008, \pasp, 120, 531

\bibitem[{{Du} {et~al.}(2015){Du}, {Bergin}, \& {Hogerheijde}}]{du2015}
{Du}, F., {Bergin}, E.~A., \& {Hogerheijde}, M.~R. 2015, \apjl, 807, L32

\bibitem[{{Du} {et~al.}(2017){Du}, {Bergin}, {Hogerheijde}, {van Dishoeck},
  {Blake}, {Bruderer}, {Cleeves}, {Dominik}, {Fedele}, {Lis}, {Melnick},
  {Neufeld}, {Pearson}, \& {Yildiz}}]{du2017}
{Du}, F., {Bergin}, E.~A., {Hogerheijde}, M., {et~al.} 2017, ArXiv e-prints,
  arXiv:1705.00799

\bibitem[{{Dutrey} {et~al.}(1997){Dutrey}, {Guilloteau}, \&
  {Guelin}}]{Dutrey1997}
{Dutrey}, A., {Guilloteau}, S., \& {Guelin}, M. 1997, \aap, 317, L55

\bibitem[{{Dutrey} {et~al.}(2003){Dutrey}, {Guilloteau}, \&
  {Simon}}]{Dutrey2003}
{Dutrey}, A., {Guilloteau}, S., \& {Simon}, M. 2003, \aap, 402, 1003

\bibitem[{{Espaillat} {et~al.}(2014){Espaillat}, {Muzerolle}, {Najita},
  {Andrews}, {Zhu}, {Calvet}, {Kraus}, {Hashimoto}, {Kraus}, \&
  {D'Alessio}}]{Espaillat2014}
{Espaillat}, C., {Muzerolle}, J., {Najita}, J., {et~al.} 2014, Protostars and
  Planets VI, 497

\bibitem[{{Favre} {et~al.}(2013){Favre}, {Cleeves}, {Bergin}, {Qi}, \&
  {Blake}}]{Favre2013}
{Favre}, C., {Cleeves}, L.~I., {Bergin}, E.~A., {Qi}, C., \& {Blake}, G.~A.
  2013, \apjl, 776, L38

\bibitem[{{Fedele} {et~al.}(2010){Fedele}, {van den Ancker}, {Henning},
  {Jayawardhana}, \& {Oliveira}}]{fedele2010}
{Fedele}, D., {van den Ancker}, M.~E., {Henning}, T., {Jayawardhana}, R., \&
  {Oliveira}, J.~M. 2010, \aap, 510, A72

\bibitem[{{Feiden}(2016)}]{feiden16}
{Feiden}, G.~A. 2016, \aap, 593, A99

\bibitem[{{Feigelson} \& {Babu}(2012)}]{feigelson2012}
{Feigelson}, E.~D., \& {Babu}, G.~J. 2012, {Modern Statistical Methods for
  Astronomy}

\bibitem[{{Foreman-Mackey} {et~al.}(2013){Foreman-Mackey}, {Hogg}, {Lang}, \&
  {Goodman}}]{mcmc2013}
{Foreman-Mackey}, D., {Hogg}, D.~W., {Lang}, D., \& {Goodman}, J. 2013, \pasp,
  125, 306

\bibitem[{{France} {et~al.}(2014){France}, {Herczeg}, {McJunkin}, \&
  {Penton}}]{france2014}
{France}, K., {Herczeg}, G.~J., {McJunkin}, M., \& {Penton}, S.~V. 2014, \apj,
  794, 160

\bibitem[{{Gaia Collaboration} {et~al.}(2016){Gaia Collaboration}, {Brown},
  {Vallenari}, {Prusti}, {de Bruijne}, {Mignard}, {Drimmel}, \&
  {co-authors}}]{Gaia2016}
{Gaia Collaboration}, {Brown}, A.~G.~A., {Vallenari}, A., {et~al.} 2016, ArXiv
  e-prints, arXiv:1609.04172

\bibitem[{{Galli} {et~al.}(2013){Galli}, {Bertout}, {Teixeira}, \&
  {Ducourant}}]{galli2013}
{Galli}, P.~A.~B., {Bertout}, C., {Teixeira}, R., \& {Ducourant}, C. 2013,
  \aap, 558, A77

\bibitem[{{Gorti} {et~al.}(2011){Gorti}, {Hollenbach}, {Najita}, \&
  {Pascucci}}]{gorti2011}
{Gorti}, U., {Hollenbach}, D., {Najita}, J., \& {Pascucci}, I. 2011, \apj, 735,
  90

\bibitem[{{Haisch} {et~al.}(2001){Haisch}, {Lada}, \& {Lada}}]{haisch2001}
{Haisch}, Jr., K.~E., {Lada}, E.~A., \& {Lada}, C.~J. 2001, \apjl, 553, L153

\bibitem[{{Harris} {et~al.}(2012){Harris}, {Andrews}, {Wilner}, \&
  {Kraus}}]{harris12}
{Harris}, R.~J., {Andrews}, S.~M., {Wilner}, D.~J., \& {Kraus}, A.~L. 2012,
  \apj, 751, 115

\bibitem[{{Hartmann} {et~al.}(1998){Hartmann}, {Calvet}, {Gullbring}, \&
  {D'Alessio}}]{Hartmann1998}
{Hartmann}, L., {Calvet}, N., {Gullbring}, E., \& {D'Alessio}, P. 1998, \apj,
  495, 385

\bibitem[{{Hendler} {et~al.}(2017){Hendler}, {Mulders}, {Pascucci},
  {Greenwood}, {Kamp}, {Henning}, {Menard}, {Dent}, \& {Evans}}]{hendler2017}
{Hendler}, N., {Mulders}, G.~D., {Pascucci}, I., {et~al.} 2017, ArXiv e-prints,
  arXiv:1705.01952

\bibitem[{{Henning} \& {Semenov}(2013)}]{henning2013}
{Henning}, T., \& {Semenov}, D. 2013, Chemical Reviews, 113, 9016

\bibitem[{{Hern{\'a}ndez} {et~al.}(2008){Hern{\'a}ndez}, {Hartmann}, {Calvet},
  {Jeffries}, {Gutermuth}, {Muzerolle}, \& {Stauffer}}]{hernandez2008}
{Hern{\'a}ndez}, J., {Hartmann}, L., {Calvet}, N., {et~al.} 2008, \apj, 686,
  1195

\bibitem[{{Howard} {et~al.}(2010){Howard}, {Marcy}, {Johnson}, {Fischer},
  {Wright}, {Isaacson}, {Valenti}, {Anderson}, {Lin}, \& {Ida}}]{howard2010}
{Howard}, A.~W., {Marcy}, G.~W., {Johnson}, J.~A., {et~al.} 2010, Science, 330,
  653

\bibitem[{{Howard} {et~al.}(2012){Howard}, {Marcy}, {Bryson}, {Jenkins},
  {Rowe}, {Batalha}, {Borucki}, {Koch}, {Dunham}, {Gautier}, {Van Cleve},
  {Cochran}, {Latham}, {Lissauer}, {Torres}, {Brown}, {Gilliland}, {Buchhave},
  {Caldwell}, {Christensen-Dalsgaard}, {Ciardi}, {Fressin}, {Haas}, {Howell},
  {Kjeldsen}, {Seager}, {Rogers}, {Sasselov}, {Steffen}, {Basri},
  {Charbonneau}, {Christiansen}, {Clarke}, {Dupree}, {Fabrycky}, {Fischer},
  {Ford}, {Fortney}, {Tarter}, {Girouard}, {Holman}, {Johnson}, {Klaus},
  {Machalek}, {Moorhead}, {Morehead}, {Ragozzine}, {Tenenbaum}, {Twicken},
  {Quinn}, {Isaacson}, {Shporer}, {Lucas}, {Walkowicz}, {Welsh}, {Boss},
  {Devore}, {Gould}, {Smith}, {Morris}, {Prsa}, {Morton}, {Still}, {Thompson},
  {Mullally}, {Endl}, \& {MacQueen}}]{howard2012}
{Howard}, A.~W., {Marcy}, G.~W., {Bryson}, S.~T., {et~al.} 2012, \apjs, 201, 15

\bibitem[{{Hughes} {et~al.}(2011){Hughes}, {Wilner}, {Andrews}, {Qi}, \&
  {Hogerheijde}}]{hughes11}
{Hughes}, A.~M., {Wilner}, D.~J., {Andrews}, S.~M., {Qi}, C., \& {Hogerheijde},
  M.~R. 2011, \apj, 727, 85

\bibitem[{{Isella} {et~al.}(2007){Isella}, {Testi}, {Natta}, {Neri}, {Wilner},
  \& {Qi}}]{isella2007}
{Isella}, A., {Testi}, L., {Natta}, A., {et~al.} 2007, \aap, 469, 213

\bibitem[{{Isella} {et~al.}(2016){Isella}, {Guidi}, {Testi}, {Liu}, {Li}, {Li},
  {Weaver}, {Boehler}, {Carperter}, {De Gregorio-Monsalvo}, {Manara}, {Natta},
  {P{\'e}rez}, {Ricci}, {Sargent}, {Tazzari}, \& {Turner}}]{isella2016}
{Isella}, A., {Guidi}, G., {Testi}, L., {et~al.} 2016, Physical Review Letters,
  117, 251101

\bibitem[{{Johnson} {et~al.}(2010){Johnson}, {Aller}, {Howard}, \&
  {Crepp}}]{Johnson2010}
{Johnson}, J.~A., {Aller}, K.~M., {Howard}, A.~W., \& {Crepp}, J.~R. 2010,
  \pasp, 122, 905

\bibitem[{{Kama} {et~al.}(2016{\natexlab{a}}){Kama}, {Bruderer}, {Carney},
  {Hogerheijde}, {van Dishoeck}, {Fedele}, {Baryshev}, {Boland}, {G{\"u}sten},
  {Aikutalp}, {Choi}, {Endo}, {Frieswijk}, {Karska}, {Klaassen}, {Koumpia},
  {Kristensen}, {Leurini}, {Nagy}, {Perez Beaupuits}, {Risacher}, {van der
  Marel}, {van Kempen}, {van Weeren}, {Wyrowski}, \& {Y{\i}ld{\i}z}}]{kama2016}
{Kama}, M., {Bruderer}, S., {Carney}, M., {et~al.} 2016{\natexlab{a}}, \aap,
  588, A108

\bibitem[{{Kama} {et~al.}(2016{\natexlab{b}}){Kama}, {Bruderer}, {van
  Dishoeck}, {Hogerheijde}, {Folsom}, {Miotello}, {Fedele}, {Belloche},
  {G{\"u}sten}, \& {Wyrowski}}]{kama2016b}
{Kama}, M., {Bruderer}, S., {van Dishoeck}, E.~F., {et~al.} 2016{\natexlab{b}},
  \aap, 592, A83

\bibitem[{{Kamp} \& {Dullemond}(2004)}]{kamp04}
{Kamp}, I., \& {Dullemond}, C.~P. 2004, \apj, 615, 991

\bibitem[{{Kelly}(2007)}]{Kelly2007}
{Kelly}, B.~C. 2007, \apj, 665, 1489

\bibitem[{{Kim} {et~al.}(2009){Kim}, {Watson}, {Manoj}, {Furlan}, {Najita},
  {Forrest}, {Sargent}, {Espaillat}, {Calvet}, {Luhman}, {McClure}, {Green}, \&
  {Harrold}}]{Kim2009}
{Kim}, K.~H., {Watson}, D.~M., {Manoj}, P., {et~al.} 2009, \apj, 700, 1017

\bibitem[{{Kraus} {et~al.}(2017){Kraus}, {Herczeg}, {Rizzuto}, {Mann},
  {Slesnick}, {Carpenter}, {Hillenbrand}, \& {Mamajek}}]{kraus2017}
{Kraus}, A.~L., {Herczeg}, G.~J., {Rizzuto}, A.~C., {et~al.} 2017, \apj, 838,
  150

\bibitem[{{Kraus} {et~al.}(2012){Kraus}, {Ireland}, {Hillenbrand}, \&
  {Martinache}}]{kraus12}
{Kraus}, A.~L., {Ireland}, M.~J., {Hillenbrand}, L.~A., \& {Martinache}, F.
  2012, \apj, 745, 19

\bibitem[{{Krijt} {et~al.}(2016){Krijt}, {Ciesla}, \& {Bergin}}]{krijt2016}
{Krijt}, S., {Ciesla}, F.~J., \& {Bergin}, E.~A. 2016, \apj, 833, 285

\bibitem[{{Kusaka} {et~al.}(1970){Kusaka}, {Nakano}, \& {Hayashi}}]{kusaka1970}
{Kusaka}, T., {Nakano}, T., \& {Hayashi}, C. 1970, Progress of Theoretical
  Physics, 44, 1580

\bibitem[{{Lopez Mart{\'{\i}}} {et~al.}(2013){Lopez Mart{\'{\i}}}, {Jimenez
  Esteban}, {Bayo}, {Barrado}, {Solano}, \& {Rodrigo}}]{Lopez2013}
{Lopez Mart{\'{\i}}}, B., {Jimenez Esteban}, F., {Bayo}, A., {et~al.} 2013,
  \aap, 551, A46

\bibitem[{{Luhman}(2004)}]{Luhman2004}
{Luhman}, K.~L. 2004, \apj, 602, 816

\bibitem[{{Luhman} \& {Mamajek}(2012)}]{LuhmanMamajek2012}
{Luhman}, K.~L., \& {Mamajek}, E.~E. 2012, \apj, 758, 31

\bibitem[{{Luhman} {et~al.}(2008){Luhman}, {Allen}, {Allen}, {Gutermuth},
  {Hartmann}, {Mamajek}, {Megeath}, {Myers}, \& {Fazio}}]{Luhman2008}
{Luhman}, K.~L., {Allen}, L.~E., {Allen}, P.~R., {et~al.} 2008, \apj, 675, 1375

\bibitem[{{Lyons} \& {Young}(2005)}]{lyons05}
{Lyons}, J.~R., \& {Young}, E.~D. 2005, \nat, 435, 317

\bibitem[{{Manara} {et~al.}(2016{\natexlab{a}}){Manara}, {Fedele}, {Herczeg},
  \& {Teixeira}}]{manara2016a}
{Manara}, C.~F., {Fedele}, D., {Herczeg}, G.~J., \& {Teixeira}, P.~S.
  2016{\natexlab{a}}, \aap, 585, A136

\bibitem[{{Manara} {et~al.}(2016{\natexlab{b}}){Manara}, {Rosotti}, {Testi},
  {Natta}, {Alcal{\'a}}, {Williams}, {Ansdell}, {Miotello}, {van der Marel},
  {Tazzari}, {Carpenter}, {Guidi}, {Mathews}, {Oliveira}, {Prusti}, \& {van
  Dishoeck}}]{manara2016b}
{Manara}, C.~F., {Rosotti}, G., {Testi}, L., {et~al.} 2016{\natexlab{b}}, \aap,
  591, L3

\bibitem[{{Manara} {et~al.}(2017){Manara}, {Testi}, {Herczeg}, {Pascucci},
  {Alcala}, {Natta}, {Antoniucci}, {Fedele}, {Mulders}, {Henning}, {Mohanty},
  {Prusti}, \& {Rigliaco}}]{manara17}
{Manara}, C.~F., {Testi}, L., {Herczeg}, G.~J., {et~al.} 2017, ArXiv e-prints,
  arXiv:1704.02842

\bibitem[{{Manoj} {et~al.}(2011){Manoj}, {Kim}, {Furlan}, {McClure}, {Luhman},
  {Watson}, {Espaillat}, {Calvet}, {Najita}, {D'Alessio}, {Adame}, {Sargent},
  {Forrest}, {Bohac}, {Green}, \& {Arnold}}]{manoj2011}
{Manoj}, P., {Kim}, K.~H., {Furlan}, E., {et~al.} 2011, \apjs, 193, 11

\bibitem[{{Mayor} {et~al.}(2011){Mayor}, {Marmier}, {Lovis}, {Udry},
  {S{\'e}gransan}, {Pepe}, {Benz}, {Bertaux}, {Bouchy}, {Dumusque}, {Lo Curto},
  {Mordasini}, {Queloz}, \& {Santos}}]{Mayor2011}
{Mayor}, M., {Marmier}, M., {Lovis}, C., {et~al.} 2011, ArXiv e-prints,
  arXiv:1109.2497

\bibitem[{{McClure} {et~al.}(2016){McClure}, {Bergin}, {Cleeves}, {van
  Dishoeck}, {Blake}, {Evans}, {Green}, {Henning}, {{\"O}berg}, {Pontoppidan},
  \& {Salyk}}]{McClure2016}
{McClure}, M., {Bergin}, T., {Cleeves}, I., {et~al.} 2016, ArXiv e-prints,
  arXiv:1608.07817

\bibitem[{{Miotello} {et~al.}(2014){Miotello}, {Bruderer}, \& {van
  Dishoeck}}]{miotello2014}
{Miotello}, A., {Bruderer}, S., \& {van Dishoeck}, E.~F. 2014, \aap, 572, A96

\bibitem[{{Miotello} {et~al.}(2016){Miotello}, {van Dishoeck}, {Kama}, \&
  {Bruderer}}]{miotello2016}
{Miotello}, A., {van Dishoeck}, E.~F., {Kama}, M., \& {Bruderer}, S. 2016,
  \aap, 594, A85

\bibitem[{{Miotello} {et~al.}(2017){Miotello}, {van Dishoeck}, {Williams},
  {Ansdell}, {Guidi}, {Hogerheijde}, {Manara}, {Tazzari}, {Testi}, {van der
  Marel}, \& {van Terwisga}}]{miotello2016b}
{Miotello}, A., {van Dishoeck}, E.~F., {Williams}, J.~P., {et~al.} 2017, \aap,
  599, A113

\bibitem[{{Mohanty} {et~al.}(2013){Mohanty}, {Greaves}, {Mortlock}, {Pascucci},
  {Scholz}, {Thompson}, {Apai}, {Lodato}, \& {Looper}}]{mohanty2013}
{Mohanty}, S., {Greaves}, J., {Mortlock}, D., {et~al.} 2013, \apj, 773, 168

\bibitem[{{Mordasini} {et~al.}(2012){Mordasini}, {Alibert}, {Benz}, {Klahr}, \&
  {Henning}}]{Mordasini2012}
{Mordasini}, C., {Alibert}, Y., {Benz}, W., {Klahr}, H., \& {Henning}, T. 2012,
  \aap, 541, A97

\bibitem[{{Mulders} {et~al.}(2015{\natexlab{a}}){Mulders}, {Pascucci}, \&
  {Apai}}]{Mulders2015}
{Mulders}, G.~D., {Pascucci}, I., \& {Apai}, D. 2015{\natexlab{a}}, \apj, 798,
  112

\bibitem[{{Mulders} {et~al.}(2015{\natexlab{b}}){Mulders}, {Pascucci}, \&
  {Apai}}]{muldes2015b}
---. 2015{\natexlab{b}}, \apj, 814, 130

\bibitem[{{Murphy} {et~al.}(2013){Murphy}, {Lawson}, \& {Bessell}}]{Murphy2013}
{Murphy}, S.~J., {Lawson}, W.~A., \& {Bessell}, M.~S. 2013, \mnras, 435, 1325

\bibitem[{{Nguyen} {et~al.}(2012){Nguyen}, {Brandeker}, {van Kerkwijk}, \&
  {Jayawardhana}}]{Nguyen2012}
{Nguyen}, D.~C., {Brandeker}, A., {van Kerkwijk}, M.~H., \& {Jayawardhana}, R.
  2012, \apj, 745, 119

\bibitem[{{{\"O}berg} {et~al.}(2011){{\"O}berg}, {Murray-Clay}, \&
  {Bergin}}]{oberg11}
{{\"O}berg}, K.~I., {Murray-Clay}, R., \& {Bergin}, E.~A. 2011, \apjl, 743, L16

\bibitem[{{Obermeier} {et~al.}(2016){Obermeier}, {Koppenhoefer}, {Saglia},
  {Henning}, {Bender}, {Kodric}, {Deacon}, {Riffeser}, {Burgett}, {Chambers},
  {Draper}, {Flewelling}, {Hodapp}, {Kaiser}, {Kudritzki}, {Magnier},
  {Metcalfe}, {Price}, {Sweeney}, {Wainscoat}, \& {Waters}}]{obermeier2016}
{Obermeier}, C., {Koppenhoefer}, J., {Saglia}, R.~P., {et~al.} 2016, \aap, 587,
  A49

\bibitem[{{Pascucci} {et~al.}(2016){Pascucci}, {Testi}, {Herczeg}, {Long},
  {Manara}, {Hendler}, {Mulders}, {Krijt}, {Ciesla}, {Henning}, {Mohanty},
  {Drabek-Maunder}, {Apai}, {Sz{\H u}cs}, {Sacco}, \&
  {Olofsson}}]{pascucci2016}
{Pascucci}, I., {Testi}, L., {Herczeg}, G.~J., {et~al.} 2016, \apj, 831, 125

\bibitem[{{P{\'e}rez} {et~al.}(2016){P{\'e}rez}, {Carpenter}, {Andrews},
  {Ricci}, {Isella}, {Linz}, {Sargent}, {Wilner}, {Henning}, {Deller},
  {Chandler}, {Dullemond}, {Lazio}, {Menten}, {Corder}, {Storm}, {Testi},
  {Tazzari}, {Kwon}, {Calvet}, {Greaves}, {Harris}, \& {Mundy}}]{Perez2016}
{P{\'e}rez}, L.~M., {Carpenter}, J.~M., {Andrews}, S.~M., {et~al.} 2016,
  Science, 353, 1519

\bibitem[{{Perez} {et~al.}(2015){Perez}, {Casassus}, {M{\'e}nard}, {Roman},
  {van der Plas}, {Cieza}, {Pinte}, {Christiaens}, \& {Hales}}]{perez2015}
{Perez}, S., {Casassus}, S., {M{\'e}nard}, F., {et~al.} 2015, \apj, 798, 85

\bibitem[{{Pericaud} {et~al.}(2016){Pericaud}, {Di Folco}, {Dutrey},
  {Guilloteau}, \& {Pietu}}]{pericaud17}
{Pericaud}, J., {Di Folco}, E., {Dutrey}, A., {Guilloteau}, S., \& {Pietu}, V.
  2016, ArXiv e-prints, arXiv:1612.06582

\bibitem[{{Petigura} {et~al.}(2013){Petigura}, {Howard}, \&
  {Marcy}}]{petigura2013}
{Petigura}, E.~A., {Howard}, A.~W., \& {Marcy}, G.~W. 2013, Proceedings of the
  National Academy of Science, 110, 19273

\bibitem[{{Pollack} {et~al.}(1996){Pollack}, {Hubickyj}, {Bodenheimer},
  {Lissauer}, {Podolak}, \& {Greenzweig}}]{pollack1996}
{Pollack}, J.~B., {Hubickyj}, O., {Bodenheimer}, P., {et~al.} 1996, \icarus,
  124, 62

\bibitem[{{Rodriguez} {et~al.}(2015){Rodriguez}, {van der Plas}, {Kastner},
  {Schneider}, {Faherty}, {Mardones}, {Mohanty}, \& {Principe}}]{rodriguez2015}
{Rodriguez}, D.~R., {van der Plas}, G., {Kastner}, J.~H., {et~al.} 2015, \aap,
  582, L5

\bibitem[{{Rosenfeld} {et~al.}(2013){Rosenfeld}, {Andrews}, {Wilner},
  {Kastner}, \& {McClure}}]{Rosenfeld2013}
{Rosenfeld}, K.~A., {Andrews}, S.~M., {Wilner}, D.~J., {Kastner}, J.~H., \&
  {McClure}, M.~K. 2013, \apj, 775, 136

\bibitem[{{Schegerer} {et~al.}(2006){Schegerer}, {Wolf}, {Voshchinnikov},
  {Przygodda}, \& {Kessler-Silacci}}]{Schegerer2006}
{Schegerer}, A., {Wolf}, S., {Voshchinnikov}, N.~V., {Przygodda}, F., \&
  {Kessler-Silacci}, J.~E. 2006, \aap, 456, 535

\bibitem[{{Schmidt} {et~al.}(2013){Schmidt}, {Vogt}, {Neuh{\"a}user},
  {Bedalov}, \& {Roell}}]{Schmidt2013}
{Schmidt}, T.~O.~B., {Vogt}, N., {Neuh{\"a}user}, R., {Bedalov}, A., \&
  {Roell}, T. 2013, \aap, 557, A80

\bibitem[{{Schwarz} {et~al.}(2016){Schwarz}, {Bergin}, {Cleeves}, {Blake},
  {Zhang}, {{\"O}berg}, {van Dishoeck}, \& {Qi}}]{schwarz16}
{Schwarz}, K.~R., {Bergin}, E.~A., {Cleeves}, L.~I., {et~al.} 2016, \apj, 823,
  91

\bibitem[{{Smith} {et~al.}(2009){Smith}, {Pontoppidan}, {Young}, {Morris}, \&
  {van Dishoeck}}]{smith2009}
{Smith}, R.~L., {Pontoppidan}, K.~M., {Young}, E.~D., {Morris}, M.~R., \& {van
  Dishoeck}, E.~F. 2009, \apj, 701, 163

\bibitem[{{van der Marel} {et~al.}(2016){van der Marel}, {van Dishoeck},
  {Bruderer}, {Andrews}, {Pontoppidan}, {Herczeg}, {van Kempen}, \&
  {Miotello}}]{vandermarel2016}
{van der Marel}, N., {van Dishoeck}, E.~F., {Bruderer}, S., {et~al.} 2016,
  \aap, 585, A58

\bibitem[{{van der Marel} {et~al.}(2015){van der Marel}, {van Dishoeck},
  {Bruderer}, {P{\'e}rez}, \& {Isella}}]{vandermarel2015}
{van der Marel}, N., {van Dishoeck}, E.~F., {Bruderer}, S., {P{\'e}rez}, L., \&
  {Isella}, A. 2015, \aap, 579, A106

\bibitem[{{van der Marel} {et~al.}(2013){van der Marel}, {van Dishoeck},
  {Bruderer}, {Birnstiel}, {Pinilla}, {Dullemond}, {van Kempen}, {Schmalzl},
  {Brown}, {Herczeg}, {Mathews}, \& {Geers}}]{vandermarel2013}
{van der Marel}, N., {van Dishoeck}, E.~F., {Bruderer}, S., {et~al.} 2013,
  Science, 340, 1199

\bibitem[{{van der Plas} {et~al.}(2016){van der Plas}, {M{\'e}nard},
  {Ward-Duong}, {Bulger}, {Harvey}, {Pinte}, {Patience}, {Hales}, \&
  {Casassus}}]{vanderPlas2016}
{van der Plas}, G., {M{\'e}nard}, F., {Ward-Duong}, K., {et~al.} 2016, \apj,
  819, 102

\bibitem[{{van Dishoeck} \& {Black}(1988)}]{vanDishoeck1988}
{van Dishoeck}, E.~F., \& {Black}, J.~H. 1988, \apj, 334, 771

\bibitem[{{van Kempen} {et~al.}(2009){van Kempen}, {van Dishoeck},
  {G{\"u}sten}, {Kristensen}, {Schilke}, {Hogerheijde}, {Boland}, {Menten}, \&
  {Wyrowski}}]{vankempen2009}
{van Kempen}, T.~A., {van Dishoeck}, E.~F., {G{\"u}sten}, R., {et~al.} 2009,
  \aap, 507, 1425

\bibitem[{{van Zadelhoff} {et~al.}(2001){van Zadelhoff}, {van Dishoeck}, {Thi},
  \& {Blake}}]{vanZadlehoff2001}
{van Zadelhoff}, G.-J., {van Dishoeck}, E.~F., {Thi}, W.-F., \& {Blake}, G.~A.
  2001, \aap, 377, 566

\bibitem[{{Walsh} {et~al.}(2012){Walsh}, {Nomura}, {Millar}, \&
  {Aikawa}}]{walsh12}
{Walsh}, C., {Nomura}, H., {Millar}, T.~J., \& {Aikawa}, Y. 2012, \apj, 747,
  114

\bibitem[{{Wang} \& {Henning}(2006)}]{wang2006}
{Wang}, H., \& {Henning}, T. 2006, \apj, 643, 985

\bibitem[{{Weidenschilling}(1977)}]{weidenschilling1977}
{Weidenschilling}, S.~J. 1977, \mnras, 180, 57

\bibitem[{{Wichmann} {et~al.}(1997){Wichmann}, {Krautter}, {Covino}, {Alcala},
  {Neuhaeuser}, \& {Schmitt}}]{wichmann1997}
{Wichmann}, R., {Krautter}, J., {Covino}, E., {et~al.} 1997, \aap, 320, 185

\bibitem[{{Williams} \& {Best}(2014)}]{WilliamsBest2014}
{Williams}, J.~P., \& {Best}, W.~M.~J. 2014, \apj, 788, 59

\bibitem[{{Williams} \& {Cieza}(2011)}]{WilliamsCieza2011}
{Williams}, J.~P., \& {Cieza}, L.~A. 2011, \araa, 49, 67

\bibitem[{{Wilson} \& {Rood}(1994)}]{wilson1994}
{Wilson}, T.~L., \& {Rood}, R. 1994, \araa, 32, 191

\bibitem[{{Winn} \& {Fabrycky}(2015)}]{WinnFabrycky2015}
{Winn}, J.~N., \& {Fabrycky}, D.~C. 2015, \araa, 53, 409

\bibitem[{{Woitke} {et~al.}(2009){Woitke}, {Kamp}, \& {Thi}}]{woitke2009}
{Woitke}, P., {Kamp}, I., \& {Thi}, W.-F. 2009, \aap, 501, 383

\bibitem[{{Xu} {et~al.}(2017){Xu}, {Bai}, \& {{\"O}berg}}]{xu2016}
{Xu}, R., {Bai}, X.-N., \& {{\"O}berg}, K. 2017, \apj, 835, 162

\bibitem[{{Youdin} \& {Goodman}(2005)}]{youdin2005}
{Youdin}, A.~N., \& {Goodman}, J. 2005, \apj, 620, 459

\bibitem[{{Youdin} \& {Shu}(2002)}]{youdin2002}
{Youdin}, A.~N., \& {Shu}, F.~H. 2002, \apj, 580, 494

\bibitem[{{Yu} {et~al.}(2017){Yu}, {Evans}, {Dodson-Robinson}, {Willacy}, \&
  {Turner}}]{yu2017}
{Yu}, M., {Evans}, N.~J., {Dodson-Robinson}, S.~E., {Willacy}, K., \& {Turner},
  N.~J. 2017, ArXiv e-prints, arXiv:1704.05508

\bibitem[{{Yu} {et~al.}(2016){Yu}, {Willacy}, {Dodson-Robinson}, {Turner}, \&
  {Evans}}]{yu2016}
{Yu}, M., {Willacy}, K., {Dodson-Robinson}, S.~E., {Turner}, N.~J., \& {Evans},
  II, N.~J. 2016, \apj, 822, 53

\bibitem[{{Zhang} {et~al.}(2014){Zhang}, {Isella}, {Carpenter}, \&
  {Blake}}]{zhang2014}
{Zhang}, K., {Isella}, A., {Carpenter}, J.~M., \& {Blake}, G.~A. 2014, \apj,
  791, 42

\end{thebibliography}

\clearpage
\appendix

\section{Distances to the Cha I and Lupus Star Forming Regions} \label{gaia}
We calculated a distance of $188\pm12$ pc to Cha~I from the average Gaia DR1 TGAS parallax to Cha~I members HD 97048, HD 97300, CV Cha, CR Cha, and DI Cha \citep{Gaia2016}. The uncertainty includes a systematic uncertainty in parallax of $\sim 0.3$ mas/yr and a standard deviation of $0.2$ mas/yr in the measurements and excludes uncertainty introduced by the decision to ignore candidate members  HD 93828 and HD 96675. The star HD 93828 is spatially coincident with the projected location of Cha I with a distance of $206\pm8$ pc (excluding the systematic error of $\sim 0.3$ mas/yr), but has a proper motion consistent with Cha II \citep{Lopez2013}. The star HD 96675 is located at $161\pm7$ pc (again excluding systematic errors), a statistically significant outlier from the stochastic uncertainty in the Cha I distance.  This Cha I distance is consistent with the $180 \pm 10$ pc measured by Voirin et al.~(2017, submitted) using Gaia parallaxes combined with the distribution of reddening along the line of sight.

We adopt a distance of $157\pm10$ pc to Lupus, calculated from the average Gaia parallax to Lupus members with disks (Sz 68, Sz 82, RU Lup, HD 142527, RY Lup, and 2MASS J16083070-3828268). These objects have been previously identified as members of distinct regions Lupus I, Lupus II, Lupus III, and Lupus IV \citep{comeron2008}. The Gaia parallaxes establishes that these sub-regions of Lupus are all located at a similar distance, in contrast to previous assumptions based on uncertain Hipparcos parallaxes. The uncertainty includes a systematic uncertainty in parallax of $\sim 0.3$ mas/yr and a standard deviation of $0.24$ mas/yr in the measurements. Several likely members\footnote{RX J1511.0-3252, HD 135127, RXJ1518.4-3738, RXJ1524.5-3652, RX J1529.7-3628, RXJ1531.3-3329, RX J1546.6-3618, RX J1547.6-4018, RXJ1549.9-3629, RXJ1605.7-3905, RXJ1610.0-4016, HD147402} were identified in a ROSAT X-ray survey \citep{wichmann1997} and have proper motions consistent with Lupus membership \citep{galli2013,Gaia2016}, but have an average distance of 139 pc. Since their initial identifications and the brightness-limited Gaia measurements may be biased to nearby objects, they are excluded from our distance analysis. These diskless members may be an older population, similar to the halo of older stars that surrounds the Taurus Molecular Cloud complexes \citep{kraus2017}.

\newpage

\section{Maps and comments for CO Detections}
\label{app-zoo}

\setcounter{figure}{0}
\renewcommand{\thefigure}{B\arabic{figure}}

\begin{figure}[!t]
\centerline{\includegraphics[scale=0.9,trim=0 50 0 50]{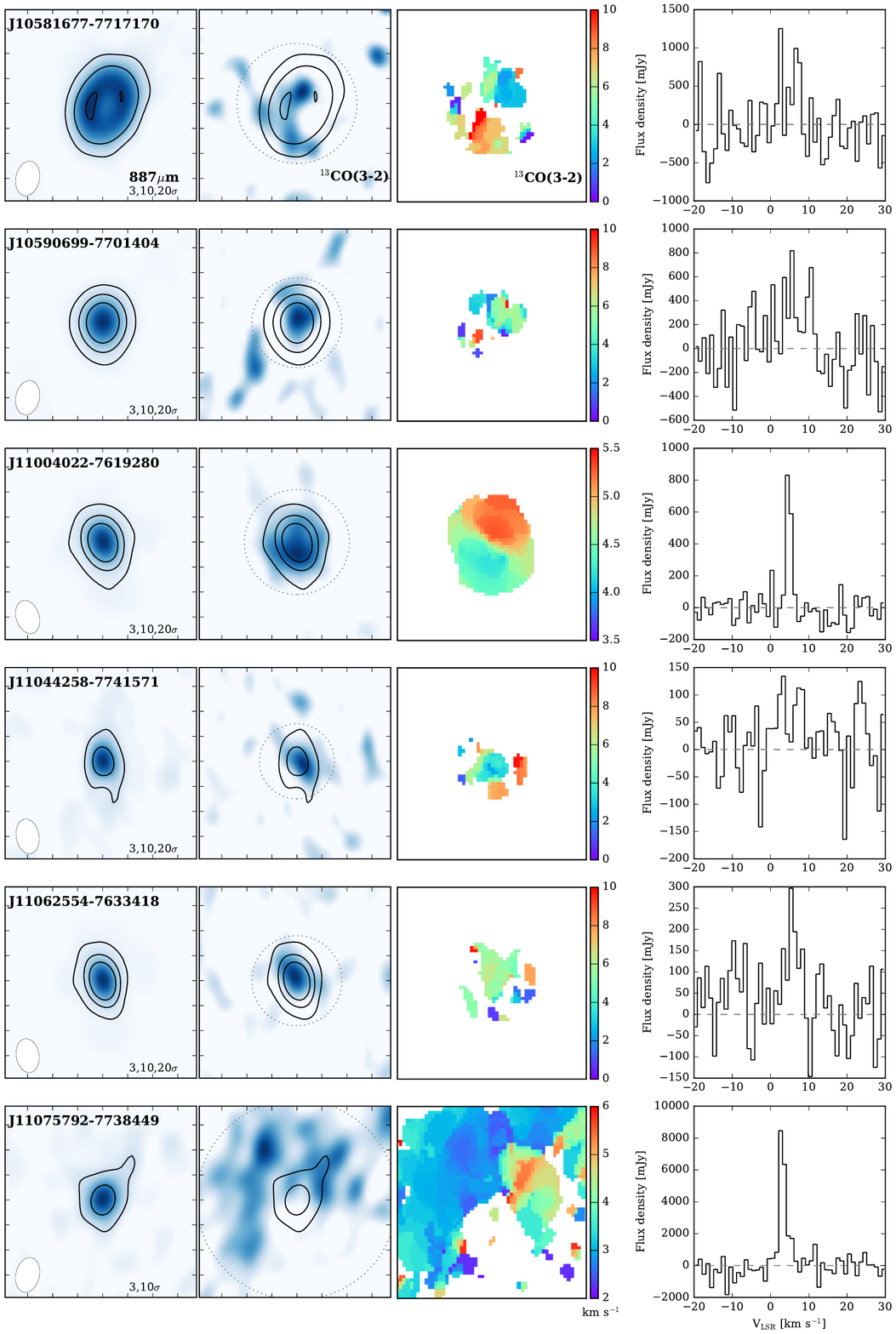}}
\captcont{From left to right: ALMA 887 $\mu$m continuum image, $^{13}$CO moment 0 map, $^{13}$CO moment 1 map with 3$\sigma$ cutoff, and $^{13}$CO spectrum with extraction region shown in dash circle in moment 0 map for $^{13}$CO detections in Cha I  observed by our ALMA Cycle 2 program. Images are $4^{\prime\prime}\times4^{\prime\prime}$ in size. Synthesized beams are shown in the lower right corner of the continuum images. \label{fig-zoo}}
\end{figure}

\begin{figure}[!ht]
\centerline{\includegraphics[scale=0.9,trim=0 50 0 50]{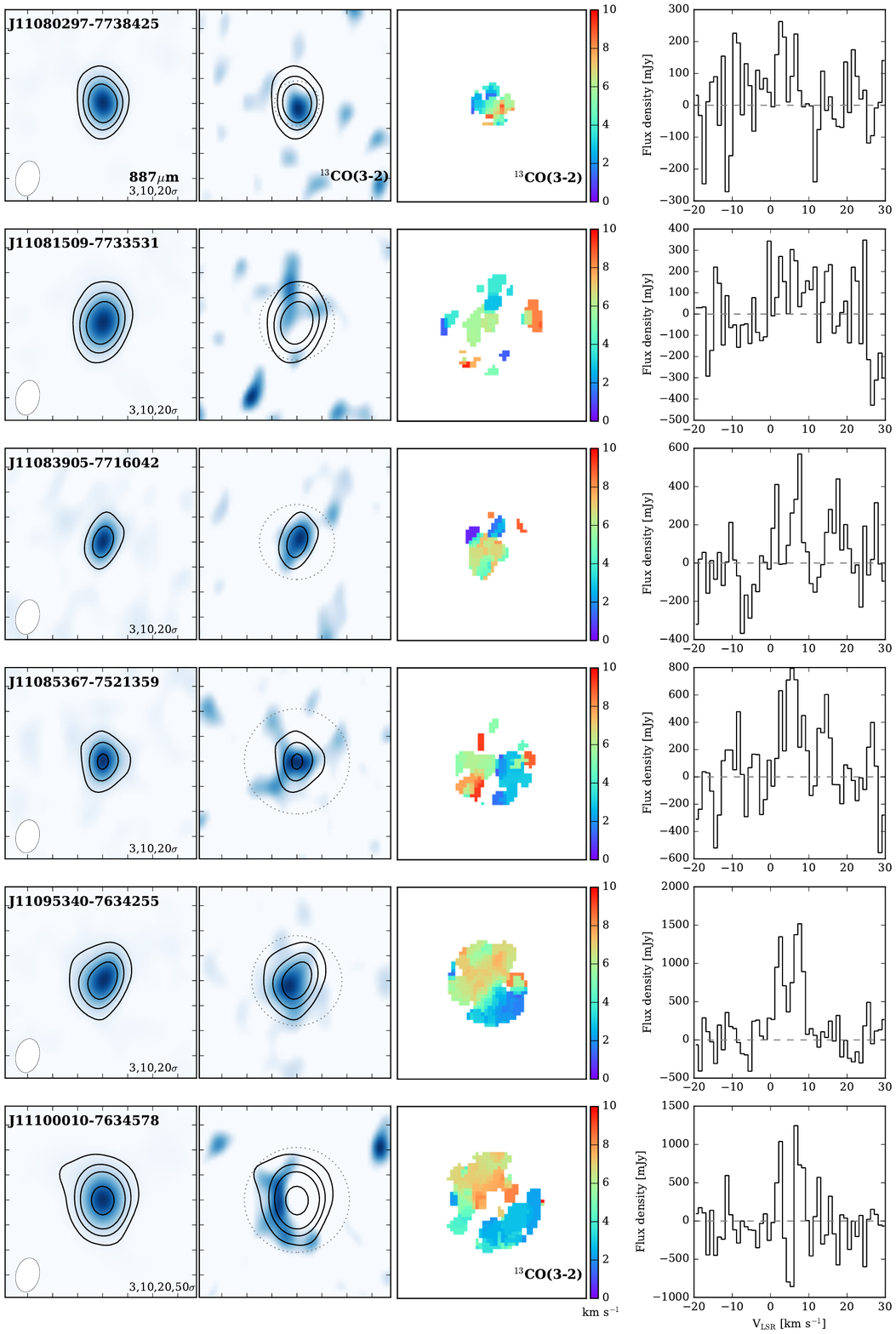}}
\captcont{Cont.}
\end{figure}

\begin{figure}[!ht]
\centerline{\includegraphics[scale=0.9,trim=0 50 0 50]{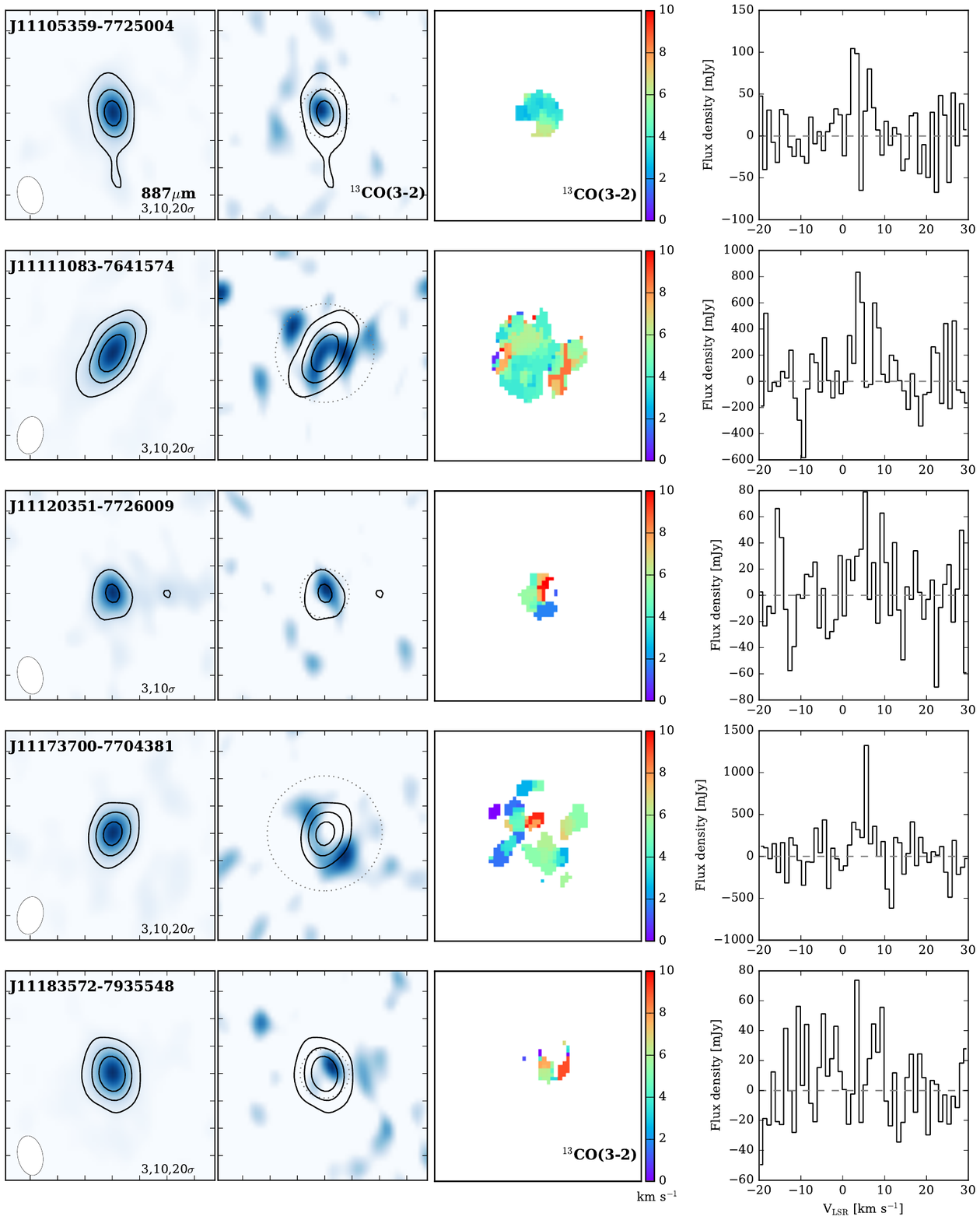}}
\captcont{Cont.}
\end{figure}

\pagebreak

\subsection{Comments on individual detections}\label{sec:indldetections}
In this subsection, we discuss the CO emission properties of the two sources 2MASS J11100010-7634578 and 2MASS J11075792-7738449 in detail.

\setcounter{figure}{1}
\renewcommand{\thefigure}{B\arabic{figure}}

\begin{figure*}
 \centering
  \includegraphics[scale=0.5]{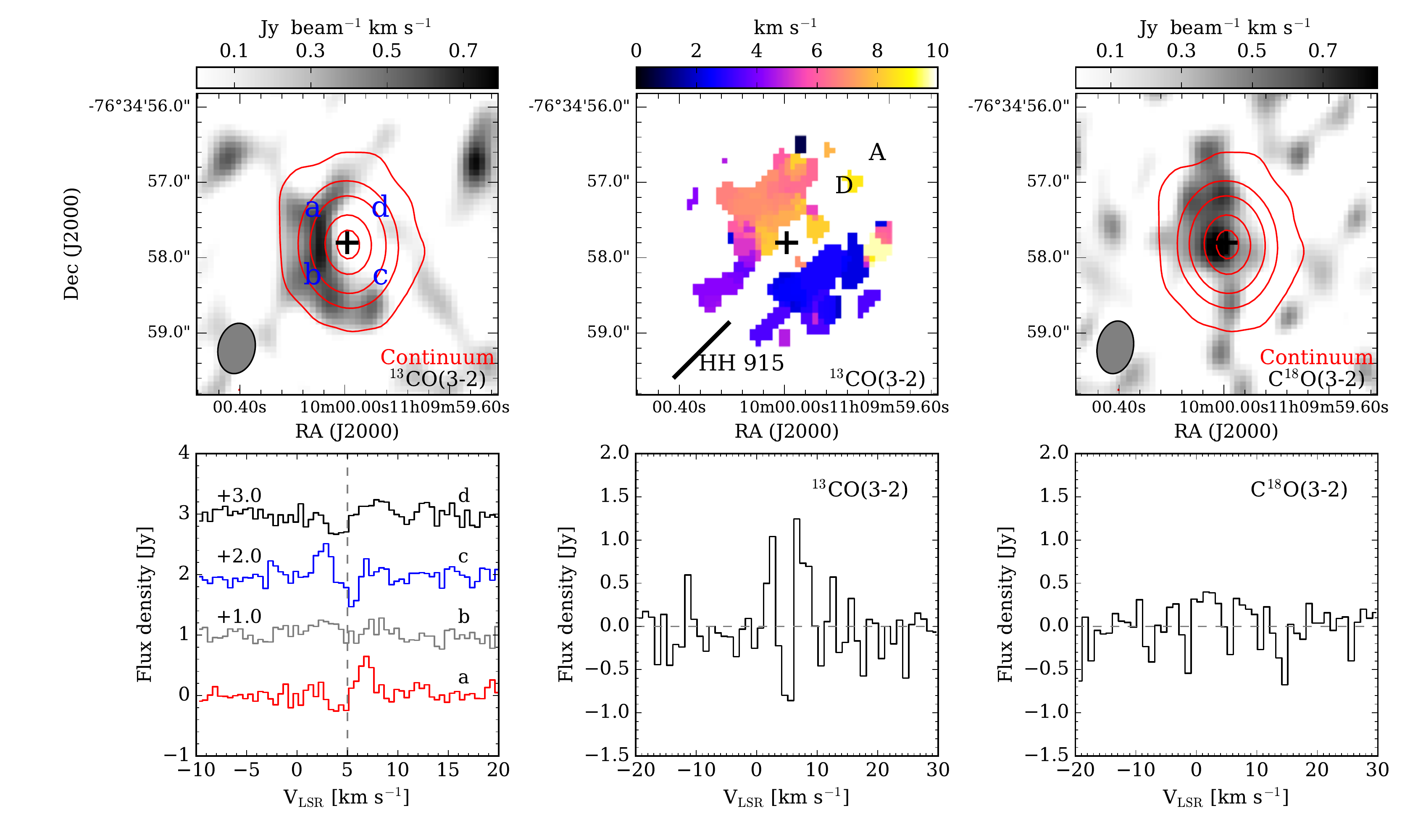}
\caption{2MASS J11100010-7634578: Integrated Intensity maps (Moment 0 map) of $^{13}$CO and C$^{18}$O are shown in the upper left and upper right panels, respectively, with 887 $\mu$m continuum contours at [3,30,100,300,500]$\sigma$, with 1$\sigma$ $\sim$ 1mJy beam$^{-1}$. The velocity map (Moment 1 map) of $^{13}$CO is in the upper middle panel. The HH object position associated with this source is noted in the velocity map, as well as the two H$_2$ emission knots, A and D. Four spectra are extracted from the a, b, c, d position with $0\farcs5$ radius aperture from $^{13}$CO intensity map. Line spectra for $^{13}$CO and C$^{18}$O are also shown as comparison, extracted from determined aperture sizes of 1.05 and 0.75 arcsec, respectively. \label{fig:J11100010}}
\end{figure*}

\subsubsection{2MASS J11100010-7634578 } \label{sec:J11100010}
2MASS J11100010-7634578 displays an incomplete and asymmetric ring structure in $^{13}$CO emission and is the only source detected in C$^{18}$O.
The $^{13}$CO intensity map, created by integrating the flux from 0 -- 10 \kms, shows a lack of $^{13}$CO emission at the continuum center and to the NW of the disk (Figure \ref{fig:J11100010}, upper left). To understand this peculiar distribution of $^{13}$CO emission, we extract four spectra from $0\farcs5$ radius aperture in the marked positions of intensity map. Positions a and c correspond to the red- and blue-shifted components to the NE and SW directions, respectively, indicating a Keplerian-like disk structure (also see the velocity map in Figure \ref{fig:J11100010}). The flux difference in SW and NE directions is attributed to the apparent negative flux around 5 km s$^{-1}$ in the blue-shifted component (marker c). The reason for negative flux in spectra c and d is unclear and may be caused by either random noise fluctuations or perhaps foreground absorption in Cha I cloud. The C$^{18}$O emission is more compact towards the source center than $^{13}$CO and is not affected by the negative fluxes.

The flux ratio of $^{13}$CO to C$^{18}$O emission is $\sim 1$, when extracting fluxes over the 0--10 km/s spectral range and the same aperture of $1.05$ arcsec.  If the velocity range from 4--6 km/s is excluded from both spectra, the $^{13}$CO to C$^{18}$O flux ratio increases to $\sim 2$. If the $^{13}$CO emission is optically thick where the sub-mm continuum is produced, then the continuum emission under the line would be absorbed and lower the line flux after continuum subtraction \citep[see supplementary material in][]{isella2016}.  At any given location in the disk, the line would be most opaque within the turbulent velocity of $\sim 0.5$ km s$^{-1}$ \citep[e.g.][]{hughes11}.  Correcting the $^{13}$CO flux for this effect would increase the line flux by $\sim 30\%$.

2MASS J11100010-7634578 is thought to drive the highly collimated jet HH 915, with a P.A. $\sim$ 135$\degr$ \citep{Schegerer2006}. Two near-infrared H$_2$ emission knots (marked as A and D in the velocity map in Figure \ref{fig:J11100010}), located to the NW of the source location, were suggested as counterparts of the HH 915 object \citep{wang2006}. The jet direction is perpendicular to the red and blue-shifted component direction (in NE--SW),
as depicted in the velocity map. Therefore, the asymmetric $^{13}$CO emission likely traces the Keplerian rotation of disk materials, with the NW part of disk contaminated by high noise fluctuation or strong absorption.

This source has a stellar companion discovered from near-IR interferometry with a projected separation of 1 AU \citep{anthonioz2015}, which is too close to affect the disk at the large spatial scales detectable in our ALMA observations.  However, the presence of a companion may affect disk dispersal timescale and therefore the disk mass and size \citep[e.g.][]{kraus12,harris12}.

\setcounter{figure}{2}
\renewcommand{\thefigure}{B\arabic{figure}}

\begin{figure*}[t]
 \centering
  \includegraphics[width=0.95\textwidth]{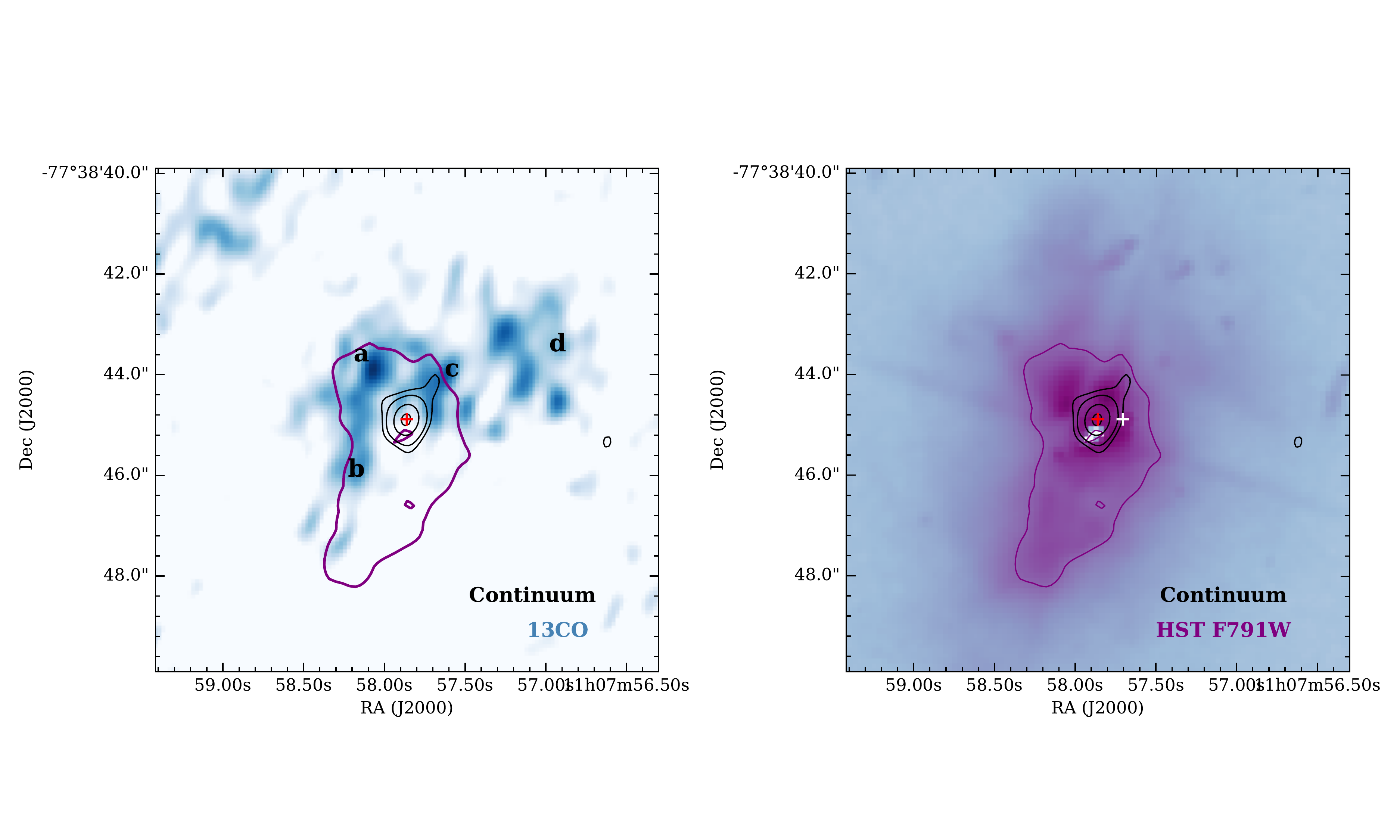}
\caption{2MASS J11075792-7738449. Left: The $^{13}$CO intensity map, integrated from 2 - 6 km s$^{-1}$. Continuum emission is also shown in black contours, with contour levels of [3,5,10,15]$\sigma$. The source position and the newly found 0.5 arcsec companion are marked with red and white cross, respectively. Four emission components are labeled as a, b ,c, d. Right: Circumstellar disc/reflection nebula in HST filter F791W. The purple contour depicts the nebula edge at 3$\%$ peak level. \label{fig:J11075792} }
\end{figure*}

\subsubsection{2MASS J11075792-7738449} \label{sec:J11075792}
While sub-mm continuum emission is clearly detected at the source position of 2MASS J11075792-7738449, the associated $^{13}$CO emission is completely offset from its continuum center, as shown in Figure \ref{fig:J11075792} (see also the channel map in Figure \ref{fig:channel}). The $^{13}$CO emission peaks to the NE of the source position by $\sim$ 1$\arcsec$ (component a) and extends to the south by 3$\arcsec$ (component b).

High-resolution optical imaging with {\it HST} reveals a reflection nebulosity associated with the object \citep{Schmidt2013}, shown as the background in the right panel of Figure \ref{fig:J11075792}. \citet{Schmidt2013} argues that the reflection nebulosity is illuminated by the Herbig A0 star HD 97048, separated by $\sim$ 37$\arcsec$ (0.034 pc for a Cha I distance of 189 pc) to the SE. The extended $^{13}$CO emission (components a and b) is spatially consistent with the reflection nebulosity. The measured $^{13}$CO flux increases significantly when including visibilities with baselines $<$ 40 k$\lambda$. Since short baselines trace extended emission,  the increased flux may originate from the nebula or nearby clouds.  This origin may also be the most likely cause for component d, since no optical or sub-mm source is associated with it.

\citet{Schmidt2013} discovered an $\sim$M4.5 companion, which is located $0\farcs5$ ($\sim 94$ AU) W of the primary star and has a proper motion suggesting either an edge-on or a highly eccentric orbit.  Since the dust disk has survived, the lack of $^{13}$CO emission in the dust disk therefore should not be related to any disk dispersal caused by the companion. The detected $^{13}$CO emission may not trace the disk but the nearby clouds instead. More sensitive observations are needed to understand the $^{13}$CO emission from and around 2MASS J11075792-7738449.

\setcounter{figure}{3}
\renewcommand{\thefigure}{B\arabic{figure}}

\begin{figure*}[h]
    \centering
        \includegraphics[scale=0.9]{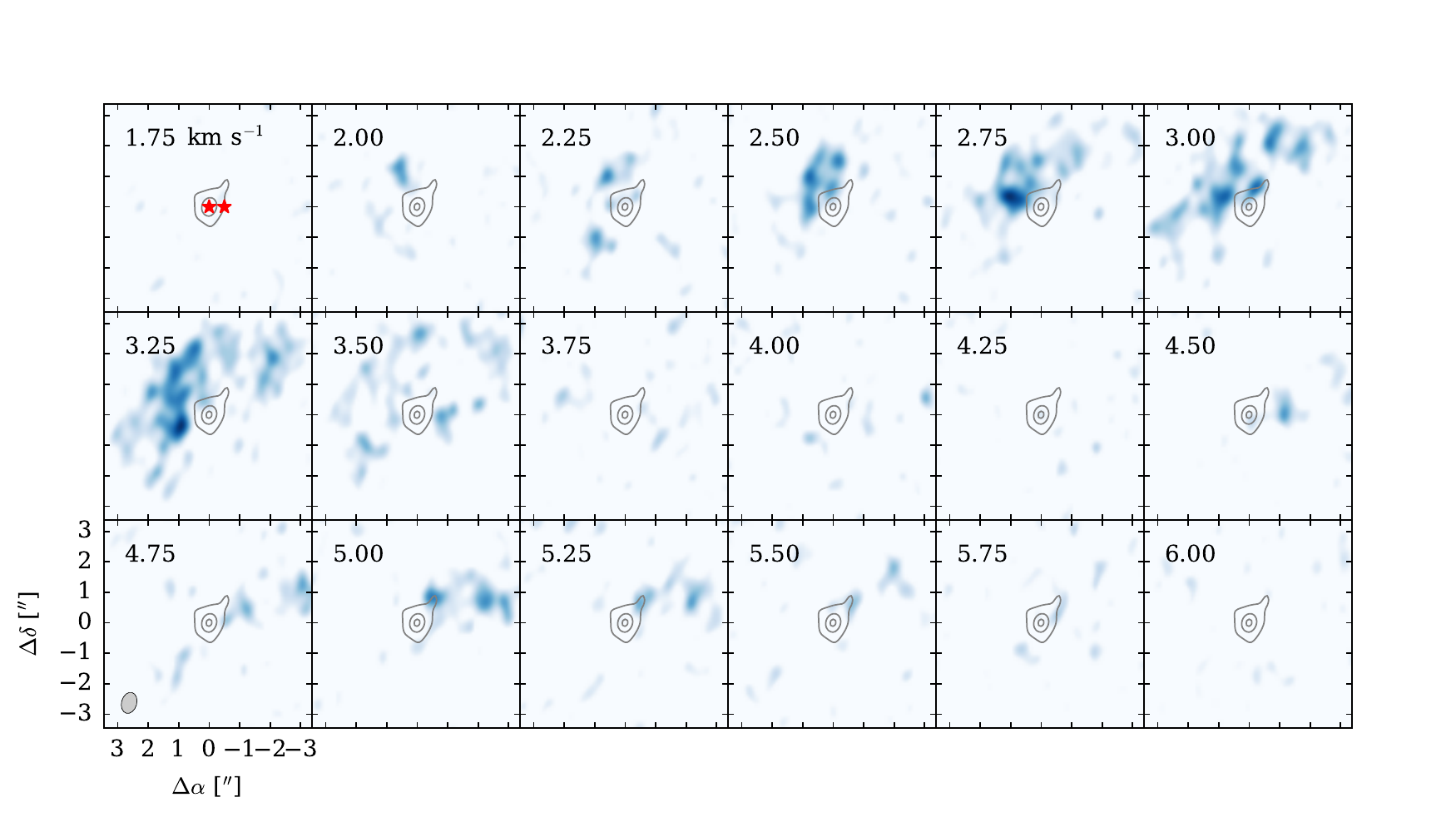}
\caption{2MASS J11075792-7738449 channel map with channel velocity shown in the upper left corner. Continuum emission is shown in black contours, with contour levels of [3,10,15]$\sigma$. The source position and the newly found 0.5 arcsec companion are marked with red stars. \label{fig:channel}}
\end{figure*}

\clearpage

\begin{deluxetable}{lcccccccc}
\tabletypesize{\tiny}
\tablecaption{Source Properties and Measured CO Fluxes \label{tab:co_flux_all}}
\tablewidth{0pt}
\tablehead{
\colhead{2MASS} & \colhead{SpTy}  &\colhead{log(M$_{\ast}$)} & \colhead{F$_{887{\mu}m}$} & \multicolumn{3}{c}{$^{13}$CO J = 3-2} & \multicolumn{2}{c}{C$^{18}$O J = 3-2} \\
\cline{5-7}
\cline{8-9}
\colhead{ } &\colhead{ } &\colhead{ } &\colhead{ } & \colhead{Flux} & \multicolumn{2}{c}{S/N} & \colhead{Flux} & \colhead{S/N}  \\
\colhead{ } &\colhead{ } & \colhead{ [M$_{\sun}$] } & \colhead{[mJy] } &  \colhead{[mJy km $s^{-1}$]} & \colhead{$0\farcs6$} & \colhead{$0\farcs3$} & \colhead{[mJy km $s^{-1}$]} & \colhead{$0\farcs3$} }

\startdata
J10533978-7712338 &    M2 &  -0.41  &     4.6   $\pm$   0.79  &  240  $\pm$ 289  &  0.8  &  1.5  &   95  $\pm$ 119  &  0.8 \\
J10555973-7724399 &    K7 &  -0.13  &    34.1   $\pm$   1.32  & -525  $\pm$ 295  & -1.8  & -1.6  &  -39  $\pm$ 117  & -0.3 \\
J10561638-7630530 &  M6.5 &  -0.96  &    3.99   $\pm$   0.16  &  116  $\pm$  80  &  1.4  &  2.2  &  -57  $\pm$  45  & -1.3 \\
J10563044-7711393 &    K7 &  -0.07  &  117.58   $\pm$    1.1  &  508  $\pm$ 269  &  1.9  &  1.0  &  144  $\pm$ 134  &  1.1 \\
J10574219-7659356 &    M3 &  -0.52  &    9.12   $\pm$   0.83  &  265  $\pm$ 269  &  1.0  &  0.8  &   13  $\pm$ 123  &  0.1 \\
J10580597-7711501 &  M5.5 &  -0.96  &    2.68   $\pm$   0.16  &  102  $\pm$  78  &  1.3  &  0.5  &   71  $\pm$  43  &  1.7 \\
J10581677-7717170 &    K2 &    0.1  &  310.18   $\pm$    1.0  & 1091  $\pm$ 271  &  4.0  &  3.5  &   10  $\pm$ 129  &  0.1 \\
J10590108-7722407 &    K7 &  -0.07  &   65.34   $\pm$    1.7  &  224  $\pm$ 289  &  0.8  &  0.2  &  111  $\pm$ 125  &  0.9 \\
J10590699-7701404 &    K0 &   0.23  &  442.18   $\pm$   0.76  & 1361  $\pm$ 278  &  4.9  &  5.7  &   17  $\pm$ 118  &  0.1 \\
J11004022-7619280 &    M4 &  -0.62  &   69.75   $\pm$   0.17  &  602  $\pm$  82  &  7.3  &  7.4  &   99  $\pm$  44  &  2.3 \\
J11022491-7733357 &    K2 &   0.13  &  225.68   $\pm$   0.74  &  740  $\pm$ 267  &  2.8  &  2.4  &  296  $\pm$ 130  &  2.3 \\
J11023265-7729129 &    M3 &  -0.52  &   -0.21   $\pm$   0.82  &  466  $\pm$ 280  &  1.7  &  1.1  &   76  $\pm$ 129  &  0.6 \\
J11025504-7721508 &  M4.5 &  -0.74  &    1.16   $\pm$   0.16  &  -55  $\pm$  77  & -0.7  & -0.4  &   -9  $\pm$  44  & -0.2 \\
J11040425-7639328 &  M4.5 &  -0.74  &    2.77   $\pm$   0.16  &  180  $\pm$  77  &  2.3  &  2.4  &   10  $\pm$  49  &  0.2 \\
J11040909-7627193 &    K5 &  -0.06  &  104.78   $\pm$    0.6  & -259  $\pm$ 281  & -0.9  & -1.4  &  -43  $\pm$ 125  & -0.3 \\
J11044258-7741571 &    M4 &  -0.62  &    4.15   $\pm$   0.16  &  336  $\pm$  77  &  4.3  &  4.9  &   45  $\pm$  43  &  1.0 \\
J11045701-7715569 &    M3 &  -0.53  &    2.54   $\pm$   0.81  &  284  $\pm$ 274  &  1.0  &  1.5  & -124  $\pm$ 122  & -1.0 \\
J11062554-7633418 &  M5.5 &  -0.91  &   46.05   $\pm$   0.15  &  485  $\pm$  82  &  5.9  &  6.4  &   21  $\pm$  43  &  0.5 \\
J11062942-7724586 &    M6 &  -1.12  &    0.25   $\pm$   0.16  & -107  $\pm$  79  & -1.3  & -1.4  &   43  $\pm$  40  &  1.1 \\
J11063276-7625210 &  M6.5 &  -1.13  &   -0.01   $\pm$   0.16  &   18  $\pm$  81  &  0.2  &  1.0  &   81  $\pm$  43  &  1.9 \\
J11063945-7736052 &    M5 &  -0.78  &    0.37   $\pm$   0.16  &   98  $\pm$  81  &  1.2  &  0.7  &   34  $\pm$  41  &  0.8 \\
J11064180-7635489 &    M5 &  -0.78  &    0.97   $\pm$   0.16  & -214  $\pm$  84  & -2.5  & -4.1  &   -6  $\pm$  43  & -0.1 \\
J11064510-7727023 &    K6 &  -0.03  &    0.53   $\pm$   0.82  &  297  $\pm$ 303  &  1.0  &  0.4  &   -4  $\pm$ 117  & -0.0 \\
J11065906-7718535 &  M4.5 &  -0.71  &   24.28   $\pm$   0.35  &  124  $\pm$  77  &  1.6  &  2.2  &   32  $\pm$  41  &  0.8 \\
J11065939-7530559 &  M5.5 &  -0.97  &    3.11   $\pm$   0.16  &  -13  $\pm$  84  & -0.2  &  0.4  &  138  $\pm$  41  &  3.4 \\
J11070925-7718471 &    M3 &  -0.52  &    0.06   $\pm$   0.82  &   35  $\pm$ 259  &  0.1  & -0.2  &   92  $\pm$ 127  &  0.7 \\
J11071181-7625501 &  M5.5 &  -0.97  &    0.03   $\pm$   0.16  &  -50  $\pm$  80  & -0.6  & -0.8  &   65  $\pm$  45  &  1.4 \\
J11071206-7632232 &    M0 &  -0.23  &    4.23   $\pm$   0.81  &  195  $\pm$ 288  &  0.7  &  1.5  &   53  $\pm$ 127  &  0.4 \\
J11071330-7743498 &    M4 &  -0.63  &    0.42   $\pm$   0.81  & -117  $\pm$ 268  & -0.4  & -0.6  &   18  $\pm$ 110  &  0.2 \\
J11071860-7732516 &  M5.5 &  -0.92  &    0.93   $\pm$   0.16  &    1  $\pm$  87  &  0.0  &  0.6  &   25  $\pm$  43  &  0.6 \\
J11072074-7738073 &    K0 &   0.29  &   26.36   $\pm$   1.46  & -597  $\pm$ 262  & -2.3  & -2.5  &    6  $\pm$ 117  &  0.1 \\
J11072825-7652118 &    M3 &  -0.53  &     1.5   $\pm$   0.81  & -463  $\pm$ 275  & -1.7  & -2.4  &   10  $\pm$ 119  &  0.1 \\
J11074245-7733593 &  M5.5 &  -0.88  &    2.37   $\pm$   0.41  & -110  $\pm$  77  & -1.4  & -1.7  &  -58  $\pm$  43  & -1.4 \\
J11074366-7739411 &    M1 &  -0.31  &  107.27   $\pm$   0.56  &  644  $\pm$ 280  &  2.3  &  1.7  &  110  $\pm$ 113  &  1.0 \\
J11074656-7615174 &  M6.5 &  -1.15  &    2.18   $\pm$   0.16  &  163  $\pm$  82  &  2.0  &  2.3  &   22  $\pm$  45  &  0.5 \\
J11075730-7717262 &  M1.2 &  -0.31  &    6.47   $\pm$    0.8  &  458  $\pm$ 268  &  1.7  &  2.0  &  -50  $\pm$ 119  & -0.4 \\
J11075792-7738449 &    K5 &  -0.01  &   19.85   $\pm$   1.48  & 1435  $\pm$ 268  &  5.6  &  3.8  &  -28  $\pm$ 114  & -0.3 \\
J11075809-7742413 &    M3 &  -0.51  &    6.45   $\pm$   0.79  & -459  $\pm$ 256  & -1.8  & -1.1  &  -78  $\pm$ 131  & -0.6 \\
J11080002-7717304 &    K7 &  -0.18  &   -0.69   $\pm$    0.8  &    0  $\pm$ 263  &  0.0  & -0.4  &  254  $\pm$ 109  &  2.3 \\
J11080148-7742288 &    K7 &   -0.2  &   44.37   $\pm$   0.82  &  355  $\pm$ 251  &  1.4  &  1.9  &   50  $\pm$ 122  &  0.4 \\
J11080297-7738425 &    M1 &   -0.2  &  102.24   $\pm$   0.58  &  803  $\pm$ 268  &  3.0  &  4.1  &   33  $\pm$ 115  &  0.3 \\
J11081509-7733531 &    K0 &   0.12  &  209.29   $\pm$   0.43  &  740  $\pm$ 263  &  2.8  &  2.5  &  195  $\pm$ 119  &  1.6 \\
J11081850-7730408 &  M6.5 &  -1.14  &    0.26   $\pm$   0.16  &   50  $\pm$  80  &  0.6  &  0.5  &  -54  $\pm$  45  & -1.2 \\
J11082238-7730277 &  M5.5 &   -0.9  &    0.23   $\pm$   0.16  &   78  $\pm$  79  &  1.0  &  0.9  &   -2  $\pm$  44  & -0.1 \\
J11082570-7716396 &    M8 &  -1.51  &    0.23   $\pm$   0.15  &   68  $\pm$  74  &  0.9  &  0.6  &  -36  $\pm$  42  & -0.8 \\
J11082650-7715550 &  M5.5 &  -0.96  &   -0.24   $\pm$   0.16  &  -73  $\pm$  79  & -0.9  & -0.7  &   13  $\pm$  45  &  0.3 \\
J11083905-7716042 &    K7 &  -0.08  &   14.11   $\pm$   0.79  & 1174  $\pm$ 252  &  4.7  &  6.0  &  264  $\pm$ 111  &  2.4 \\
J11083952-7734166 &  M6.5 &  -0.99  &    0.02   $\pm$   0.16  &   25  $\pm$  81  &  0.3  & -0.3  &   38  $\pm$  41  &  0.9 \\
J11085090-7625135 &  M5.5 &   -0.9  &   -0.04   $\pm$   0.16  &  -30  $\pm$  88  & -0.3  & -1.0  &   -3  $\pm$  44  & -0.1 \\
J11085367-7521359 &    M1 &  -0.28  &    24.6   $\pm$   1.37  & 1344  $\pm$ 280  &  4.8  &  5.0  &    3  $\pm$ 122  &  0.0 \\
J11085464-7702129 &  M0.5 &  -0.18  &     3.9   $\pm$   0.79  & -506  $\pm$ 255  & -2.0  & -1.4  &   -2  $\pm$ 127  & -0.0 \\
J11085497-7632410 &  M5.5 &  -0.91  &    0.46   $\pm$   0.16  &  138  $\pm$  89  &  1.6  &  1.8  &  100  $\pm$  44  &  2.3 \\
J11091812-7630292 &    M0 &  -0.18  &     1.3   $\pm$   0.79  &  283  $\pm$ 267  &  1.1  &  1.7  & -155  $\pm$ 118  & -1.3 \\
J11092266-7634320 &    M1 &  -0.25  &    3.85   $\pm$   0.78  &  420  $\pm$ 274  &  1.5  &  0.9  &  102  $\pm$ 108  &  0.9 \\
J11092379-7623207 &  M0.5 &  -0.29  &  123.11   $\pm$   0.57  &  192  $\pm$ 241  &  0.8  &  1.2  &   40  $\pm$ 118  &  0.3 \\
J11094260-7725578 &    M5 &  -0.77  &    0.37   $\pm$   0.16  &   38  $\pm$  77  &  0.5  &  1.0  &    6  $\pm$  43  &  0.1 \\
J11094621-7634463 &    M3 &  -0.47  &    4.73   $\pm$   0.79  &  213  $\pm$ 272  &  0.8  &  0.2  &  105  $\pm$ 114  &  0.9 \\
J11094742-7726290 &    M1 &  -0.22  &  147.85   $\pm$   0.86  &  293  $\pm$ 255  &  1.1  &  0.4  &    8  $\pm$ 114  &  0.1 \\
J11095215-7639128 &  M6.2 &   -1.2  &    0.37   $\pm$   0.16  &   14  $\pm$  76  &  0.2  &  0.4  &  -21  $\pm$  39  & -0.5 \\
J11095336-7728365 &  M5.5 &  -0.96  &    0.29   $\pm$   0.16  &   28  $\pm$  76  &  0.4  & -0.9  &   24  $\pm$  43  &  0.6 \\
J11095340-7634255 &    K7 &  -0.12  &    76.1   $\pm$   1.83  & 3526  $\pm$ 270  & 13.0  & 12.0  &   62  $\pm$ 116  &  0.5 \\
J11095407-7629253 &    M1 &  -0.21  &   30.49   $\pm$   1.24  &   51  $\pm$ 279  &  0.2  & -0.8  &  117  $\pm$ 117  &  1.0 \\
J11095873-7737088 &  M0.5 &  -0.29  &   20.81   $\pm$   0.57  & -259  $\pm$ 275  & -0.9  & -0.1  &   96  $\pm$ 110  &  0.9 \\
J11100010-7634578 &    K0 &   0.21  & 1363.47   $\pm$   0.82  &  450  $\pm$ 272  &  1.7  & -0.8  &  476  $\pm$ 119  &  4.0 \\
J11100369-7633291 &    M0 &  -0.14  &    9.83   $\pm$   0.79  &  254  $\pm$ 274  &  0.9  &  0.3  &   43  $\pm$ 111  &  0.4 \\
J11100469-7635452 &    K7 &  -0.08  &    7.73   $\pm$   0.78  & -165  $\pm$ 276  & -0.6  & -0.5  & -120  $\pm$ 117  & -1.0 \\
J11100704-7629376 &    K7 &  -0.14  &    7.17   $\pm$   0.78  & -104  $\pm$ 246  & -0.4  &  0.3  &   14  $\pm$ 114  &  0.1 \\
J11100785-7727480 &  M5.5 &  -0.89  &    0.49   $\pm$   0.16  &  148  $\pm$  75  &  2.0  &  0.6  &   38  $\pm$  41  &  0.9 \\
J11101141-7635292 &    K5 &    0.0  &   73.82   $\pm$    1.4  &  252  $\pm$ 251  &  1.0  & -0.2  &   69  $\pm$ 114  &  0.6 \\
J11103801-7732399 &    K4 &   0.05  &    5.37   $\pm$   0.78  &  426  $\pm$ 257  &  1.7  &  1.6  & -162  $\pm$ 126  & -1.3 \\
J11104141-7720480 &  M5.5 &  -0.96  &    -0.0   $\pm$   0.16  &   22  $\pm$  84  &  0.3  &  0.4  &   75  $\pm$  39  &  1.9 \\
J11104959-7717517 &    M2 &  -0.38  &   58.37   $\pm$   1.45  &  543  $\pm$ 244  &  2.2  &  2.0  &  210  $\pm$ 115  &  1.8 \\
J11105333-7634319 &    M3 &  -0.51  &   31.02   $\pm$   1.29  &  122  $\pm$ 275  &  0.4  &  1.4  &   99  $\pm$ 109  &  0.9 \\
J11105359-7725004 &    M5 &  -0.81  &    7.88   $\pm$   0.34  &  180  $\pm$  79  &  2.3  &  4.1  &   55  $\pm$  43  &  1.3 \\
J11105597-7645325 &  M6.5 &  -0.98  &    2.23   $\pm$   0.22  &  -83  $\pm$  70  & -1.2  & -2.9  &   12  $\pm$  43  &  0.3 \\
J11111083-7641574 &    M1 &  -0.31  &   54.27   $\pm$   1.75  & 1127  $\pm$ 266  &  4.2  &  3.6  &  110  $\pm$ 121  &  0.9 \\
J11113965-7620152 &  M3.5 &  -0.59  &   21.48   $\pm$    0.8  &  177  $\pm$ 254  &  0.7  &  1.0  &  -27  $\pm$ 125  & -0.2 \\
J11114632-7620092 &    K2 &   0.09  &    35.2   $\pm$   1.26  &  462  $\pm$ 285  &  1.6  &  0.7  &  278  $\pm$ 119  &  2.3 \\
J11120351-7726009 &  M5.5 &  -0.89  &    2.95   $\pm$   0.16  &  167  $\pm$  81  &  2.0  &  4.0  &   97  $\pm$  42  &  2.3 \\
J11120984-7634366 &    M5 &  -0.78  &    4.44   $\pm$   0.22  &  130  $\pm$  79  &  1.6  &  2.4  &    0  $\pm$  41  &  0.0 \\
J11122441-7637064 &    K2 &   0.04  &    0.19   $\pm$   0.78  &  210  $\pm$ 270  &  0.8  &  0.5  &  -70  $\pm$ 122  & -0.6 \\
J11122772-7644223 &    K0 &    0.2  &   59.05   $\pm$   1.29  &  468  $\pm$ 246  &  1.9  &  2.9  &  148  $\pm$ 114  &  1.3 \\
J11123092-7644241 &  M0.5 &  -0.17  &   12.18   $\pm$   0.83  & -413  $\pm$ 262  & -1.6  & -0.7  &  -71  $\pm$ 117  & -0.6 \\
J11124268-7722230 &    K0 &    0.2  &   -0.02   $\pm$   0.79  &  218  $\pm$ 284  &  0.8  & -0.5  &  -22  $\pm$ 115  & -0.2 \\
J11124861-7647066 &  M4.5 &  -0.69  &    -0.1   $\pm$   0.16  &   56  $\pm$  83  &  0.7  &  1.3  &  -71  $\pm$  41  & -1.7 \\
J11132446-7629227 &    M4 &  -0.62  &    8.07   $\pm$   0.79  &  223  $\pm$ 256  &  0.9  &  0.8  &  244  $\pm$ 125  &  1.9 \\
J11142454-7733062 &  M4.5 &  -0.71  &    7.43   $\pm$   0.34  &   19  $\pm$  81  &  0.2  &  0.4  &    5  $\pm$  39  &  0.1 \\
J11160287-7624533 &    K8 &  -0.19  &   12.83   $\pm$   1.68  &  488  $\pm$ 251  &  1.9  &  1.6  & -152  $\pm$ 121  & -1.3 \\
J11173700-7704381 &  M0.5 &  -0.28  &   28.26   $\pm$   1.29  & 1044  $\pm$ 258  &  4.0  &  1.9  &   60  $\pm$ 117  &  0.5 \\
J11175211-7629392 &  M4.5 &  -0.69  &   -0.31   $\pm$   0.16  &  -18  $\pm$  71  & -0.2  & -0.3  &   36  $\pm$  41  &  0.9 \\
J11183572-7935548 &    M5 &  -0.77  &   14.52   $\pm$   0.35  &  142  $\pm$  76  &  1.9  &  3.2  &   16  $\pm$  37  &  0.4 \\
J11241186-7630425 &  M5.5 &   -0.9  &    1.47   $\pm$   0.16  &  133  $\pm$  79  &  1.7  &  0.9  &  -26  $\pm$  41  & -0.6 \\
J11432669-7804454 &  M5.5 &  -0.86  &    1.36   $\pm$    0.5  &   28  $\pm$  82  &  0.3  &  1.0  &    6  $\pm$  40  &  0.2 \\
\enddata
\tablecomments{The spectral type, stellar mass and continuum flux are adopted from \citet{pascucci2016}.}
\end{deluxetable}

\newpage

\begin{deluxetable}{lcccrrrrrr}
\tabletypesize{\tiny}
\tablecaption{CO Fluxes and Gas Masses\label{tab:flux_mass}}
\tablewidth{0pt}
\tablehead{
\colhead{2MASS} &\colhead{F$_{13CO}$} & \colhead{radius}  & \colhead{F$_{C18O}$}  & \colhead{M$_{\rm gas}$} &\colhead{M$_{min}$ } & \colhead{M$_{max}$ }  & \colhead{M$_{\rm gas}$} &\colhead{M$_{min}$ } & \colhead{M$_{max}$}  \\
\cline{5-7}
\cline{8-10}
\colhead{ } & \colhead{} & \colhead{} & \colhead{} &\multicolumn{3}{c}{MvD16} &\multicolumn{3}{c}{WB14}  \\
\colhead{ } & \colhead{[mJy km $s^{-1}$]} & \colhead{[]$\arcsec$]} & \colhead{[mJy km $s^{-1}$]} & \multicolumn{6}{c}{[M$_{Jup}$]  } }
\startdata
J10533978-7712338   &    $\textless$    867           &   0.6   &    $\textless   $ 357          &  0.34 &    --  &    -- &  10.48  &    --  &     -- \\
J10555973-7724399   &    $\textless$    885           &   0.6   &    $\textless   $ 351          &  0.38 &    --  &    -- &  10.48  &    --  &     -- \\
J10561638-7630530   &    $\textless$    240           &   0.6   &    $\textless   $ 135          &  0.08 &    --  &    -- &   1.05  &    --  &     -- \\
J10563044-7711393   &    $\textless$    807           &   0.6   &    $\textless   $ 402          &  0.46 &    --  &    -- &  31.43  &    --  &     -- \\
J10574219-7659356   &    $\textless$    807           &   0.6   &    $\textless   $ 369          &  0.46 &    --  &    -- &  10.48  &    --  &     -- \\
J10580597-7711501   &    $\textless$    234           &   0.6   &    $\textless   $ 129          &  0.07 &    --  &    -- &   1.05  &    --  &     -- \\
J10581677-7717170   &       2535 $\pm$   576  &   1.2   &    $\textless   $ 387          &  0.84 &  0.19  &  5.05 &   1.57  &  1.05  &   3.14 \\
J10590108-7722407   &    $\textless$    867           &   0.6   &    $\textless   $ 375          &  0.34 &    --  &    -- &  31.43  &    --  &     -- \\
J10590699-7701404   &       1867 $\pm$   442  &   0.9   &    $\textless   $ 354          &  0.41 &  0.13  &  4.63 &    1.3  &  1.05  &  10.48 \\
J11004022-7619280   &       1004 $\pm$    86  &  1.05   &    $\textless   $ 132          &  0.21 &  0.08  &  1.71 &    0.6  &  0.31  &   3.14 \\
J11022491-7733357   &    $\textless$    801           &   0.6   &    $\textless   $ 390          &  0.45 &    --  &    -- &  31.43  &    --  &     -- \\
J11023265-7729129   &    $\textless$    840           &   0.6   &    $\textless   $ 387          &   0.5 &    --  &    -- &  31.43  &    --  &     -- \\
J11025504-7721508   &    $\textless$    230           &   0.6   &    $\textless   $ 132          &  0.07 &    --  &    -- &   1.05  &    --  &     -- \\
J11040425-7639328   &    $\textless$    230           &   0.6   &    $\textless   $ 147          &  0.07 &    --  &    -- &   1.05  &    --  &     -- \\
J11040909-7627193   &    $\textless$    843           &   0.6   &    $\textless   $ 375          &   0.5 &    --  &    -- &  31.43  &    --  &     -- \\
J11044258-7741571   &    $ $    413 $\pm$   108  &  0.75   & $\textless   $ 129          &  0.08 &  0.02  &  0.19 &    0.3  &   0.1  &   3.14 \\
J11045701-7715569   &    $\textless$    822           &   0.6   &    $\textless   $ 366          &  0.48 &    --  &    -- &  10.48  &    --  &     -- \\
J11062554-7633418   &        668 $\pm$   143  &   0.9   &    $\textless   $ 129          &  0.13 &  0.04  &  0.48 &   0.42  &  0.31  &   3.14 \\
J11062942-7724586   &    $\textless$    237           &   0.6   &    $\textless   $ 120          &  0.07 &    --  &    -- &   1.05  &    --  &     -- \\
J11063276-7625210   &    $\textless$    243           &   0.6   &    $\textless   $ 129          &  0.08 &    --  &    -- &   1.05  &    --  &     -- \\
J11063945-7736052   &    $\textless$    243           &   0.6   &    $\textless   $ 123          &  0.08 &    --  &    -- &   1.05  &    --  &     -- \\
J11064180-7635489   &    $\textless$    252           &   0.6   &    $\textless   $ 129          &  0.08 &    --  &    -- &   3.14  &    --  &     -- \\
J11064510-7727023   &    $\textless$    909           &   0.6   &    $\textless   $ 351          &  0.45 &    --  &    -- &  10.48  &    --  &     -- \\
J11065906-7718535   &    $\textless$    230           &   0.6   &    $\textless   $ 123          &  0.07 &    --  &    -- &   1.05  &    --  &     -- \\
J11065939-7530559   &    $\textless$    252           &   0.6   &    $\textless   $ 123          &  0.08 &    --  &    -- &   3.14  &    --  &     -- \\
J11070925-7718471   &    $\textless$    777           &   0.6   &    $\textless   $ 381          &  0.42 &    --  &    -- &  10.48  &    --  &     -- \\
J11071181-7625501   &    $\textless$    240           &   0.6   &    $\textless   $ 135          &  0.08 &    --  &    -- &   1.05  &    --  &     -- \\
J11071206-7632232   &    $\textless$    863           &   0.6   &    $\textless   $ 381          &  0.33 &    --  &    -- &  31.43  &    --  &     -- \\
J11071330-7743498   &    $\textless$    804           &   0.6   &    $\textless   $ 330          &  0.45 &    --  &    -- &  10.48  &    --  &     -- \\
J11071860-7732516   &    $\textless$    261           &   0.6   &    $\textless   $ 129          &  0.08 &    --  &    -- &   3.14  &    --  &     -- \\
J11072074-7738073   &    $\textless$    786           &   0.6   &    $\textless   $ 351          &  0.43 &    --  &    -- &  10.48  &    --  &     -- \\
J11072825-7652118   &    $\textless$    825           &   0.6   &    $\textless   $ 357          &  0.48 &    --  &    -- &  10.48  &    --  &     -- \\
J11074245-7733593   &    $\textless$    230           &   0.6   &    $\textless   $ 129          &  0.07 &    --  &    -- &   1.05  &    --  &     -- \\
J11074366-7739411   &    $\textless$    840           &   0.6   &    $\textless   $ 339          &   0.5 &    --  &    -- &  10.48  &    --  &     -- \\
J11074656-7615174   &    $\textless$    246           &   0.6   &    $\textless   $ 135          &  0.08 &    --  &    -- &   3.14  &    --  &     -- \\
J11075730-7717262   &    $\textless$    804           &   0.6   &    $\textless   $ 357          &  0.45 &    --  &    -- &  10.48  &    --  &     -- \\
J11075792-7738449   &      [11640 $\pm$   604]  &   2.0   &    $\textless   $ 342          & -- &    --  &    -- & --  & --  & -- \\
J11075809-7742413   &    $\textless$    768           &   0.6   &    $\textless   $ 393          &  0.41 &    --  &    -- &  10.48  &    --  &     -- \\
J11080002-7717304   &    $\textless$    789           &   0.6   &    $\textless   $ 327          &  0.44 &    --  &    -- &  10.48  &    --  &     -- \\
J11080148-7742288   &    $\textless$    753           &   0.6   &    $\textless   $ 366          &   0.4 &    --  &    -- &  10.48  &    --  &     -- \\
J11080297-7738425   &        624 $\pm$   188  &  0.45   &    $\textless   $ 345          &  0.12 &  0.03  &  0.48 &   0.59  &   0.1  &  10.48 \\
J11081509-7733531   &        978 $\pm$   326  &  0.75   &    $\textless   $ 357          &   0.2 &  0.05  &  4.17 &   0.91  &  0.31  &  10.48 \\
J11081850-7730408   &    $\textless$    240           &   0.6   &    $\textless   $ 135          &  0.08 &    --  &    -- &   1.05  &    --  &     -- \\
J11082238-7730277   &    $\textless$    237           &   0.6   &    $\textless   $ 132          &  0.07 &    --  &    -- &   1.05  &    --  &     -- \\
J11082570-7716396   &    $\textless$    221           &   0.6   &    $\textless   $ 126          &  0.07 &    --  &    -- &   1.05  &    --  &     -- \\
J11082650-7715550   &    $\textless$    237           &   0.6   &    $\textless   $ 135          &  0.07 &    --  &    -- &   1.05  &    --  &     -- \\
J11083905-7716042   &       1252 $\pm$   342  &  0.75   &    $\textless   $ 333          &  0.26 &  0.08  &  4.36 &   1.09  &  0.31  &  10.48 \\
J11083952-7734166   &    $\textless$    243           &   0.6   &    $\textless   $ 123          &  0.08 &    --  &    -- &   1.05  &    --  &     -- \\
J11085090-7625135   &    $\textless$    264           &   0.6   &    $\textless   $ 132          &  0.08 &    --  &    -- &   3.14  &    --  &     -- \\
J11085367-7521359   &       2634  $\pm$  527  &  1.05   &    $\textless   $ 366          &  0.89 &  0.21  &  4.78 &   1.53  &  1.05  &   3.14 \\
J11085464-7702129   &    $\textless$    765           &   0.6   &    $\textless   $ 381          &  0.41 &    --  &    -- &  10.48  &    --  &     -- \\
J11085497-7632410   &    $\textless$    267           &   0.6   &    $\textless   $ 132          &  0.08 &    --  &    -- &   3.14  &    --  &     -- \\
J11091812-7630292   &    $\textless$    801           &   0.6   &    $\textless   $ 354          &  0.45 &    --  &    -- &  10.48  &    --  &     -- \\
J11092266-7634320   &    $\textless$    822           &   0.6   &    $\textless   $ 324          &  0.48 &    --  &    -- &  10.48  &    --  &     -- \\
J11092379-7623207   &    $\textless$    723           &   0.6   &    $\textless   $ 354          &  0.37 &    --  &    -- &  10.48  &    --  &     -- \\
J11094260-7725578   &    $\textless$    230           &   0.6   &    $\textless   $ 129          &  0.07 &    --  &    -- &   1.05  &    --  &     -- \\
J11094621-7634463   &    $\textless$    816           &   0.6   &    $\textless   $ 342          &  0.47 &    --  &    -- &  10.48  &    --  &     -- \\
J11094742-7726290   &    $\textless$    765           &   0.6   &    $\textless   $ 342          &  0.41 &    --  &    -- &  10.48  &    --  &     -- \\
J11095215-7639128   &    $\textless$    227           &   0.6   &    $\textless   $ 117          &  0.07 &    --  &    -- &   1.05  &    --  &     -- \\
J11095336-7728365   &    $\textless$    227           &   0.6   &    $\textless   $ 129          &  0.07 &    --  &    -- &   1.05  &    --  &     -- \\
J11095340-7634255   &       5373 $\pm$   450  &   0.9   &    $\textless   $ 348          &  4.55 &  0.36  &104.76 &  15.98  &  3.14  & 104.76 \\
J11095407-7629253   &    $\textless$    837           &   0.6   &    $\textless   $ 351          &  0.49 &    --  &    -- &  10.48  &    --  &     -- \\
J11095873-7737088   &    $\textless$    825           &   0.6   &    $\textless   $ 330          &  0.48 &    --  &    -- &  10.48  &    --  &     -- \\
J11100010-7634578   &       1614 $\pm$   503  &  1.05   &       1581 $\pm$ 343 &  18.01 &   2.14  & 76.22 &  51.43  &  3.14  & 104.76 \\
J11100369-7633291   &    $\textless$    822           &   0.6   &    $\textless   $ 333          &  0.48 &    --  &    -- &  10.48  &    --  &     -- \\
J11100469-7635452   &    $\textless$    828           &   0.6   &    $\textless   $ 351          &  0.48 &    --  &    -- &  10.48  &    --  &     -- \\
J11100704-7629376   &    $\textless$    738           &   0.6   &    $\textless   $ 342          &  0.38 &    --  &    -- &  10.48  &    --  &     -- \\
J11100785-7727480   &    $\textless$    224           &   0.6   &    $\textless   $ 123          &  0.07 &    --  &    -- &   1.05  &    --  &     -- \\
J11101141-7635292   &    $\textless$    753           &   0.6   &    $\textless   $ 342          &   0.4 &    --  &    -- &  10.48  &    --  &     -- \\
J11103801-7732399   &    $\textless$    771           &   0.6   &    $\textless   $ 378          &  0.42 &    --  &    -- &  10.48  &    --  &     -- \\
J11104141-7720480   &    $\textless$    252           &   0.6   &    $\textless   $ 117          &  0.08 &    --  &    -- &   3.14  &    --  &     -- \\
J11104959-7717517   &    $\textless$    732           &   0.6   &    $\textless   $ 345          &  0.37 &    --  &    -- &  10.48  &    --  &     -- \\
J11105333-7634319   &    $\textless$    825           &   0.6   &    $\textless   $ 327          &  0.48 &    --  &    -- &  10.48  &    --  &     -- \\
J11105359-7725004   &        173 $\pm$    54  &  0.45   &    $\textless   $ 129          &  0.03 &  0.01  &  0.07 &   0.14  &   0.1  &   1.05 \\
J11105597-7645325   &    $\textless$    210           &   0.6   &    $\textless   $ 129          &  0.07 &    --  &    -- &   1.05  &    --  &     -- \\
J11111083-7641574   &       1913 $\pm$   440  &   0.9   &    $\textless   $ 363          &  0.42 &  0.14  &  4.75 &   1.31  &  1.05  &  10.48 \\
J11113965-7620152   &    $\textless$    762           &   0.6   &    $\textless   $ 375          &  0.41 &    --  &    -- &  10.48  &    --  &     -- \\
J11114632-7620092   &    $\textless$    855           &   0.6   &    $\textless   $ 357          &  0.51 &    --  &    -- &  10.48  &    --  &     -- \\
J11120351-7726009   &        165  $\pm$   54  &  0.45   &    $\textless   $ 126          &  0.03 &  0.01  &  0.07 &   0.13  &   0.1  &   1.05 \\
J11120984-7634366   &    $\textless$    237           &   0.6   &    $\textless   $ 123          &  0.07 &    --  &    -- &   1.05  &    --  &     -- \\
J11122441-7637064   &    $\textless$    810           &   0.6   &    $\textless   $ 366          &  0.46 &    --  &    -- &  10.48  &    --  &     -- \\
J11122772-7644223   &    $\textless$    738           &   0.6   &    $\textless   $ 342          &  0.38 &    --  &    -- &  10.48  &    --  &     -- \\
J11123092-7644241   &    $\textless$    786           &   0.6   &    $\textless   $ 351          &  0.43 &    --  &    -- &  10.48  &    --  &     -- \\
J11124268-7722230   &    $\textless$    851           &   0.6   &    $\textless   $ 345          &  0.51 &    --  &    -- &  10.48  &    --  &     -- \\
J11124861-7647066   &    $\textless$    249           &   0.6   &    $\textless   $ 123          &  0.08 &    --  &    -- &   3.14  &    --  &     -- \\
J11132446-7629227   &    $\textless$    768           &   0.6   &    $\textless   $ 375          &  0.41 &    --  &    -- &  10.48  &    --  &     -- \\
J11142454-7733062   &    $\textless$    243           &   0.6   &    $\textless   $ 117          &  0.08 &    --  &    -- &   1.05  &    --  &     -- \\
J11160287-7624533   &    $\textless$    753           &   0.6   &    $\textless   $ 363          &   0.4 &    --  &    -- &  10.48  &    --  &     -- \\
J11173700-7704381   &       2170 $\pm$   476  &  1.05   &    $\textless   $ 351          &  0.48 &  0.16  &  4.59 &   1.35  &  1.05  &   3.14 \\
J11175211-7629392   &    $\textless$    212           &   0.6   &    $\textless   $ 123          &  0.07 &    --  &    -- &   1.05  &    --  &     -- \\
J11183572-7935548   &        164 $\pm$    50  &  0.45   &    $\textless   $ 110          &  0.03 &  0.01  &  0.07 &   0.13  &   0.1  &   1.05 \\
J11241186-7630425   &    $\textless$    237           &   0.6   &    $\textless   $ 123          &  0.07 &    --  &    -- &   1.05  &    --  &     -- \\
J11432669-7804454   &    $\textless$    246           &   0.6   &    $\textless   $ 120          &  0.08 &    --  &    -- &   3.14  &    --  &     -- \\
\enddata
\tablecomments{The aperture radii for $^{13}$CO flux calculations are listed in the third column. Upper limits for C$^{18}$O fluxes are calculated from the  $0\farcs3$ radius apertures, with the only exception, 2MASS J11100010-7634578, adopted with a $0\farcs75$ radius aperture. }
\end{deluxetable}

\newpage

\section{Corner plots of fit} \label{sec:corner_plot}
We show in Figure \ref{fig:corner} the corner plots, with slope, the intercept and intrinsic scatter for the fit of the $\log{F_{13CO}}$ vs. $\log{M_*}$ relation (see \S \ref{sec:co flux relation} and Figure \ref{fig:Fco_vs}). This is done using model $F_{13CO}=\beta\cdot M_* ^\alpha$, with non-detections applying calculated fluxes instead of 3$\sigma$ upper limits.

\setcounter{figure}{0}
\renewcommand{\thefigure}{C\arabic{figure}}

\begin{figure}[h]
    \centering
        \includegraphics[scale=0.8]{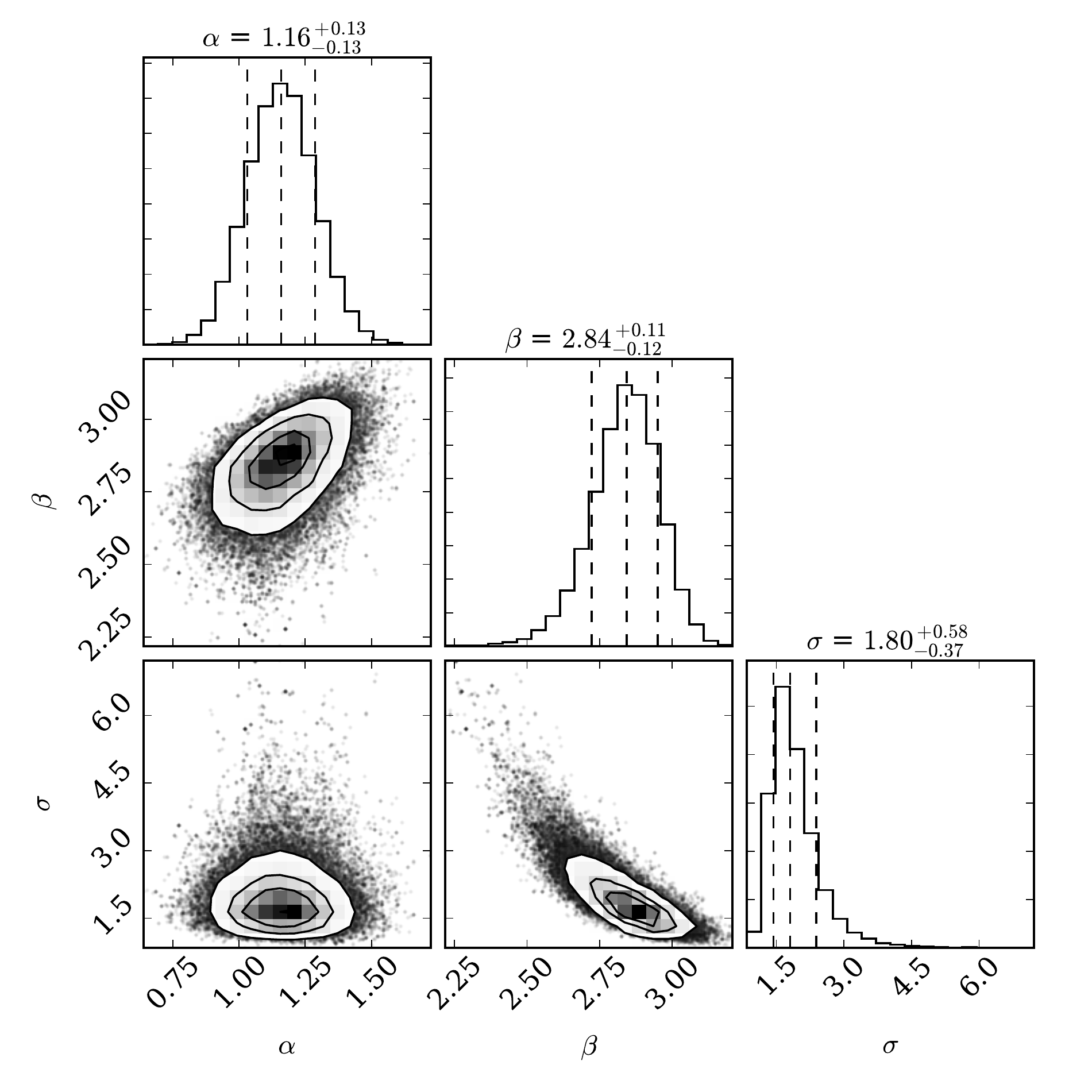}
\caption{Corner plot of the $\log{F_{13CO}}$ vs. $\log{M_*}$ fit with model $F_{13CO}=\beta\cdot M_* ^\alpha$. \label{fig:corner}}
\end{figure}

\newpage

\section{Line Luminosity versus Disk mass} \label{sec:MvD16_fit}
With the model grids from MvD16, we fit simple functions for line luminosities and disk gas masses using the median $^{13}$CO and C$^{18}$O J = 3 - 2 line luminosities in each gas mass bin. Similar to the methods of \citet{miotello2016,miotello2016b}, We fit a linear relation for line luminosity and gas mass in the low mass regime and a logarithmic relation in the higher mass end.  Compared to the recent analysis of \citet{miotello2016b}, published during the final phases of preparation of this paper, we fit a linear relation in the log-log plane and choose a different transition mass.  The fitting coefficients are listed in Table \ref{tab:mass_fit}. We calculate gas masses for our $^{13}$CO detections using our fitted coefficients and the values provided in \citet{miotello2016b}, in which models with disk inclinations of 10$\degr$ and 80$\degr$ are fitted separately. As shown in Figure \ref{fig:fit_comparison}, the derived gas masses are almost consistent.  In most cases, the gas mass derived using our fitted functions should be reduced by factors of 1.6 (for models with inclination angle 10$\degr$) and 1.2 (for 80$\degr$) for consistency with the results in \citet{miotello2016b}.  The discrepancies between the two methods are slightly larger for the few brightest sources in our sample.



\setcounter{figure}{0}
\renewcommand{\thefigure}{D\arabic{figure}}

\begin{figure}[h]
    \centering
        \includegraphics[width=0.95\textwidth]{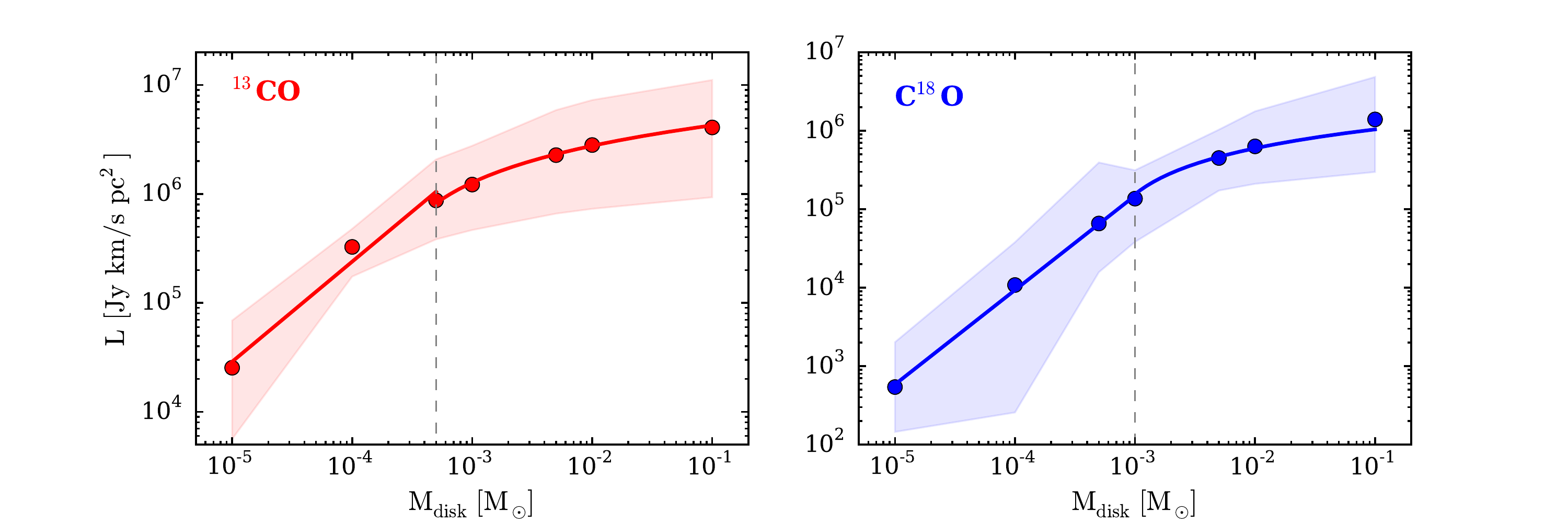}
\caption{Fitting the median line luminosity in each disk mass bin with disk gas mass for $^{13}$CO (Left) and C$^{18}$O(Right).  The transition masses for $^{13}$CO and C$^{18}$O fitting functions are shown in dash gray lines.  The data used here were obtained from \citet{miotello2016}. \label{fig:MvD16_fit}}
\end{figure}

\setcounter{figure}{1}
\renewcommand{\thefigure}{D\arabic{figure}}

\begin{figure}[h]
    \centering
        \includegraphics[width=0.7\textwidth]{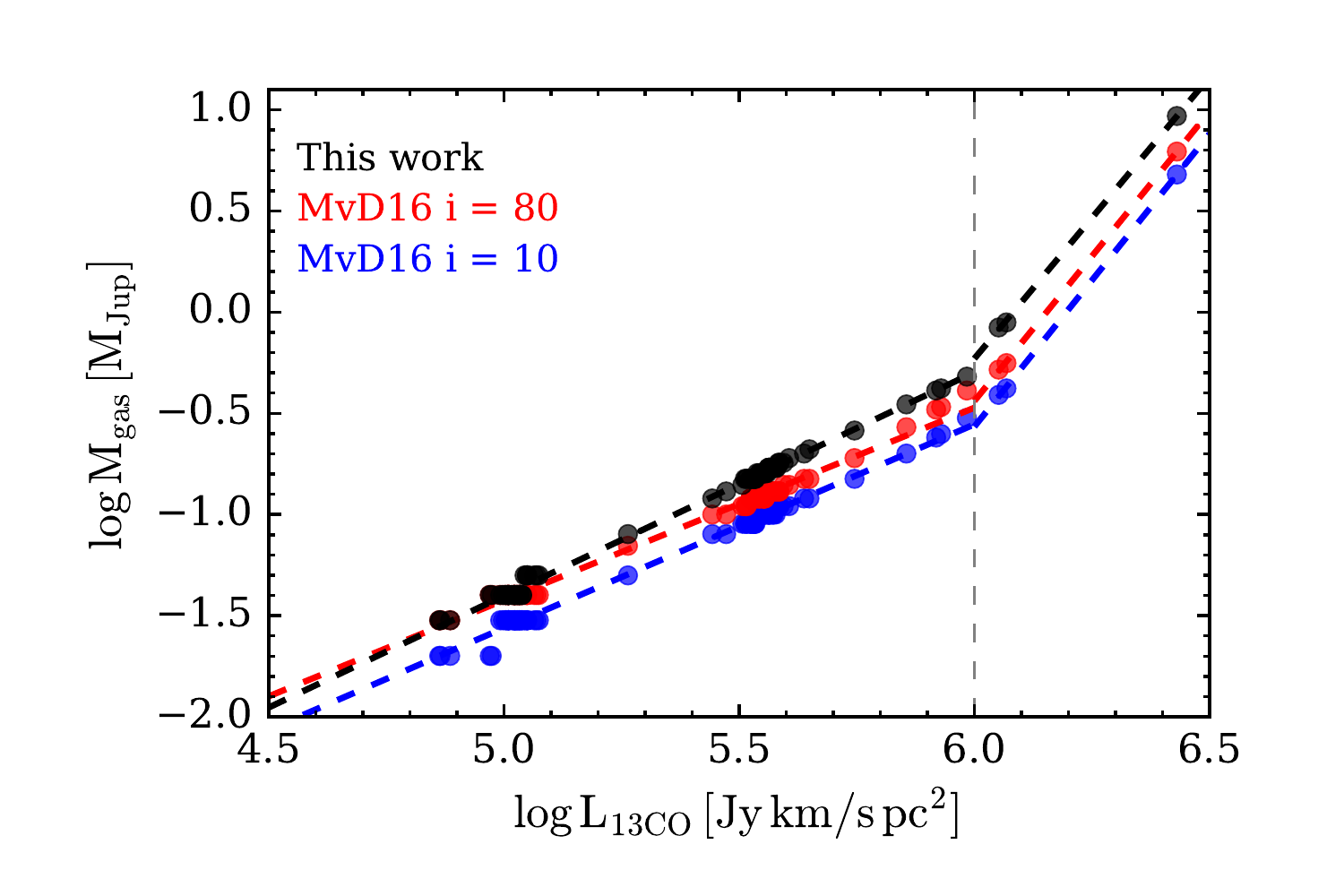}
\caption{The comparison of gas masses for our Cha I sample, derived using different fitting functions. Scaling relation breaks at the $^{13}$CO luminosity $\sim 10^6 L_{\odot}$.\label{fig:fit_comparison}}
\end{figure}

\begin{deluxetable}{rccccc}
\tablecaption{Coefficients in line luminosity and gas mass fitting\label{tab:mass_fit}}
\tablehead{
\colhead{} & \colhead{A} & \colhead{B} & \colhead{C} & \colhead{D} & \colhead{M$_{tr}$[M$_{\odot}$]} }
\startdata
$^{13}$CO   & 9.054 & 0.919 & 5.765$\cdot$10$^6$ & 1.498$\cdot$10$^6$ &  5$\cdot$10$^{-4}$ \\
C$^{18}$O   & 8.783 & 1.203 & 1.484$\cdot$10$^6$ & 4.428$\cdot$10$^5$ & 1$\cdot$10$^{-3}$ \\
\enddata
\tablecomments{The fitting functions between $^{13}$CO and C$^{18}$O line luminosity and disk gas mass using model data in MvD16: log(L$_{CO}$)= A + B$\cdot$log(M$_{gas}$) when M$_{gas}$ $\leq$  M$_{tr}$; L$_{CO}$= C + D$\cdot$log(M$_{gas}$) when M$_{gas}$ $\geq$ M$_{tr}$}
\end{deluxetable}

\end{CJK*}
\end{document}